\documentclass[11pt]{article}

\usepackage[utf8]{inputenc}
\usepackage{mathtools}       
\usepackage{graphicx}
\usepackage[colorinlistoftodos]{todonotes}
\usepackage[margin=0.9in]{geometry}
\usepackage{multicol}
\usepackage{graphicx} 
\usepackage{amssymb}
\usepackage{amsmath}
\usepackage{mathrsfs}
\usepackage{physics}
\usepackage{slashed}
\usepackage{hyperref}
\usepackage{cite}

\newcommand{\half}{{\tfrac{1}{2}}}
\newcommand{\wt}{\widetilde}
\newcommand{\NS}{{\rm NS}}
\newcommand{\RR}{{\rm R}}
\newcommand{\G}{{\cal G}}
\newcommand{\HH}{{\cal H}}

\newcommand{\XX}{{\cal X}}
\newcommand{\hd}{{\boldsymbol{\cdot} }}

\newcommand{\ben}{\begin{eqnarray}\displaystyle}
	\newcommand{\een}{\end{eqnarray}}
\newcommand{\bea}{\begin{eqnarray}}
	\newcommand{\eea}{\end{eqnarray}}

\newcommand{\be}{\begin{equation}}   
	\newcommand{\ee}{\end{equation}}
  
\newcommand{\refb}[1]{(\ref{#1})}   
\def\d{{\rm d}}  
\def\e{{\epsilon}}

\newcommand{\sectiono}[1]{\section{#1}\setcounter{equation}{0}}

\begin{document}
	\begin{titlepage}
		\rightline{}
		\rightline\today 
		\rightline{MIT-CTP/5865} 
		\begin{center}
			\vskip 1.1cm
			
			{\Large \bf {Type II RR string fields and exotic diffeomorphisms}}\\

			\vskip 1.5cm
			
			{\large\bf {Raji Ashenafi Mamade and Barton Zwiebach}}
			\vskip 1cm

			\vskip .3cm
			
			{\it   Center for Theoretical Physics -- a Leinweber Institute \\
				Massachusetts Institute of Technology, \\
				Cambridge MA 02139, USA}\\
			\vskip .1cm
			
			\vskip .4cm
			raji@mit.edu, zwiebach@mit.edu

			\vskip 1.9cm
		\end{center}

		\begin{quote} 	
			
			\centerline{\bf Abstract} 
			\bigskip
			
			\medskip
			
We study  the  theory of massless fields of type II 
strings arising from the string field theory that uses two string fields, a physical one and an extra one that allows the writing of an action, but whose degrees of freedom ultimately decouple.   The mechanism allowing the description of the 
self-dual five-form of type IIB, anticipated by Sen, is used by the SFT to describe all Ramond-Ramond forms in type IIB and IIA in a manifestly duality-invariant way.  We find explicit expressions for the leading terms in the gauge transformation of the RR fields 
and focus on diffeomorphisms, which are exotic for both the physical and the extra fields, perhaps as needed to describe propagating degrees of freedom that do not gravitate.  The algebra of diffeomorphisms includes field-dependent structure
constants and only closes on-shell, as predicted by the type II SFT gauge algebra.
			
		\end{quote} 
		\vfill
		\setcounter{footnote}{0}
		
		\setcounter{tocdepth}{2}  
		
	\end{titlepage}

	\baselineskip 15pt

	\tableofcontents
	
	
	
	\sectiono{Introduction and summary}
	
	Type IIB supergravity has a four-form gauge potential whose five-form
	field strength is self dual.  It has been long known that there is
	no simple Lorentz 
	 invariant action from which the equations
	of motion for this field can be derived~\cite{Marcus:1982yu}.
	This complication was seen as
	an obstruction to the possible construction of a gauge-invariant
	Type IIB superstring
	field theory.  An intriguing solution to this problem was given by
	Sen~\cite{Sen:2015uaa}
	who, in addition to the expected string field, introduced an extra
	string field with non-conventional picture number 
	that helps write an action that reproduces the correct equations of motion.  
	The extra string field describes a full copy of the states of physical
	string, but these states decouple from the states of the physical string, whose 
	scattering amplitudes are the correct ones.  The extra string field appears
	only quadratically in the action, and the associated propagating but 
	non-interacting degrees of freedom have negative norms.   For references to 
	other approaches to the formulation of type II SFT, see the 
	recent review~\cite{Sen:2024nfd}.

	The new string field theory for type II has been studied in some detail~\cite{Sen:2017szq}, and has been used for some  
	computations~\cite{Cho:2018nfn,Alexandrov:2021shf,Alexandrov:2021dyl}.
	Perhaps surprisingly, not even the free field theories have been written down at the component level in the literature. 
	Instead, anticipating the mechanisms at play in the string field theory,  an effective action for the self-dual five form showing how an extra four-form gauge field allows for a Lorentz invariant action was given in~\cite{Sen:2015nph}.  The degrees of freedom in the four-form decouple.   The construction was used to give a 
	covariant action for type IIB supergravity.  This mechanism was further applied to other chiral theories as well, and its various properties including dimensional reduction and partition function have been studied ~\cite{Sen:2019qit, Andriolo:2021gen, Vanichchapongjaroen:2024tkj, Evnin:2022kqn, Hull:2023dgp, Phonchantuek:2023iao, Barbagallo:2022kbt, Chakrabarti:2022jcb, Andriolo:2023ltv}.
	
	From the viewpoint of the string field theory, a few facts are puzzling.  While the
	degrees of freedom carried by the extra fields ultimately decouple\footnote{This decoupling is seen
		in the gauge-fixed version of the theory; the decoupling fields being a linear combination of the extra fields and the physical fields.}, the quadratic terms
	mix them with the physical fields.  As a result the gauge transformations of the extra fields are as nontrivial as those of the physical ones.  Since the extra degrees of freedom decouple from physical fields, they decouple from gravity. Nevertheless, diffeomorphism invariance-- a part of the gauge invariance of the string field theory-- holds exactly.
	This statement appears to be in tension with the understanding that all degrees of freedom gravitate in a theory that includes general relativity.

Diffeomorphisms in string field theory have been studied for a while but
still remain somewhat mysterious.  The associated linearized
transformations of the
graviton  were seen as part of the gauge symmetry
of the free closed string field theory~\cite{Siegel:1985tw}.  
		It was later shown by~\cite{Ghoshal:1991pu} 
		that using redefinitions of the SFT gauge parameters
		one could in principle 
		recover the standard diffeomorphism algebra, 
		when including cubic terms in the action.  
			This was carried out explicitly in the context of 
		double field theory~\cite{Hull:2009mi} to leading order in derivatives, and
		leading $\alpha'$ corrections  to the algebra of
		gauge transformations were further studied in~\cite{Hohm:2014xsa}. 
		There are no results beyond cubic order in fields. 
		A recent investigation~\cite{Mazel:2025fxj}
		set up a formalism to achieve a
		manifest description of diffeomorphisms in bosonic
		string theory.
		
		 In our companion
		paper~\cite{RMBZ} we derived the gauge algebra of type II SFT 
		and computed explicitly the algebra of diffeomorphisms, an off-shell
		calculation that includes 
		insertions of picture-changing operators. 
		 To leading order in derivatives, the type II
		algebra of gauge transformations does not involve the off-shell 
		data of the three-string vertex, and is thus {\em universal}.   The algebra, as expected,
		also includes field-dependent structure constants and trivial gauge transformations-- those for which the variation of the field vanishes on-shell. 
		
		A focus in this paper is the transformations of the fields under
		diffeomorphisms.  These are exotic:  even to leading order they do not coincide with Lie derivatives,
		a fact that was clear in the description of the self-dual five form 
		in~\cite{Sen:2015nph}.  In fact, the transformation of {\em all} the physical
		RR forms, to leading order, is not a Lie derivative.  The transformation
		of the extra fields is also startling:  to leading order they transform into the 		physical forms.
		These unusual transformations produce a general coordinate invariance in a theory in which some propagating degrees of freedom do not couple to gravity!\footnote{Less exotic, but still novel diffeomorphisms have appeared in double field theory, where transformations use generalized Lie derivatives that result in a gauge algebra defined by the Courant bracket~\cite{Hohm:2010pp}.}

\medskip  
	
	The purpose of this paper is to further investigate the II superstring field theory, focusing on the RR sector and its couplings to the NSNS fields.   We have done as follows:
	
	\begin{enumerate}
		
		\item   We wrote the 
		general massless RR string fields and  evaluated the quadratic terms in the action. After a set of field redefinitions we obtain the action \refb{quadSRRIIB}
		in a simplified form where the five-form field is described in the way anticipated by~\cite{Sen:2015nph}.  We carry the computations for the case of IIB, and give the results for IIA.  
		\item  We also give the quadratic action for the NSNS massless 
		sector of the type II theories (\S\ref{kinnsnsn}). 
		This action coincides with the quadratic action of bosonic string field theory.
		
		\item While other RR form fields admit conventional actions in field theory, the string field theory describes  them with the same mechanism used for the self-dual five-form.  In fact, for type IIB, while the naive kinetic term $(Q^{(5)} , Q^{(5)})$ vanishes because of 
		self-duality of $Q^{(5)}$, for the one-form and its dual nine-form, for example,
		the combination $(Q^{(1)} , Q^{(1)})+ (Q^{(9)} , Q^{(9)})$ also vanishes when using the  duality relations.
By including {\em all} forms in the action,  described with the extra-field mechanism and not the naive terms shown above, the duality relations are manifest.  This quadratic action is shown in equation~\refb{finaction}. 
 Such action is in the spirit of the so-called `democratic' formulation of type II theories, where forms and their duals are included in the action~(see, for example, \cite{Hohm:2011dv}).
		
		\item We use the SFT to compute cubic interactions.  We do not examine the NSNS cubic terms, that are expected to be similar to those of bosonic strings when truncated to two derivatives.   We determine all IIB cubic couplings of the massless 
		NSNS fields to two RR fields, to leading order in derivatives.  All terms we've calculated for the IIB action are collected in~\refb{finXXXXaction}.   For the IIA action the corresponding result is in~\refb{finactionA}.

		\item  We use the SFT to compute the gauge transformations of
		Ramond Ramond fields.  The RR sector gauge parameters vanish for the massless physical fields,
		which use the language of field strengths.  The gauge parameters 
		associated to the extra RR fields do not vanish but generate expected inhomogeneous
		transformations.
		We focus on the nontrivial gauge transformations induced by NSNS gauge parameters.
		These transformations affect both the physical and extra RR fields, have no field-independent terms, and rotate the extra RR fields into the physical RR fields.  
		Our results, for the case of the self-dual five-form, agree with those anticipated
		in~\cite{Sen:2015nph}.

		\item  The RR physical forms and extra fields all have exotic transformations under diffeomorphisms: they do not transform with Lie derivatives, as one would have in familiar formulations (the Kalb-Ramond transformations are also unusual).  

\item 
		Given the exotic diffeomorphisms of the RR fields, we compute the algebra of these transformations (using some partial information from the 
		effective field theory~\cite{Sen:2015nph}).   In the resulting gauge algebra~\refb{galgeffft}, in addition to the expected Lie bracket, we see terms corresponding to field dependent structure constants as well as trivial gauge transformations, in full agreement with the
		SFT gauge algebra computation.  By field redefinitions of the diffeomorphism parameters we confirm that the somewhat different
		 bracket of the type II SFT turns into the
		Lie bracket.

\item 
We establish, to first order in the metric fluctuation around flat space, 
the relation between the RR physical fields $Q^{(k)}$
of the IIB SFT and    
 the conventional supergravity field strengths $F^{(k)}$, that arise
from potentials (see~\refb{redefs} 
and~\refb{redefsFQ}).  In this way we extend the result of~\cite{Sen:2015nph} which gave
this relation for the case of the five forms.		
		
\end{enumerate}

The calculations in this paper involve a number of conventions and identities 
needed to get results with reliable signs and multiplicative constants. 
Many of those are given in \S\ref{230rhf}, which discuss Grassmanality
of operators, GSO projections, spin fields, and OPE's, among others.   When treating
differential forms we use an inner product in the space of forms to build actions.
In order to manipulate efficiently the relevant objects, in addition to the exterior
derivative $\d$ and its adjoint $\d ^\dagger$,  we introduce the adjoint $i_X^\dagger$ 
of the contraction operator $i_X$, with $X$ a vector field as well as the adjoint
${\cal L}_X^\dagger$ of the Lie derivative ${\cal L}_X$.  Moreover,  we have defined
a degree-zero action of a symmetric tensor on forms (\refb{haction}) which turns
to have interesting identities that simplify a number of computations. When not explicitly
stated we follow the conventions of~\cite{Polchinski:1998rr}.  We have also included
a number of technical appendices.

	\sectiono{Type II SFT}
	
	In this section we begin by reviewing the structure of type II string theory
	in the framework~\cite{Sen:2015uaa} that uses two string fields;
	a field $\Psi$ with conventional picture numbers, and a field $\wt\Psi$
	with unusual picture numbers in all sectors except the NSNS sector.  We quickly
	restrict ourselves to the NSNS and RR sectors of the theory, thus focusing 
	on the spacetime bosons,  and discuss the
	expansion of the action up to cubic order in string fields.  
	We then turn to
	the gauge algebra, which we compute and later discuss how
	it applies for the NSNS and RR gauge parameters.  We conclude with some
	discussion of technicalities (GSO projections, Grassmanality, inner product, 
	picture changing operators, OPE's for spin fields, differential forms) needed for the explicit computation of the action.   
	
	\subsection{Action and gauge transformations}

	The type II string field theory makes use of
	picture $(-1, -1)$ states for the NSNS sector, and 
	of {\em both} picture $(-{1\over 2}, - {1\over 2})$ and  
	$(-{3\over 2} - {3\over 2})$ states in the RR sector. 
	With $\HH_{p,q}$    
	denoting the space of string states of anti-holomorphic picture number $p$ and 
	holomorphic picture number $q$ the string field theory has     two  
	string fields, $\Psi$ and $\wt\Psi$, each a direct sum over the four
	sectors of the theory: NSNS, NSR, RNS, and RR.  We have
	\be
	\label{sschisa}
	\begin{split} 
		\hbox{Type II string fields:} \  \ 
		& \Psi\in \HH_c \equiv \HH_{-1,-1}\oplus \HH_{-1,-{1\over 2}}\oplus \HH_{-{1\over 2},-1}\oplus \HH_{-{1\over 2},-{1\over 2}}, \\[0.6ex] 
		& \wt\Psi\in \wt\HH_c \equiv
		\HH_{-1,-1}\oplus \HH_{-1,-{3\over 2}}\oplus \HH_{-{3\over 2},-1}\oplus \HH_{-{3\over 2},-{3\over 2}}\, .
	\end{split}
	\ee
	The BPZ inner product pairs states in $\HH_c$ and $\wt\HH_c$.  
	Both $\Psi$ and $\wt\Psi$ are Grassmann even and have ghost number two.  
	As usual for closed strings, they
	are both annihilated by $L_0^-= L_0- \bar L_0$ and $b_0^- =b_0 - \bar b_0$.   The classical action
	is  
	\be\label{ebvii}
	S=- \tfrac{1}{2}  \langle \wt \Psi ,  
	Q \, \G \, \wt\Psi\rangle  
	+ \langle \wt \Psi ,   
	Q \, \Psi\rangle +\sum_{n=3}^\infty {1\over n!} \{\Psi^n\}\, .
	\ee
	Here $\langle \cdot \,, \cdot \rangle$ is the closed string inner product, the BPZ inner product with a $c_0^-$ insertion, $Q$ is the BRST operator, and $\{ \Psi^n\}$ is a tree-level multilinear function, representing interactions.   Zero modes of the picture changing operators
	$\XX(z)$ and $\bar\XX (\bar z)$ are defined by 
	\be
	\label{zmpcos}
	\XX_0=\ointop {dz\over z} \XX(z), \qquad \bar\XX_0=\ointop {d\bar z\over \bar z} \bar\XX(\bar z)\, , 
	\ee
	with the contour integrals normalized as $\oint {dz/z} = 1$ and $\oint {d\bar z/\bar z} = 1$. The operator $\G:  \wt\HH_c \to \HH_c$ by changing the picture number  is defined by:
	\be
	\label{Gdefpc}
	\G\equiv \begin{cases} \hbox{${\bf 1}\ $   on $\HH_{-1,-1}$}\, ,\cr \hbox{$\XX_0$ on $\HH_{-1,-3/2}$}\, ,
		\cr \hbox{$\bar\XX_0$ on $\HH_{-3/2,-1}$}\, ,\cr \hbox{$\XX_0\bar\XX_0$ on $\HH_{-3/2,-3/2}$.}
	\end{cases}
	\ee
	Note that $\wt\Psi$ 
	describes a free field, it does not appear in the interactions.  The consistency of 
	the action requires the following vanishing commutators, which hold:
	\be
	[Q, {\cal X}_0 ] =  [Q, \bar {\cal X}_0 ] = 0 \,,  \ \ \
	[b_0^\pm , \mathcal{X}_0 ] =  [b_0^\pm , \bar{\cal X}_0 ] = 0\,. 
	\ee
	The gauge symmetry of the classical theory, with Grassmann odd gauge parameters $(\Lambda, \wt\Lambda) \in (\HH_c, \wt\HH_c)$ at ghost number one are given by the transformations
	\be
	\label{gaugetrsft}
	\begin{split}
		\delta_{\Lambda, \wt\Lambda}  |\widetilde\Psi\rangle  =& \  Q |\widetilde \Lambda\rangle  + \sum_{n=1}^\infty  {1\over n!}  \, [ \Lambda \Psi^n ]\,, \\ 
		\delta_{\Lambda, \wt\Lambda}  |\Psi\rangle  =& \  Q | \Lambda\rangle  + \sum_{n=1}^\infty  {1\over n!}  \, {\cal G} \,  [ \Lambda \Psi^n ]\,. 
	\end{split}
	\ee
	The string field products here are all genus zero, as befits the classical action. 
	
	\bigskip
	Our focus in this paper is on the NSNS and RR sectors
	of the theory.  To keep the notation from being cumbersome,
	on string fields and gauge parameters 
	we use subscripts R and NS for the RR and NSNS sectors, respectively.
	Since subtleties do not occur in the NSNS sector, it is possible to set
	\be 
	\label{wtnsns}
	\wt\Psi_{{}_\NS} =\Psi_{{}_\NS}  \,, 
	\ee 
	without changing the interacting part of the theory. 
	Thus the string fields will take the form 
	\be
	\label{sschisaX}
	\begin{split} 
		& \Psi = \Psi_{{}_\NS} + \Psi_{{}_\RR} \in  \HH_{-1,-1} \oplus \HH_{-1/2,-1/2}, \\[0.6ex] 
		& \wt\Psi = \Psi_{{}_\NS} + \wt\Psi_{{}_\RR}    \in
		\HH_{-1,-1}\oplus  \HH_{-3/2,-3/2}\, .
	\end{split}
	\ee
	We do not disturb the equality~\refb{wtnsns} when choosing $\wt\Lambda_{{}_\NS}  =\Lambda_{{}_\NS}$.  Consistent with this, the gauge parameters decompose as follows
	\be
	\label{sschisaXY}
	\begin{split} 
		& \Lambda = \Lambda_{{}_\NS} + \Lambda_{{}_\RR} \in  \HH_{-1,-1} \oplus \HH_{-1/2,-1/2}, \\[0.6ex] 
		& \wt\Lambda =  \Lambda_{{}_\NS} + \wt\Lambda_{{}_\RR}    \in
		\HH_{-1,-1}\oplus  \HH_{-3/2,-3/2}\, .
	\end{split}
	\ee

	It follows from the expression for the action $S$, the definition of ${\cal G}$, and
	the above expansions of $\Psi$ and $\wt\Psi$ that the quadratic action $S_2$ plus the cubic action $S_3$,  restricted to NSNS and RR sectors is
	given by
	\be 
	\label{S2+S3} 
	\begin{split}
		S_2 + S_3 = & \ - \tfrac{1}{2}\, \langle \widetilde \Psi_{{}_\RR} , Q {\cal X}_0\Bar{{\cal X}}_0 \widetilde\Psi_{{}_\RR}  \rangle + \langle \Psi_{{}_\RR} , Q \widetilde  \Psi_{{}_\RR}  \rangle
		+ \tfrac{1}{2}\langle \Psi_{{}_\NS} , Q   \Psi_{{}_\NS}  \rangle\\[1.5ex]
		& \  + \tfrac{1}{3!}  \{  \Psi_{{}_\NS} , \Psi_{{}_\NS} , \Psi_{{}_\NS}  \}   
		+   \tfrac{1}{2!}  \{  \Psi_{{}_\NS} , \Psi_{{}_\RR} , \Psi_{{}_\RR}  \} \,. 
	\end{split}
	\ee
	The quadratic terms defining $S_2$ are on the first line, the cubic terms defining $S_3$
	are on the second line.  The action $S_2$ has linearized gauge transformation that can be read from~\refb{gaugetrsft}:
	\be
	\label{lingt3948r}
	\delta \Psi_{{}_\RR}  = \  Q \Lambda_{{}_\RR} \,,  \hskip20pt 
	\delta \widetilde\Psi_{{}_\RR}  = \  Q \widetilde \Lambda_{{}_\RR}, \hskip20pt  
	\delta \Psi_{{}_\NS}  = \  Q \Lambda_{{}_\NS} \,.
	\ee
	The gauge parameter $\Lambda_{{}_\NS}$ 
	includes the parameters for diffeomorphisms. 
	The linearized equations of motion obtained by varying $\tilde\Psi_{{}_\RR}$,  $\Psi_{{}_\RR}$, 
	and $\Psi_{{}_\NS}$ in $S_2$ are, respectively, 
	\be
	\begin{split}
		Q {\cal X}_0\Bar{{\cal X}}_0 \widetilde \Psi_{{}_\RR} -  Q \Psi_{{}_\RR} = \,  0\,, \ \ \ \ 
		Q \wt \Psi_{{}_\RR} = \,   0\, , \ \ \ \   Q \Psi_{{}_\NS} = \,   0\, .
	\end{split}
	\ee
	The second equation implies that the first term in the first equation vanishes. The 
	linearized equations of motion are, equivalently, 
	\be \label{eom}
	\begin{split}
		Q \Psi_{{}_\RR} = 0\,, \ \ \ \  Q \wt \Psi_{{}_\RR} = 0\,, \ \ \ \ Q  \Psi_{{}_\NS} = 0\,.
	\end{split}
	\ee
	These equations, together with the gauge invariances~\refb{lingt3948r} show that the
	spectrum is given by the BRST cohomology classes in $\Psi_{{}_\RR} , \widetilde\Psi_{{}_\RR}$,
	and $\Psi_{{}_\NS}$.  As it is well-known, the RR spectrum is doubled.

	\bigskip
	
	The string products are only relevant to the string fields $\Psi$ and $\Lambda$ -- no such product
	ever involves a $\wt\Psi$ or $\wt\Lambda$ field.  
	In order to apply the definition~\refb{Gdefpc} of the operator
	${\cal G}$  concretely one must know 
	the picture number of the string products.  For a product  with $n$ inputs 
	and assuming the fields are either NSNS or RR and we write
	$n = n_{{}_\NS} + n_{{}_\RR}$.  Those NSNS and RR string fields are of 
	pictures $(-1,-1)$ and $(-\half, -\half)$, respectively. 
	
	The picture numbers of products 
	are defined such that inner products (or multilinear functions) involving an arbitrary
	number of NSNS fields and an even number of RR fields can be nonzero. 
	Thus, depending if $n_\RR$ is odd or even, we have nonvanishing 
	\be
	\bigl\langle \Psi_{{}_\RR} \, ,  \bigl[ \, \underbrace{\Psi_{{}_\NS} , \cdots , \Psi_{{}_\NS } }_{n_{{}_\NS}} \,, 
	\underbrace{  \Psi_{{}_\RR} , \cdots, \Psi_{{}_\RR} }_{n_{{}_\RR} \in \mathbb{Z}_{\rm odd} } \bigr]
	\bigr\rangle \,, \  \  \ \hbox{and} \ \  \   
	\bigl\langle \Psi_{{}_\NS} \, ,  \bigl[ \, \underbrace{\Psi_{{}_\NS} , \cdots , \Psi_{{}_\NS } }_{n_{{}_\NS}} \,, 
	\underbrace{  \Psi_{{}_\RR} , \cdots, \Psi_{{}_\RR} }_{n_{{}_\RR} \in \mathbb{Z}_{\rm even} } \bigr]
	\bigr\rangle\,. 
	\ee
	Since the total picture number in correlators must be $(-2, -2)$, the picture 
	numbers of the products must be
	\be
	\label{picinsertionX}
	(p, \bar p)  \bigl(\, \bigl[ \, \underbrace{\Psi_{{}_\NS} , \cdots , \Psi_{{}_\NS } }_{n_{{}_\NS}} \,, 
	\underbrace{  \Psi_{{}_\RR} , \cdots, \Psi_{{}_\RR} }_{n_{{}_\RR}} \bigr] \bigr) =  
	\begin{cases}  (-\tfrac{3}{2}, -\tfrac{3}{2}) 
		\,, \ \ n_{{}_\RR} \in \mathbb{Z}_{\rm odd} \, ,  \\[1.0ex]
		(-1, -1)  \,, \ \  \ n_{{}_\RR} \in \mathbb{Z}_{\rm even} \, . \end{cases}  
	\ee
	It now follows that ${\cal G} = {\cal X}_0 \bar{\cal X}_0$ when acting on a product with
	an odd number of RR fields, and ${\cal G}= 1$ otherwise.  
	For convenience we will define
	\be
	{\cal G}_0  \equiv  {\cal X}_0 \bar{\cal X}_0\,. 
	\ee
	Finally, knowing the picture number of a product tells us how many units of picture number $\Delta p = \Delta \bar p$ must
	be supplied by explicit insertion of PCO's.  We quickly see that   
	\be
	\Delta p  ( [ \Psi_{{}_\NS}, \Psi_{{}_\NS}]) = 1 \,, \ \ \ 
	\Delta p  ( [ \Psi_{{}_\NS}, \Psi_{{}_\RR}]) =  \Delta p  ( [ \Psi_{{}_\RR}, \Psi_{{}_\RR}]) = 0 \,.
	\ee
	Only the two-product of NSNS fields requires an explicit insertion of a picture changing operator.  More generally, one can verify that $ \Delta p ( n_{{}_\NS} ,   n_{{}_\RR} )  =  -1 + n_{{}_\NS} + \lfloor \tfrac{n_{{}_\RR}}{2} \rfloor $, where the floor function gives the largest integer less than or equal to the argument.

	\subsection{Gauge algebra and expansions}

	The gauge algebra of the type II SFT has been evaluated in~\cite{RMBZ},
	using~\cite{Firat:2024dwt} to deal with the two-field situation.  To review
	the result, we must give some definitions. 
	One defines primed products involving the string field $\Psi$ (and not $\wt\Psi$),  
	and ${\cal E}$, which is a linear combination of the classical equations
	of motion for $\Psi$ and $\wt\Psi$ and thus vanishes on shell:
	\be
	\label{primedproducts2049u2}
	\begin{split}
		[A_1,\cdots,A_n]' \equiv  & \ \sum_{p=0}^{\infty}\tfrac{1}{p!}[A_1,\cdots,A_n,\Psi^p]\,, \ \ n \geq 1 \,, \\
		{\cal E} \equiv & \  Q\Psi +\sum_{n=2}^{\infty}\tfrac{1}{n!} \, {\cal G} [\Psi^{n}]
		\,. 
	\end{split}
	\ee
	As shown in~\cite{RMBZ} the commutator of gauge transformation 
	is a gauge
	transformation plus terms vanishing on-shell
	\be \label{gaugealgebraSFT}
	\begin{split}
		& [\delta_{\Lambda_2, \widetilde \Lambda_2} \, , 
		\delta_{\Lambda_1, \widetilde \Lambda_1}]\, \Psi 
		=\  \delta_{\Lambda_{12}, \widetilde \Lambda_{12}} \Psi + \mathcal{G}[\Lambda_1,\Lambda_2,  \mathcal{E}]'\,, \\[1.0ex]
		& [\delta_{\Lambda_2, \widetilde \Lambda_2} \, , 
		\delta_{\Lambda_1, \widetilde \Lambda_1}]\, \wt\Psi = \ 
		\delta_{\Lambda_{12}, \widetilde \Lambda_{12}} \wt\Psi +[\Lambda_1,\Lambda_2,\mathcal{E}]'\,. 
	\end{split}
	\ee
	The gauge parameters on the right-hand side are given by  
	\be
	\label{galgori}
	\Lambda_{12} = \mathcal{G}[\Lambda_1, \Lambda_2]' \,, \ \ \ 
	\wt\Lambda_{12} = [\Lambda_1, \Lambda_2]' \,.
	\ee
	A few facts are noteworthy:
	\begin{itemize}
		\item The parameters $\Lambda_{12}, \wt \Lambda_{12}$ 
		are field dependent beyond their leading
		field-independent terms. 
		\item  The gauge algebra bracket is encoded in $[\Lambda_1, \Lambda_2]' $:  it equals $\wt\Lambda_{12}$ and acted by ${\cal G}$ gives $\Lambda_{12}$. 
		\item  
		$\wt \Lambda_1$ and $\wt \Lambda_2$ in $ \delta_{\Lambda_1, \widetilde \Lambda_1}$ and $ \delta_{\Lambda_2, \widetilde \Lambda_2}$
		do not appear in the expressions
		for $\Lambda_{12}$ and $\wt \Lambda_{12}$. 
	\end{itemize}
	Separating out the field independent part of the gauge parameters $\Lambda_{12}$ and 
	$\wt\Lambda_{12}$ we have, over the  
	NSNS and RR sectors,  
	\be
	\label{leadingalge} 
	\begin{split}
		\Lambda_{12,\NS}
		=    & \  [\Lambda_{1,\NS}\,, \, \Lambda_{2,\NS} ] + [\Lambda_{1,\RR}\,, \, 
		\Lambda_{2,\RR} ] +  \, {\cal O} (\Psi) \,, \\[0.5ex]
		\Lambda_{12,\RR}
		\  =    & \ \,  {\cal G}_0\,  \bigl( [\Lambda_{1,\NS}\,, \, \Lambda_{2,\RR} ] - [\Lambda_{2,\NS}\,, \, 
		\Lambda_{1,\RR} ] \bigr)+  \, {\cal O} (\Psi)\,,   \\[0.5ex]
		{\wt\Lambda}_{12,\RR}\ 
		=    & \  [\Lambda_{1,\NS}\,, \, \Lambda_{2,\RR}] - [\Lambda_{2,\NS}\,, \, \Lambda_{1,\RR} ]+  \, {\cal O} (\Psi)\,. 
	\end{split}
	\ee

	\bigskip
	The gauge transformations given in~\refb{gaugetrsft} are expanded with
	leading terms as follows: 
	\be
	\label{gtexpanded} 
	\begin{split}
		\delta_{\Lambda, \wt\Lambda}  \widetilde\Psi  =& \  Q \widetilde \Lambda  + \  \, [ \Lambda , \Psi ]  \  \, + \half \ [\Lambda, \Psi, \Psi ]  +   \cdots \,, \\[1.5ex]
		\delta_{\Lambda, \wt\Lambda}  \Psi  =& \  Q \Lambda +  \, {\cal G} \,  [ \Lambda ,  \Psi ] 
		+ \half {\cal G}  [\Lambda, \Psi, \Psi ]  + \cdots \,.
	\end{split}
	\ee
	Let us focus on the massless sector.  We will see later that there is no
	gauge string field  $\Lambda_{{}_\RR}$ associated with the variation of  $\Psi_{{}_\RR}$.
	Thus, setting $\Lambda_{{}_\RR}=0$, our expansions take the form
	\be
	\begin{split}
		\Psi =   \  \Psi_{{}_\NS}   + \Psi_{{}_\RR} \,, \ \ \ \ \  & \Lambda =   \  \Lambda_{{}_\NS}  \,,\\
		\wt\Psi =   \ \Psi_{{}_\NS}   +  \wt\Psi_{{}_\RR} \,, \ \ \ \ \  &  \wt\Lambda =   \ \Lambda_{{}_\NS}   +  \wt\Lambda_{{}_\RR}  \,.
	\end{split}
	\ee
	Back on the right-hand side of~\refb{gtexpanded} we find
	\be
	\begin{split}
		\delta \Psi_{{}_\NS}   + \delta \wt\Psi_{{}_\RR}  =& \  Q \Lambda_{{}_\NS}  
		+ Q \widetilde \Lambda_{{}_\RR}  
		+  \,  [ \Lambda_{{}_\NS}  , \Psi_{{}_\NS} ]  
		+  [ \Lambda_{{}_\NS}  , \Psi_{{}_\RR}  ] 
		\\[1.5ex]
		&  \quad  + \half [ \Lambda_{{}_\NS}, \Psi_{{}_\NS},\Psi_{{}_\NS}] + [ \Lambda_{{}_\NS}, \Psi_{{}_\NS},\Psi_{{}_\RR}]  + \half [ \Lambda_{{}_\NS}, \Psi_{{}_\RR},\Psi_{{}_\RR}]+ \cdots \, , 
		\\[1.8ex]
		\delta\Psi_{{}_\NS}   + \delta\Psi_{{}_\RR}  =& \ Q  \Lambda_{{}_\NS}  
		+  \, \, {\cal G}  [ \Lambda_{{}_\NS}  , \Psi_{{}_\NS} ]  
		+ \, {\cal G}  [ \Lambda_{{}_\NS}  , \Psi_{{}_\RR}  ] 
		\\[1.5ex]
		&  \quad  + \half\, {\cal G} [ \Lambda_{{}_\NS}, \Psi_{{}_\NS},\Psi_{{}_\NS}] 
		+\, {\cal G} [ \Lambda_{{}_\NS}, \Psi_{{}_\NS},\Psi_{{}_\RR}]  
		+ \half\, {\cal G} [ \Lambda_{{}_\NS}, \Psi_{{}_\RR},\Psi_{{}_\RR}]+ \cdots \, , 
	\end{split}
	\ee
	We separate out the NS and R sectors, noting that both equations, as expected,
	give the same result for $\delta \Psi_{{}_\NS} $.  Recalling that ${\cal G}= {\bf 1}$ except when acting on a product with an odd number of RR
	fields, where ${\cal G} ={\cal G}_0$, we have
	\be
	\label{30erdlk}
	\begin{split}
		\delta \Psi_{{}_\NS}    =& \  Q  \Lambda_{{}_\NS}   
		+  \,  [ \Lambda_{{}_\NS}  , \Psi_{{}_\NS} ]  
		+ \half [ \Lambda_{{}_\NS}, \Psi_{{}_\NS},\Psi_{{}_\NS}]  + \half [ \Lambda_{{}_\NS}, \Psi_{{}_\RR},\Psi_{{}_\RR}]
		+ \cdots \,, \\[1.5ex]
		\delta \wt\Psi_{{}_\RR}  \ =& \   Q \widetilde \Lambda_{{}_\RR}  
		+  [ \Lambda_{{}_\NS}  , \Psi_{{}_\RR}  ] 
		+ [ \Lambda_{{}_\NS}, \Psi_{{}_\NS},\Psi_{{}_\RR}] 
		+ \cdots \,, \\[1.5ex]
		\delta\Psi_{{}_\RR} \  =& \   
		\, {\cal G}_0  [ \Lambda_{{}_\NS}  , \Psi_{{}_\RR}  ] 
		+\, {\cal G}_0  [ \Lambda_{{}_\NS}, \Psi_{{}_\NS},\Psi_{{}_\RR}] 
		+ \cdots \, , \\ 
	\end{split}
	\ee
	With $\Lambda_{{}_\RR}$ parameters equal to zero,  the field independent part of
	the gauge algebra in~\refb{leadingalge} 
	simplifies considerably
	\be
	\label{leadingalgeX} 
	\Lambda_{12,\NS}
	=    \  [\Lambda_{1,\NS}\,, \, \Lambda_{2,\NS} ]\, +  \, {\cal O} (\Psi)\,, \ \ \ 
	{\wt\Lambda}_{12,\RR}\ 
	=      \, {\cal O} (\Psi) \,, 
	\ee
	and note that $\Lambda_{12,\RR}  = 0$ because there are no candidate states just like for $\Lambda_{\RR} $. 
	The gauge algebra above does not include a {\em field-independent } RR gauge transformation.
	We can actually see, however,  the algebra gives  
	an field dependent RR gauge transformation.  Indeed, from the second
	equation in~~\refb{galgori} we have 
	\be
	\wt\Lambda_{12} =  [ \Lambda_1, \Lambda_2 ] +  [ \Lambda_1, \Lambda_2 , \Psi]  +  {\cal O} (\Psi^2) \,. 
	\ee
	Taking the RR sector of this equation, and using the vanishing of the leading term
	on the right-hand side, given~\refb{leadingalgeX}
	\be
	\wt\Lambda_{12, \RR}  =  [ \Lambda_{1,\NS}, \Lambda_{2,\NS} , \Psi_{{}_\RR} ] +  {\cal O} (\Psi^2)\,. 
	\ee
	Thus, when computing the closure of the algebra on $\wt\Psi_{{}_\RR}$  we 
	find, to first order in fields, the transformation in~\refb{30erdlk} with the relevant
	$\Lambda_{12}, \wt\Lambda_{12}$:  
	\be
	\label{3oirc,w48i}
	\delta_{\Lambda_{12}, \wt\Lambda_{12}} \wt\Psi_{{}_\RR}  =   Q   [ \Lambda_{1,\NS}, \Lambda_{2,\NS} , \Psi_{{}_\RR} ]    +  [ \Lambda_{12,\NS}  , \Psi_{{}_\RR}  ]  
	+ [\Lambda_{1,\NS}, \Lambda_{2,\NS} \,, Q\Psi_{{}_\RR}  ] + {\cal O} (\Psi^2)  \,, 
	\ee
	since to leading order in the string field ${\cal E} = Q \Psi$. 
	We will confirm this structure explicitly in the effective
	field theory description (see~\refb{galgeffft}).

	\subsection{GSO, Grassmanality, BRST,  PCO's and OPE's}\label{230rhf}
	
	We begin with reviewing GSO parity, a necessary ingredient of the theory because
	in type II
	string theory all vertex operators must be GSO even {\em both}
	in the antiholomorphic and the holomorphic sectors. 
	The GSO operator $(-1)^F$ associated to the worldsheet fermion number operator $F$ commutes with GSO even operators and anticommutes with GSO odd operators. 
	For the antiholomorphic sector we have a GSO operator $(-1)^{\bar F}$, also
	associated with the fermion number operator.   
	We have: 
	\be
	\label{lists}
	\begin{split}
		& \text{GSO odd fields:} \  \ \beta, \ \gamma,  \  e^{\pm \phi}, \ \psi^\mu, T_F. \\
		& \text{GSO even fields:} \hskip-2pt \  \  X^\mu, \ b,  \, c, \  \eta, \ \xi, \  j_B, \  {\cal X}  \,. 
	\end{split}
	\ee
	In general exponentials of $\phi$ have GSO property as follows:
	\be
	\hbox{GSO}  \ ( e^{q\phi})  = \begin{cases}  (-1)^q\,,  \ \ \ \ \ \hbox{if} \  q \in \mathbb{Z},
		\\[1.0ex]
		(-1)^{(q+{1\over 2} )}  \ \ \hbox{if}  \  q\in \mathbb{Z} + \tfrac{1}{2} \,. \end{cases}
	\ee
	For the spin operators 
	$(\Theta_a, \Theta_{\dot a})$ and $(\bar\Theta_a, \bar\Theta_{\dot a})$
	of the holomorphic and antiholomorphic Ramond sectors (see Appendix \ref{bispnrsdcmp}), we have
	the following assignments of GSO parity
	\be
	\label{gsoassign}
	\begin{split}
		\hbox{IIB}:  \ & \  \  \ ( \Theta_a , \bar \Theta_a) \ \hbox{are GSO even} \,,  \ \ \ (\Theta_{\dot a}, \bar \Theta_{\dot a})  \ \hbox{ are GSO odd}  \,, \\[0.5ex] 
		\hbox{IIA}:  \ & \  \  \ ( \Theta_a , \bar \Theta_{\dot a}) \ \hbox{are GSO even} \,,  \ \ \ (\Theta_{\dot a}, \bar \Theta_{a})  \ \hbox{ are GSO odd}  \,.
	\end{split}
	\ee
	For states, the GSO parity requires the definition of the parity of the vacuum.
	The SL$(2, \mathbb{R})$ vacuum $|0\rangle $ is GSO even:
	$(-1)^F  |0\rangle = +|0\rangle$.
	It corresponds to the identity operator, which is in the NS sector (the NS vacuum $|0\rangle_{{}_\NS} \equiv  c_1  e^{-\phi} |0\rangle$, however,   is GSO odd and corresponds to a would-be tachyon).   A Ramond ground state takes the form
	$e^{-\phi/2} \, \Theta_a  |0\rangle$. 
	It is a GSO even state because $ e^{-\phi/2}$, $\Theta_a$, and the state $|0\rangle$ all are
	GSO even. 
	
	Let us now consider Grassmanality, needed to move operators across each other in correlators and in OPEs.  The Grassmanality $\epsilon = \pm 1$ of each of the operators in \refb{lists}, 
	including also operators $e^{q\phi}$ with $q$ integer,   is correlated with the GSO parity
	$(-1)^F$, where $F$ is fermion worldsheet number, and the ghost number $G$
	\be
	\epsilon_{{}_\NS}  = (-1)^{F + G} \,,  \ \  \ \ 
	\epsilon_{{}_{\bar \NS}}  = (-1)^{\bar F + \bar G}  \,,
	\ee
	where we also included the antiholomorphic sector.  These formulae apply both to
	IIA and IIB theories. We added the subscript NS ($\bar{\hbox{NS}}$) because they apply to all operators in the NS sectors, all of which
	include a $e^{q\phi}$ (or $e^{q \bar\phi}$) with $q\in \mathbb{Z}$.   Note that the Grassmanality of an operator involving products of the 
	operators listed above is the product of the Grassmanalities of each of the operators, as 
	desired.  This is because of the additivity of the $F$  and $G$ quantum numbers. 
	
	It is also possible to assign Grassmanality $\epsilon = \pm $ operators in the R and $\bar R$ sectors.  For both the IIB and IIA theories we have
	\be 
	\epsilon_{{}_\RR}  =  (-1)^{F +G+1} \,, \ \ \epsilon_{{}_{\bar\RR}}  =  (-1)^{\bar F +\bar G+1} \, . 
	\ee
	Note that an R operator includes an $e^{q\phi}$ with $q\in \mathbb{Z} + \tfrac{1}{2}$ paired with a spin field ($\Theta_a$ or $\Theta_{\dot a}$), but such combination only appears once.  The above formula is thus consistent with including other type of operators
	whose Grassmanality, as we have seen,  is given by $(-1)^{F+G}$.  
	With this we identify 
	\be
	e^{-\phi/2}\Theta_{a}    \ \  \hbox{is Grassmann odd, } \ \  \epsilon= -1\,,
	\ee
	since both $e^{-\phi/2}$ and $\Theta_a$ are GSO even and the operator has $G=0$.  The
	same is true for $e^{-\bar \phi/2}\Bar{\Theta}_b$, in the anti-holomorphic sector of type IIB
	, which is also Grassmann odd.   For GSO
	even states $\epsilon_{{}_\RR}  = (-1)^{G+1}$. 
	Combining the two R sectors we have
	\be
	\epsilon_{{}_{\rm RR}} =  (-1)^{F+ \bar F}   (-1)^{G+ \bar G} \,,   
	\ee 
	While all states in the string fields must be GSO even, the above formulae for Grassmanality also apply to GSO odd operators.  The Grassmanality of such operators is often required in computations. 
	
	The holomorphic part of the BRST current of the theory is 
	\begin{equation}
		\begin{split}
			j(z) & = \ cT^m_B+\gamma T^m_F +bc\partial c
			- \tfrac{3}{4}(\partial c)\beta\gamma - \tfrac{1}{4}c(\partial \beta)\gamma+\tfrac{3}{4}c\beta\partial\gamma-b\gamma^2\\[1.0ex]
			& =\  c(T^m+T^{\eta\xi} + T^\phi) + \eta e^\phi T^m_F + b c \partial c -be^{2\phi}\eta\partial\eta + \tfrac{3}{4 }\partial(c\partial \phi) \,,   \end{split}
	\end{equation}
	with the antiholomorphic current $\bar j (\bar z)$ similarly defined.   As usual, 
	the closed string theory BRST operator $Q$ is given by
	$Q = \oint   dz \, j(z) + \oint    d\bar z \, \bar j(\bar z)$. 
	The holomorphic picture changing operator  ${\cal X}(z)$, of picture number one, ghost number zero, 
	and dimension zero, is 
	\be
	\mathcal{X}(z)\equiv  \{ Q , \xi(z) \}  = c\partial\xi + e^{\phi}T_F -\partial\eta e^{2\phi}b - \partial(\eta e^{2\phi}b)\,, \ \ \ 
	\mathcal{X}_0 \equiv  \oint {dz\over z}  \, \mathcal{X} (z) \,. 
	\ee
	The operators $\bar{\cal X}(\bar z)$ and $\bar{\cal X}_0$ are similarly defined. 
	
	\bigskip
	\noindent
	{\bf Inner product.} The closed string field theory has an inner product $\langle A , B \rangle \equiv \langle A | c_0^-
	|B\rangle$.  Here $A, B$ are vertex operators, with $|A\rangle$ and $|B\rangle$ the associated
	states, respectively.   Moreover,  
	$ \langle A |$ is the BPZ conjugate to $|A\rangle$.  Since we mostly work using the operator 
	representation of the string field, 
	the inner product $\langle A , B \rangle$ is calculated as a correlator on the $z$-sphere, with
	$B$ inserted at $z=0$,  $A$ inserted at $z_\infty \equiv \infty$,  and $c_0^-$ expressed as
	a line integral over the `equatorial' circle  $|z|=1$ of the operators $\partial c$ and $\bar \partial c$.
	Explicitly,  we have
	\be\label{cinsertion}
	\langle A , B \rangle \equiv  \Bigl\langle   A (w=0)  \, {1\over 2} \Bigl[ \int_{{}_{|z|=1}}  {dz\over z}  \partial c(z) - \int_{{}_{|z|=1}}  {d\bar z\over \bar z}  \bar\partial \bar c(\bar z)  \Bigr] \, B(z=0) \Bigr\rangle  \,, \ \ \  w = 1/z\,. 
	\ee
	Here the $B$ operator is inserted at $z=0$, and the $A$ operator is inserted at $z= z_\infty$, or at $w=0$, in the $w$ coordinate $w = 1/z$.  For $A$ a primary of dimension
	$(h, \bar h)$ we have 
	\be
	\label{conf1/z}
	A (w) =  A (z)  \Bigl( {dz\over dw} \Bigr)^h  \Bigl( {d\bar z\over d\bar w} \Bigr)^{\bar h}  
	\ \ \to \ \ 
	A (w=0) =  A (z_\infty) \,  (- z_\infty)^{2h}  (- \bar z_\infty)^{2\bar h} \,.
	\ee 
	We will encounter no phase ambiguities in evaluating the above powers because all operators we consider  have dimensions of the form $ h = m + r$ and $\bar h = n + r$ with $m,n\in \mathbb{Z}$ and $r$ real, this last contribution arising from the weight of momentum operators $e^{ipX}$.  As a result we have, unambigously,
	\be
	(- z_\infty)^{2h}  (- \bar z_\infty)^{2\bar h} =   (- z_\infty)^{2m}  (- \bar z_\infty)^{2n }
	|z_\infty|^{4r} \,, \ \  m, n \in \mathbb{Z} \,. 
	\ee
	For nonvanishing correlators the dependence on $z_\infty$ 
	must vanish. For non-primary $A$, one requires its conformal transformation 
	--the analog of~\refb{conf1/z}-- to express $A(w)$ in terms of some set of operators evaluated at $z_\infty$.  
	
	The normalization of correlators can be expressed as: 
	\be\label{normalization}
	\begin{split} 
		\expval { c\bar c(z_1) \, c\bar c(z_2)  \, c\Bar{c}(z_3) \,   e^{-2\phi} e^{-2\bar \phi} e^{ip\cdot X}}= & \ - \abs{z_{12}z_{13}z_{23}}^2(2\pi)^D \delta (p) \,. 
	\end{split}\ee
	
	\medskip
	\noindent
	{\bf OPE's for spin fields.}  We follow the conventions of~\cite{Sen:2016bwe} with minor
	translation\footnote{Denoting with primes the fields in~\cite{Sen:2016bwe}, we have: 
		$\gamma =  \tfrac{1}{2} \gamma'$,  $\beta = 2 \beta' $ and    
		$\psi =   i \sqrt{2} \psi'$.
		With these one finds that  
		$T^m = {T'}^m$ and  $T_{F}^m =  2{T'}_{F}^m$.
		The spinor indices in~\cite{Sen:2016bwe} are replaced as follows 
		$\alpha \rightarrow a$,  
		$S_\alpha \rightarrow \Theta_a$,  and $S^\alpha \rightarrow \Theta^a = C^{a\dot b}\Theta_{\dot b}$.  We use $\Gamma$ matrices with 
		index structure $\Gamma_a^{\hskip5pt \dot b}$ or $\Gamma_{\dot a}^{\hskip5pt b}$ and a charge conjugation matrix with index structure $C^{a\dot b}$ or $C^{\dot a b}$, their inverses written as $C_{\dot b c }$ and 
		$C_{b \dot c}$, respectively ($C^{a\dot b} C_{\dot b c} = \delta^a_{\ c}$ and 
		$C^{\dot a b} C_{ b \dot c} = \delta^{\dot a}_{\ \dot c}$) .  Indices are raised as follows: $\eta^b = C^{b\dot a} \eta_{\dot a}$ and $\eta^{\dot a} = C^{\dot a b} \eta_b$.  For the $\gamma$ matrices in~\cite{Sen:2016bwe} we use
		$ (\gamma^\mu)^{\alpha\beta}  \rightarrow  (C\Gamma^\mu)^{ab}$ and 
		$ (\gamma^\mu)_{\alpha\beta}  \rightarrow  (\Gamma^\mu C^{-1})_{ab}$.
	}  conventions: 
	\begin{subequations}
		\label{firsttwo}
		\begin{align}
			&\psi^\mu(z)e^{-\phi/2}\Theta_\alpha (0)\ 
			\sim \ -\, \frac{1}{\sqrt{2z\, }} \, 
			(\Gamma^\mu)_\alpha^{\hskip5pt \beta} e^{-\phi/2}\Theta_{\beta }(0) 
			+ \order{z^{1/2}}\,, 
			\label{w20rij} \\[1.0ex]
			&e^\phi \psi^\mu(z)e^{-\phi/2}\Theta_\alpha (0)\ \sim \ -\frac{1}{\sqrt{2 }} \, 
			(\Gamma^\mu)_\alpha^{\hskip5pt \beta} e^{\phi/2}\Theta_{\beta }(0) + \order{z^{1/2}}\,. 
			\label{w20rijK}
		\end{align}
	\end{subequations}
	The first equation follows from~\cite{Sen:2016bwe}[(3.12), first two equations]. 
	The second follows by OPE of the first with $e^\phi (z)$. 
	Now we consider 
	\begin{equation}
		\psi^\mu (z)  e^{-3\phi/2} \Theta_\alpha (0) = 
		\psi^\mu (z) :\hskip-2pt e^{-\phi} \underbrace{e^{-\phi/2} \Theta_\alpha (0)}: 
	\end{equation}
	and use~\refb{w20rij} to find
	the first equation below
	\begin{subequations}
		\label{firsttwoo}
		\begin{align}
			&\psi^\mu(z)e^{-3\phi/2}\Theta_\alpha (0)\ \sim \ \frac{1}{\sqrt{2z }} 
			(\Gamma^\mu)_\alpha^{\hskip5pt \beta} e^{-3\phi/2}\Theta_{\beta }(0) + \order{z^{1/2}}\,, 
			\\[1.0ex]  
			&e^\phi \psi^\mu(z)e^{-3\phi/2}\Theta_\alpha (0)\ \sim \ \frac{1}{\sqrt{2 }} \, 
			z \, 
			(\Gamma^\mu)_\alpha^{\hskip5pt \beta} e^{-\phi/2}\Theta_{\beta }(0) 
			+ \order{z^{1/2}}\,, 
		\end{align}
	\end{subequations}
	with the second equation following from the first by OPE with $e^{\phi}(z)$.
	Finally, we have two OPE's that involve two spin fields.  
	From~\cite{Sen:2016bwe}[eqns.(3.12), (3.16)] we find
	\begin{subequations}
		\begin{align}
			& e^{-\phi/2}\Theta_a(z)e^{-3\phi/2}\Theta_{\dot b}(0) \sim \frac{1}{z^{2}}\, C_{\dot b a} \, e^{-2\phi}(0)+  \cdots  \label{eours} \\[1.0ex]
			& e^{-\phi/2}\Theta_a(z)e^{-\phi/2}\Theta_{ b}(0) \sim - {1\over \sqrt{2}}
			\frac{1}{z}\, (\Gamma^\mu C^{-1})_{a b} e^{-\phi}\psi_\mu(0)+  \cdots   \,, \label{woshr}
		\end{align}
	\end{subequations}
	the second equation requiring that $\Gamma C^{-1}$ be a symmetric matrix since the two operators on the LHS are Grassmann odd. The symmetry of $\Gamma C^{-1}$
	and the defining property 
	$C \gamma^\mu C^{-1} = - \Gamma^{\mu \, T}$
	together imply that $C$ and $C^{-1}$ are antisymmetric.  We work in such a convention. 
	
	For a useful three-point function we have
	\be
	\label{useful3pt} 
	\bigl\langle e^{-\phi} \psi^\mu (z_1)  \, e^{-\phi/2} \Theta_a (z_2) \ e^{-\phi/2} \Theta_c (z_3) \bigr\rangle \ = \   {1\over \sqrt{2} }{1\over z_{12} z_{13} z_{23} } ( \Gamma^\mu C^{-1} )_{ac} \, . 
	\ee 
	This result, familiar up to constants in the literature, is fixed for the overall coefficient and sign 
	using OPE's and two-point functions.  
	
	\medskip
	\noindent
	{\bf Differential forms, exterior derivatives, and inner product.}
	The spacetime dimension will be taken to be $d=10$.  All indices are
	lowered and raised with the Minkowski metric $\eta_{\mu\nu} = \hbox{diag} (-1, 1, \cdots 1)$. 
	We have $\epsilon_{\mu_1 \cdots \mu_d}$  totally antisymmetric with
	\be
	\epsilon_{012 \cdots 9} = 1\,, \ \ \ \epsilon^{012 \cdots 9} = -1\,, \ \ \ 
	\hbox{and} \ \ \ 
	\e_{\mu_1\cdots \mu_p\, \nu_1 \cdots \nu_q} \e^{\mu_1\cdots \mu_p\, \rho_1 \cdots \rho_q} 
	=   -  \, p! \, q! \, 
	\delta_{\nu_1}^{[\rho_1}  \cdots \delta_{\nu_q}^{\rho_q]} \,,
	\ee
	with $p+q = 10$. Antisymmetrization is always with unit weight:
		$A_{[\mu_1  \cdots  \mu_k]} \equiv  {1\over k!} \sum_{\sigma\in S_k}  \epsilon (\sigma) 
		A_{[\mu_{\sigma(1)}  \cdots  \mu_{\sigma (k)} ]}$, 
		where $S_k$ is the permutation group, namely, $\sigma : \{ 1, \ldots , k\} \to \{ \sigma(1) , \ldots , \sigma(k)\}$ is a permutation, and $\epsilon(\sigma)$ is the sign of the permutation. 
		A $p$-form  
		$A^{(p)}$ will be written as follows 
		\begin{equation}
			A^{(p)} 
			= \frac{1}{p!} \,  A^{(p)}_{~\mu_1\cdots\mu_p}\d x^{\mu_1}\wedge\cdots\wedge \d x^{\mu_p}\,. 
		\end{equation}
		The exterior  derivative $\d$   
		takes a $p$-from to a $p+1$ form with components
		\begin{equation}
			({\rm d} A^{(p)})_{\mu_1\cdots\mu_{p+1}} = (p+1)\, 
			\partial_{[\mu_1}A^{(p)}_{\mu_2\cdots\mu_{p+1}]}   \,. 
		\end{equation}
		The wedge product works as follows for $A^{(p)}\wedge B^{(k)}$:
		\be
		(A^{(p)}\wedge B^{(k)})_{\mu_1 \cdots \mu_p \nu_1 \cdots \nu_k} 
		= {( p+k)!\over p! \, k!}  \, A^{(p)}_{[\mu_1 \cdots \mu_p} B^{(k)}_{\nu_1 \cdots \nu_k]} \,. 
		\ee
		The Hodge dual of a $p$ form $A^{(p)}$ gives
		a $q$ form $*A^{(p)}$, with $q= 10-p$ and with components
		\be   
		(*A^{(p)})_{\nu_1\cdots\nu_q} = \, \frac{1}{p!}~\epsilon_{\nu_1\cdots\nu_{q}}^{\hskip25pt\mu_1\cdots\mu_{p}}   
		A^{(p)}_{~\mu_1\cdots\mu_p}\,. 
		\ee
		For a general curved metric $g_{\alpha\beta}$, the covariant Hodge duality
		operator $*_g$ is defined as follows:
		\be\label{curvedhogde}
		(*_g A^{(p)})_{\nu_1 \cdots \nu_q}  \ = \  {1\over p!} \, g_{\nu_1 \rho_1} \cdots g_{\nu_q \rho_q} 
		{1\over \sqrt{g} } \ \epsilon^{\rho_1 \cdots \rho_q \,\mu_1 \cdots \mu_p} A^{(p)}_{\mu_1\cdots 
			\mu_p} \,, 
		\ee
		with $g = | \hbox{det}  (g_{\alpha\beta}) |$ and $\epsilon = \pm 1$ defined as before. This general Hodge  $*_g$ reduces to the earlier Hodge $*$ when the metric becomes Minkowskian, and we reserve the notation without subscript for the flat metric. 
		The repeated application of  Hodge duality on a $p$ form gives the identity operator
		up to a sign
		\begin{equation}
			\label{hodge-hodge1}
			*\, * = (-1)^{1+ p}\,   \ \ \hbox{on} \  \Lambda^p  \,,  \  \ \ d = \hbox{even.}  
		\end{equation}
		We define the inner product
		\be
		\label{inn-prod-def}
		\  (A^{(p)},B^{(p)}) \equiv  \int  * \, A^{(p)}\wedge  B^{(p)}
		= \frac{1}{p!} \int A^{(p)}_{\mu_1\cdots\mu_p}{B^{(p)}}^{\mu_1\cdots\mu_p}  \ \omega \,, 
		\ee
		with $\omega = \d x^0\wedge \cdots \wedge \d x^9$. The definition also holds in curved space with the $*_g$:
		\be
		\label{inn-prod-def-gg}
		\  (A^{(p)},B^{(p)})_g \equiv  \int  *_g \, A^{(p)}\wedge  B^{(p)}
		= \frac{1}{p!} \int A^{(p)}_{\mu_1\cdots\mu_p}\, 
		g^{\mu_1\nu_1} \cdots g^{\mu_p\nu_p}\, 
		B^{(p)}_{\nu_1\cdots\nu_p} \,  \sqrt{g} \, \omega \,. 
		\ee	
		The inner product is symmetric under the exchange of its arguments, and odd under simultaneous Hodge action on both arguments: 
		\be
		\label{symdadj} 
		( A^{(p)} , B^{(p)} )  =   ( B^{(p)} , A^{(p)} )  \,,  \ \ \ 
		( *A^{(p)} , * B^{(p)} )  =  -\, ( A^{(p)} , B^{(p)} ) \,. 
		\ee
		This implies that the inner product of self-dual or anti-self dual forms will vanish. 
		Using the Hodge star one has the adjoint $\d^\dagger$
		taking $p$ forms to $p-1$ forms, and satisfying $\d^\dagger \d^\dagger=0$:
		\be 
		\d^\dagger \equiv  - *  \d  *   \ \  \hbox{in} \ d=10 \,, \ \ \ 
		({\rm d}^\dagger A^{(p)})_{\mu_1\cdots\mu_{p-1}} \  
		=\, - \, \partial^{\mu}A^{(p)}_{~\mu\mu_1\cdots\mu_{p-1}}  \,. 
		\ee
		Analogous definitions hold in curved space with respect to $*_g$ giving us $\d_g ^\dagger$. The adjoint $\d^\dagger$ interacts properly with the inner product. 
		For arbitrary $A$ and $B$ forms,
		\be
		\label{adj-prop}
		(\d A^{(p-1)},B^{(p)}) = (A^{(p-1)},\d^\dagger B^{(p)}) \,, \ \ \ \
		(\d^\dagger B^{(p)}, A^{(p-1)}) = (B^{(p)}\, , \d A^{(p-1)})\,. 
		\ee 
		The contraction $i_X$ of a form with a vector $X$ is defined by
		\be
		i_X A^{(p)} =   {1\over (p-1)!}  \, X^\mu A^{(p)}_{\mu \nu_1 \cdots \nu_{p-1} } 
		\d x^{\nu_1} \wedge \cdots \wedge \d x^{\nu_{p-1}}\,. 
		\ee
		It is a (graded) derivation over the wedge product.
It is likewise convenient to define the adjoint $i_X^\dagger$ of $i_X$ for which 
		\be 
		( i_X A^{(p+1)} , B^{(p)} ) =  ( A^{(p+1)} ,  i^\dagger_X B^{(p)} )\, .
		\ee
		  One finds that on
		any differential form 
		\be
		i_X^\dagger \equiv  -    * \, i_X * = X^\sharp \wedge      \,, \ \ \ \ \hbox{on} \ \  \Lambda^{(p)} \ \ (d \,\,  \hbox{even} ) \,. 
		\ee
		where $X^\sharp \equiv X_\mu \d x^\mu$ is the one-from 
		associated to the vector field $X = X^\mu\partial_\mu$.
		Moreover, we have that on forms,  the Lie derivative ${\cal L}_X$
		 \be 
		 {\cal L}_X = \d i_X + i_X\d\,,  \ \ \ [i_X, \mathcal{L}_Y] = i_{[X,Y]}.
\ee
The Lie derivative also has an adjoint ${\cal L}_X^\dagger	$
satisfying 
\be 
		( {\cal L}_X^\dagger A^{(p)} , B^{(p)} ) =  ( A^{(p)} ,  {\cal L}_X B^{(p)} )\, , \ \ \ 
		{\cal L}^\dagger_X  \equiv \d^\dagger i_X^\dagger + i_X^\dagger \d^\dagger
		= (-1)^p  * {\cal L}_X * \,,
		\ee
		when acting on a $p$-form.

\medskip
Given a symmetric tensor $s_{\mu\nu} = s_{\nu\mu}$ it is possible to introduce
		an operator of degree zero acting on forms.  We write the operator as `$s \, \hd $'
		and define it as follows:	
		\be\label{haction}
		(s\, \hd \, A^{(k)})_{\mu_1\cdots \mu_k} \  \equiv \  
		\tfrac{1}{2}k\, s_{[\mu_1}^{\ \ \ \nu }A^{(k)}_{|\nu|\mu_2\cdots \mu_k]} -\tfrac{1}{4} s\, A^{(k)}_{\mu_1\cdots \mu_k}\, , \ \ \   s \equiv  s_\mu^{\ \mu} \,. 
		\ee 
As is clear above, the $s\,\hd$ on a form does not change its degree.   Additionally,
one quickly sees that the $s \, \hd$ action is self-adjoint relative to the inner product, and with a bit of work, it anticommutes with the Hodge operation 
\be
\label{hactidentity} 
( A^{(k)} , s\, \hd \, B^{(k)} ) = ( s\, \hd \,A^{(k)} ,  B^{(k)} ) \,, \ \ \ \ \ \
* \, s \, \hd  A^{(k)} = \, - \,  s \, \hd\,   (*A^{(k)} ) \,.   
\ee		
		We will later look at spacetimes with 
		a metric $g$ described as a perturbation around the flat Minkowski metric, $g_{\mu\nu} = \eta_{\mu\nu} + h_{\mu\nu}$. We will note here the expansion of the  
		Hodge operator $*_g$ to first order in $h_{\mu\nu}$ which can be written
		nicely in terms of the $h\, \hd$ operator.  After some calculation, one finds that   
		\be
		\label{8ginflat}
		*_g \, =\,  *(1-2 h\,  \hd\, )   + {\cal O} (h_{\mu\nu}^2) \,. 
		\ee
		Diffeomorphisms on tensor fields are given by the Lie derivative along vector fields.
		Letting $X$ be a vector field with components $X^\mu$, the transformation of the metric
		$\delta_X g = {\cal L}_X g$ implies that the fluctuation $h_{\mu\nu}$ 
		is varied as
		\be
		\delta_X h_{\mu\nu}   = \ \partial_\mu X_\nu + \partial_\nu X_\mu\,, \ \ \ 
		(X_\mu  = \eta_{\mu\nu}X^\nu)\,. 
		\ee
		Consider now the variation $\delta_X (h\, \hd\, A^{(k)} ) = \, 
		  (\delta_X h) \, \hd\,   A^{(k)} +  h\, \hd \, \delta_X  A^{(k)} $.
		  The second term depends on $A$ and is linear in $h$.  
		  The first term   
		   is independent of $h$ and has a simple form  
		\be\label{hactiondiffe}
		(\delta_X h) \, \hd\,   A^{(k)} \ = \   \tfrac{1}{2}\bigl (\d i_X A^{(k)}+ i_X \d A^{(k)}+ \d^\dagger i_X ^\dagger A^{(k)} + i_X^\dagger \d ^\dagger A^{(k)}\bigr) 
		=  \tfrac{1}{2}\bigl({\cal L}_X + {\cal L}^\dagger_X ) A^{(k)} \,. 
		\ee
		This remarkable identity (note that all derivatives on $A^{(k)}$ must cancel) is verified explicitly by expanding the terms on the right-hand side in components and comparing with the left-hand side where, since $\delta_X h_{\mu\nu}$ is symmetric, the definition~\refb{haction} applies. 
		
		We also note here that for the vector field $X = X^\mu\partial_\mu$, we have the associated one form $X^\sharp \equiv X_\mu \d x^\mu$, with the index on $X^\mu$ lowered using the flat Minkowski metric. Conversely, for the one form $\e = \e_\mu\, \d x^\mu$ we also have the associated vector field $\e^\flat  \equiv \e^\mu \partial_\mu$:
		\be 
		\begin{split} 
			X = X^\mu\partial_\mu \quad \to \quad & \ X^\sharp \equiv X_\mu \d x^\mu\,,  \ \ 
			X_\mu \equiv  \eta_{\mu\nu} X^\nu \\
			\e = \e_\mu\, \d x^\mu  \quad \to \quad & \ \e^\flat \  \equiv \e^\mu \partial_\mu\,, 
			\ \ \ \ \ \  \,  \e^\mu \equiv \eta^{\mu\nu} \e_\nu \,.
		\end{split}
		\ee
		
		\sectiono{SFT for RR and NSNS fields}
		
		In this section we begin by writing down the RR string fields: $\Psi_{{}_\RR}$
		is the physical field and $\wt\Psi_{{}_\RR}$ is the extra field.  We write the
		associated RR sector gauge parameters and compute linearized field equations
		and the RR kinetic terms.  These kinetic terms simplify considerably after some field redefinitions.   We study the RR cohomology to confirm the presence of the expected degrees of freedom as well as the extra degrees of freedom.  We then write the NSNS kinetic terms and use the string field theory to compute the cubic couplings involving the NSNS string field and two RR string fields.

		\subsection{RR fields and kinetic terms}
		We will now construct the kinetic term of the effective action for the massless fields of the IIB RR sector. Using undotted and dotted Latin indices to denote GSO even and GSO odd Ramond vertex operators, respectively, 
		we have the RR string field 
		at picture $(-\tfrac{1}{2} ,-\tfrac{1}{2})$ and
		ghost number two: 
		\begin{equation}
			\Psi_{{}_\RR} \ = \ \int\frac{d^Dp}{(2\pi)^D}\, Q^{ab}(p) \, 
			c\bar{c}\  e^{-\phi/2}\Theta_{a}  \, e^{-\bar \phi/2}\Bar{\Theta}_b\, e^{ip\cdot X}\,. 
		\end{equation}
		The component fields are encoded in the (momentum space) bispinor $Q^{ab}$.
		This string field is a GSO even operator in both holomorphic
		and antiholomorphic sectors.    There is no candidate 
		ghost number one gauge parameter at the massless level 
		corresponding to the above physical string field. Therefore 
		\be
		\Lambda_{{}_\RR} = 0 \,.
		\ee
		The lack of gauge parameters indicates that the $Q$ fields are gauge-invariant field strengths.

		The extra RR string field $\wt\Psi_{{}_\RR}$ 
		at picture $(-{3\over 2},-{3\over 2})$ and
		ghost number two is
		\be
		\label{auxRRfield2}
		\begin{split}
			\wt\Psi_{{}_\RR}  = \int\frac{d^Dp}{(2\pi)^D} \ \Big( &  \ \  N^{\Dot{a}\Dot{b}}(p)\, c \Bar{c} \  
			e^{-3\phi/2}\Theta_{\dot a}  \, e^{-3\bar \phi/2}\Bar{\Theta}_{\dot b}\, e^{ip\cdot X}\\ 
			&+ \tfrac{1}{2}P^{\Dot{a}b}(p)\, (\partial c + \bar \partial \bar c)c  \Bar{c}\, \bar \partial \bar\xi \, 
			e^{-3\phi/2}\Theta_{\dot a}  \,  e^{-5\bar \phi/2}\Bar{\Theta}_{ b}\, e^{ip\cdot X}\\
			& + \tfrac{1}{2}\Bar{P}^{a\Dot{b}}(p)\, (\partial c + \bar \partial \bar c)c  \Bar{c}\, \partial \xi\ 
			e^{-5\phi/2}\Theta_{ a}  \, e^{-3\bar \phi/2}\Bar{\Theta}_{\dot b}\, e^{ip\cdot X}\Big)\,. 
		\end{split}
		\ee
		The first operator is GSO even, because the minus signs from the dotted indices are compensated by the GSO contribution 
		$(-1)^{-{3\over 2} + {1\over 2}} = -1$ from the $\phi$ and $\bar\phi$ exponentials. Note the change of index type of the spin fields due to the presence of the GSO odd operators $e^{-5\phi/2}, e^{-5\bar \phi/2}$ on the last two states.  
		
		The gauge parameter $\wt \Lambda_{{}_\RR}$ for the $\tilde \Psi_{{}_\RR}$ string field is
		\be\label{gaugepara}
		\begin{split}
			\widetilde \Lambda_{{}_\RR} =  \int\frac{d^Dp}{(2\pi)^D}\Big(&\ n^{\Dot{a}b}(p)c  \Bar{c}\, \bar \partial \bar\xi \, 
			e^{-3\phi/2}\Theta_{\dot a}  \,  e^{-5\bar \phi/2}\Bar{\Theta}_{ b}\, e^{ip\cdot X}  + \Bar{n}^{a\Dot{b}}(p)c  \Bar{c}\, \partial \xi\ 
			e^{-5\phi/2}\Theta_{ a}  \, e^{-3\bar \phi/2}\Bar{\Theta}_{\dot b}\, e^{ip\cdot X}\\
			& + \tfrac{1}{2}l^{ab}(p)(\partial c + \bar \partial \bar c)c  \Bar{c}\, \partial \xi\ \bar \partial \bar\xi \, 
			e^{-5\phi/2}\Theta_{ a}  \,  e^{-5\bar \phi/2}\Bar{\Theta}_{ b}\, e^{ip\cdot X} \Big).
		\end{split}
		\ee
		Let us consider the RR linearized field equations in~\refb{eom}. 
		The BRST action on states given in~\refb{BRSTact01},
		allows us to get the BRST action on the states appearing in the string field $\wt\Psi_{{}_\RR}$:
		\be
		\label{brtss}
		\begin{split}
			Q(c\Bar{c} e^{-3\phi/2}\Theta_{\dot a}  \, e^{-3\bar \phi/2}\Bar{\Theta}_{\dot b}\,e^{ip\cdot X}) 
			&= \tfrac{1}{4}\, p^2 \, 
			\delta_{\dot a}^{\ \dot c}
			\delta_{\dot b}^{\ \dot d}
			\, R_{\dot c \dot d}(p)\,, \\
			Q(\tfrac{1}{2}(\partial c +\bar \partial \bar c) c\Bar{c}\,\bar \partial  \bar \xi e^{-3\phi/2}\Theta_{\dot a}  \, e^{-5\bar \phi/2}\Bar{\Theta}_{b}\,e^{ip\cdot X}) &=  \tfrac{1}{4}\,  
			\, \delta_{\dot a}^{\ \dot c}\ \slashed{p}_b^{\hskip 5pt \Dot{d}}
			\hskip7pt  R_{\dot c \dot d} (p) \,,   
			\\
			Q(\tfrac{1}{2}(\partial c +\bar \partial \bar c) c\Bar{c}\, \partial \xi e^{-5\phi/2}\Theta_{ a}  \, e^{-3\bar \phi/2}\Bar{\Theta}_{\dot b} \, e^{ip\cdot X} ) 
			&= \tfrac{1}{4}\ \slashed{p}_{a}^{\hskip 5pt \Dot{c}}\  \delta_{\dot b}^{\ \dot d}
			\hskip7pt  R_{\dot c \dot d}(p)\,, 
		\end{split}
		\ee
		with $R_{\dot c \dot d}(p) $ the operator 
		\be
		R_{\dot c \dot d} (p) \equiv    (\partial c +\bar \partial \bar c) c\Bar{c}\, 
		e^{-3\phi/2} \Theta_{\dot c}  
		\, e^{-3\bar \phi/2}\Bar{\Theta}_{\dot d}\,e^{ip\cdot X}\,.
		\ee
		We then have
		\be
		\label{QtildePsi} 
		Q \wt{\Psi}_{{}_\RR} = \int\frac{d^Dp}{(2\pi)^D}\ \tfrac{1}{4}\left(p^2N(p) 
		+ (P\slashed{p})(p)
		+ (\slashed{p}^T\Bar{P})(p)\right)^{\Dot{a}\Dot{b}}
		R_{\dot a \dot b} (p) \, .
		\ee
		The equation of motion $Q \widetilde \Psi_{{}_\RR} = 0$ thus gives a single equation
		\be\label{eom1}
		\left(p^2N(p) 
		+ (P\slashed{p})(p)
		+ (\slashed{p}^T\Bar{P})(p)\right)^{\Dot{a}\Dot{b}} = 0\,. 
		\ee
		The action of the BRST operator as given in~\refb{3irfne},  
		allows us to compute the $Q$ action on $\Psi_{{}_\RR}$ 
		\be
		\begin{split}
			Q\Psi_{{}_\RR} 
			&= \tfrac{1}{4} p^2 Q^{ab}(\partial c+\bar\partial\bar c) c \bar c e^{-\phi/2}\Theta_{a}  
			\, e^{-\bar \phi/2}\Bar{\Theta}_b\, e^{ip\cdot X}\\
			&
			\hskip7pt 
			- \tfrac{1}{2}(\slashed{p}^T Q)^{ \dot a b} c \bar c\,  \eta e^{\phi/2}\Theta_{\dot a}  \, e^{-\bar \phi/2}\Bar{\Theta}_b\, e^{ip\cdot X}\\
			&  \hskip7pt  - \tfrac{1}{2}(Q \slashed{p})^{ a \Dot{b}}c\bar c \bar \eta e^{\phi/2}\Theta_{a}  \, e^{-\bar \phi/2}\Bar{\Theta}_{\dot b} \, e^{ip\cdot X}.
		\end{split}
		\ee
		This gives us the three set of equations
		\be\label{eom3}
		p^2 Q^{ab} =0, \hskip20pt ({\slashed{p}^T Q})^{\dot a b} = 0, \hskip20pt 
		(Q \slashed{p})^{a \dot b} = 0.  
		\ee
		Let us now consider the construction of the quadratic action in~\refb{S2+S3}, restricted to the RR sector. 
		We need to evaluate $c_0^-  Q{\cal X}_0\Bar{{\cal X}}_0\wt\Psi_{{}_\RR}=
		c_0^-{\cal X}_0\Bar{{\cal X}}_0Q \wt\Psi_{{}_\RR} $. Since  $Q\wt\Psi_{{}_\RR}$ in~\refb{QtildePsi}
		involves a single operator,  we just need to find the PCO acting on it. 
		We have the action of  PCO on relevant states given in~\refb{dkjfiue=}, which
		then leads to 
		\be
		\begin{split}
			{\cal X}_0\Bar{{\cal X}}_0
			R_{\dot a \dot b} (p) 
			=  &  \ \ \ 
			\tfrac{1}{8}\slashed{p}_{\Dot{a}}^{\hskip 5pt c}
			\slashed{p}_{\Dot{b}}^{\hskip 5pt d}(\partial c +\bar \partial \bar c) c\Bar{c}\, e^{-\phi/2}\Theta_c  \, e^{-\bar \phi/2}\Bar{\Theta}_d\, e^{ip\cdot X}\\
			& -\tfrac{1}{4}\slashed{p}_{\Dot{b}}^{\hskip 5pt d}c \Bar{c}\eta\, e^{\phi/2}\Theta_{\dot a}  \, e^{-\bar \phi/2}\Bar{\Theta}_d\, e^{ip\cdot X}\\
			&  - \tfrac{1}{4}\slashed{p}_{\Dot{a}}^{\hskip 5pt c}c \Bar{c}\bar \eta e^{-\phi/2}\Theta_c  \, e^{\bar \phi/2}\Bar{\Theta}_{\dot b}\, e^{ip\cdot X}\,. 
		\end{split}
		\ee
		Since the above states have no $b$ antighosts the action of $ c_0^-$ expressed as a line integral over the equatorial circle as in equation~\refb{cinsertion} simply 
		amounts to multiplication by the operator ${1\over 2} (\partial c - \bar\partial \bar c)$. We then get:
		\be\label{cBRSTPCO}  
		\begin{split}
			c_0^-{\cal X}_0\Bar{{\cal X}}_0 Q\wt{\Psi}_{{}_\RR} = 
			\frac{1}{16}\int \hskip-5pt\frac{d^Dp}{(2\pi)^D}\ & 
			\Bigl\{  p^2\, \bigl(\, \slashed{p}^T N(p) \slashed{p}
			+ \slashed{p}^TP(p)
			+ \Bar{P}(p)\slashed{p}\bigr)^{ab} \,  \partial c\bar\partial\bar c\,c\Bar{c}\,  
			e^{-\phi/2}\Theta_a  \, e^{-\bar \phi/2}\Bar{\Theta}_b\, \\[0.5ex]
			& \hskip-80pt - \bigl(\, p^2 N(p)\slashed{p} 
			+ p^2 P(p)
			+ \slashed{p}^T\Bar{P}(p)\slashed{p}\bigr)^{\Dot{a}b}\, 
			(\partial c -\bar \partial \bar c) c\Bar{c}\eta \,e^{\phi/2}\Theta_{\dot a}  
			\,  e^{-\bar \phi/2}\Bar{\Theta}_{ b} \\[0.5ex]
			& \hskip-80pt - \bigl(\, p^2 N(p) + \slashed{p}^TP\slashed{p}(p)
			+ p^2\Bar{P}(p)\bigr)^{a\Dot{b}} 
			(\partial c -\bar \partial \bar c) 
			c\Bar{c}\bar \eta \,e^{-\phi/2}\Theta_a  
			\,  e^{\bar \phi/2}\Bar{\Theta}_{ \dot b}\, \Bigr\}  e^{ip\cdot X} \,. 
		\end{split}
		\ee
		We also need $c_0^-Q\wt{\Psi}_{{}_\RR}$ given by
		\be
		\label{cQtildePsi} 
		c_0^- Q \wt{\Psi}_{{}_\RR} 
		= \frac{1}{4}\int\frac{d^Dp}{(2\pi)^D}\bigl(\, p^2N(p) 
		+P(p)\slashed{p}
		+ \slashed{p}^T\Bar{P}(p)\bigr)^{\Dot{a}\Dot{b}}
		\, \partial c\bar\partial\bar c\,c\Bar{c}\,  
		e^{-3\phi/2}\Theta_{\dot a}  \, e^{-3\bar \phi/2}\Bar{\Theta}_{\dot b}\, e^{ip\cdot X} \, .
		\ee
		The BPZ inner products we need for the action are simply the correlators 
		\be 
		\label{twotypes}
		\bigl\langle 
		I \circ \wt{\Psi}_{{}_\RR} (0) \, c_0^-Q{\cal X}_0 \bar{\cal X}_0 \wt{\Psi}_{{}_\RR}(0)
		\bigr\rangle  \hskip20pt \text{and} \hskip20pt 
		\bigl\langle I \circ  \Psi_{{}_\RR}(0) \, c_0^-Q \wt{\Psi}_{{}_\RR}(0)\bigr\rangle \,. 
		\ee
		The first correlator above has nine different terms,  but only three contribute due to the ghost number conservation.  Only one contributes to the second correlator. 
		All of those are given in~\refb{overlap}.  With these, we can finally compute the
		quadratic action for the RR fields: 
		\begin{equation}
			\label{firstversionofaction} 
			\begin{split}
				S_2\bigl|_{\RR}  = \int\frac{d^Dp}{(2\pi)^D}
				\biggl\{  & -\tfrac{1}{32} \, 
				N_{a b}(-p) p^2
				\left[\slashed{p}^TN(p)\slashed{p} 
				+ \slashed{p}^TP(p)  + \Bar{P}(p)\slashed{p}\right]^{ab}\\
				& - \tfrac{1}{32}{P}_{a \dot b}(-p) \left[p^2\, \slashed{p}^TN(p) 
				+ \slashed{p}^TP(p)\slashed{p}
				+ p^2\Bar{P}(p)\right]^{a\Dot{b}}\\[0.5ex]
				& - \tfrac{1}{32}\Bar{P}_{\dot a b}(-p)
				\left[ p^2\, N(p)\slashed{p} +p^2P(p)  
				+ \slashed{p}^T\Bar{P}(p)\slashed{p}\, \right]^{\Dot{a}b} \\
				& +\tfrac{1}{4}Q_{\dot a \dot b}(-p)\left[\, p^2N(p)+ P(p)\slashed{p} + 
				\slashed{p}^T\Bar{P}(p) \right]^{\dot a \dot b} \biggr\} \, .
			\end{split}
		\end{equation}
		Cross terms coupling $N P,  N \bar P$ and $P\bar P$ add up. An example
		is shown in~\refb{cross-terms} for the $NP$ term. 
		The simplified action is then
		\begin{equation}
			\begin{split}
				S_2\bigl|_{\RR} = \int\frac{d^Dp}{(2\pi)^D}\biggl\{ 
				& -  \tfrac{1}{32} \, N_{ab}(-p) p^2
				\big[ \slashed{p}^TN(p)\slashed{p} + 2 \slashed{p}^TP(p) 
				+ 2 \Bar{P}(p)\slashed{p}\big]^{ab}\\
				& - \tfrac{1}{32} {P}_{a\Dot b}(-p) 
				\bigl[ \, \slashed{p}^TP(p) \slashed{p}
				+ 2 p^2\Bar{P}(p)\bigr]^{a\Dot b}- \tfrac{1}{32} \Bar{P}_{\dot a b }(-p)(\slashed{p}^T\Bar{P}\slashed{p})^{\Dot{a}b}\\[1.0ex]
				& +\tfrac{1}{4}Q_{\dot a \dot b}(-p)
				\bigl[\, 
				p^2N(p) + P(p) \slashed{p}
				+ \slashed{p}^T\Bar{P}(p) \bigr]^{\dot a \dot b} \biggr\} \,.
			\end{split}
		\end{equation}
		In position space, 
		\begin{equation}
			\begin{split} \label{actionposition}
				S_2\bigl|_{\RR} =  \int d^Dx 
				\biggl\{  &\tfrac{1}{32}\,  N_{ab}\partial^2
				\Bigl[-\slashed{\partial}^TN\overleftarrow{\slashed{\partial}} - 2i \slashed{\partial}^TP -2i\Bar{P}\overleftarrow{\slashed{\partial}}\Bigr]^{ab}\\
				& +\tfrac{1}{32}{P}_{a\dot b} 
				\left[ \slashed{\partial}^TP\overleftarrow{\slashed{\partial}}
				+ 2\partial^2\Bar{P}\right]^{a\dot b}
				+ \tfrac{1}{32}\Bar{P}_{\dot a b}(\slashed{\partial}^T\Bar{P}\overleftarrow{\slashed{\partial}})^{\Dot{a}b}\\
				& +\tfrac{1}{4}Q_{\dot a\dot b}
				\Bigl[-\partial^2N 
				-i P\overleftarrow{\slashed{\partial}} 
				-i \slashed{\partial}^T\Bar{P}\Bigr]^{\dot a \dot b} \biggr\} \,. 
			\end{split}
		\end{equation}
		We now decompose the bispinors in terms of the basis of the Clifford algebra
		\be
		\label{comopne}
		\begin{split}
			& Q^{ab} = \sum_{k\in \hat{\mathbb{Z}}_{odd}}\tfrac{1}{k!}iQ^{(k)}_{\mu_1\cdots\mu_k}(C\Gamma^{\mu_1\cdots\mu_k})^{ab}\,, \\
			& \bar P^{a\Dot{b}} = \sum_{k\in \hat{\mathbb{Z}}_{even}}\tfrac{1}{k!}\, 
			\bar P^{(k)}_{\mu_1\cdots\mu_k}(C\Gamma^{\mu_1\cdots\mu_k})^{a\Dot{b}}\,, \\
			& P^{\Dot{a}b} = \sum_{k\in \hat{\mathbb{Z}}_{even}}\tfrac{1}{k!}
			P^{(k)}_{\mu_1\cdots\mu_k}(C\Gamma^{\mu_1\cdots\mu_k})^{\Dot{a}b}\,, 
			\\
			& N^{\Dot{a}\Dot{b}} = \sum_{k\in \hat{\mathbb{Z}}_{odd}}\tfrac{1}{k!}iN^{(k)}_{\mu_1\cdots\mu_k}(C\Gamma^{\mu_1\cdots\mu_k})^{\Dot{a}\Dot{b}}\,. 
		\end{split}
		\ee
		where $C\Gamma^{\mu_1\cdots\mu_k}$ are the basis element for bispinors as explained in appendix~\ref{bispnrsdcmp} leading up to the equation~\refb{expansion} and 
		\be
		\hat{\mathbb{Z}}_{odd} = \{1,3,5,7,9\}\ , \hskip30pt \hat{\mathbb{Z}}_{even} 
		= \{0,2,4,6,8,10\}.
		\ee
		The expansion coefficients $Q^{(k)}_{\mu_1\cdots\mu_k}, P^{(k)}_{\mu_1\cdots\mu_k}, \bar P^{(k)}_{\mu_1\cdots\mu_k}$ and $N^{(k)}_{\mu_1\cdots\mu_k}$ are the component fields which can also be considered as the components of differential forms. The factor of $i$ in the expansion of $N$ and $Q$ bispinor is added to make the component fields real.

		We can go back and forth between bispinors and differential forms using the
		following identification
		\be\label{cea}
		A^{(k)} = \frac{1}{k!}(A^{(k)})_{\mu_1\cdots\mu_k}\d x^{\mu_1}\wedge\cdots\wedge\d x^{\mu_k} \ \ \longleftrightarrow \  \ \slashed{A}^{(k)} = \frac{1}{k!}(A^{(k)})_{\mu_1\cdots\mu_k}C\Gamma^{\mu_1\cdots\mu_k}\,,
		\ee
		with the Clifford algebra basis elements identified with the basis of the exterior algebra of differential forms. In general $\slashed{A}^{(k)}$ has spinor 
		indices $\slashed{A}^{(k)\alpha\beta}$
		that can be raised and lowered with the $C$ matrices.
		In this notation, for example, 
		the above bispinor decomposition of $Q^{ab}$ and $P^{\dot a b}$
		would read
		\be
		Q^{ab} =  \sum_{k\in \hat{\mathbb{Z}}_{odd}} i\, \slashed{Q}^{(k)ab}  \,, \qquad 
		P^{\Dot{a}b} = \sum_{k\in \hat{\mathbb{Z}}_{even}} 
		\slashed{P}^{(k)\dot a b}\,. 
		\ee
		The derivatives acting on the bispinors  are related to the action of $\d$ and $\d^\dagger$ on the associated differential forms.  Following~\refb{derivvvaa }, we have
		\be
		\label{diffided}
		\begin{split}
			& (\bar P\overleftarrow{\slashed{\partial}})^{ab} =  \sum_{k\in \hat{\mathbb{Z}}_{even}}\tfrac{1}{(k+1)!}\, 
			(\d \bar P^{(k)})_{\mu_1\cdots\mu_{k+1}}(C\Gamma^{\mu_1\cdots\mu_{k+1}})^{a b}+ \tfrac{1}{(k-1)!}\,(\d^\dagger \bar P^{(k)})_{\mu_1\cdots\mu_{k-1}}(C\Gamma^{\mu_1\cdots\mu_{k-1}})^{a b}\, ,\\[0.5ex]
			& (\slashed{\partial}^TP)^{a b} = \hskip-5pt \sum_{k\in \hat{\mathbb{Z}}_{even}}
			\hskip-10pt -\tfrac{1}{(k+1)!}
			(\d P^{(k)})_{\mu_1\cdots\mu_{k+1}}(C\Gamma^{\mu_1\cdots\mu_{k+1}})^{a b} + \tfrac{1}{(k-1)!}\,(\d^\dagger P^{(k)})_{\mu_1\cdots\mu_{k-1}}(C\Gamma^{\mu_1\cdots\mu_{k-1}})^{a b}\, ,  \\[0.5ex]
			& (\slashed{\partial}^TN\overleftarrow{\slashed{\partial}})^{a b} = -  \sum_{k\in \hat{\mathbb{Z}}_{odd}}\frac{1}{k!}\, i \, \bigl((\d^\dagger\d-\d\d^\dagger) N^{(k)}\bigr)_{\mu_1\cdots\mu_k}(C\Gamma^{\mu_1\cdots\mu_k})^{ab}\,,
		\end{split}
		\ee
		where the definitions of $\d$ and $\d^\dagger$ were reviewed in section~\ref{230rhf}.
		A useful identity, worked out in \refb{productsp} and \refb{innercliff},  expresses
		the integral of the product of bispinors in terms of the inner product on the 
		space of differential forms: 
		\be\label{Cliffinte}
		\int \d^Dx \,  \slashed{A}^{(k)}_{\ \alpha\beta} \, \slashed{B}^{(q)\alpha\beta}
		=\,  2^{d/2}(-1)^k \delta_{k,q}\, (A^{(k)},B^{(k)})\,. 
		\ee
		We can use these identities to write the action in terms of the component fields. Let us first look at the first term in the action: 
		\be
		\int \d^Dx \, N_{a b} \, \partial^2 
		(\slashed{\partial}^T N\overleftarrow{\slashed{\partial}})^{ab}  
		= \sum_{k,p\in \hat{\mathbb{Z}}_{odd}} \, \int \d^Dx  \slashed{N}^{(k)}_{\ a b} 
		\, \partial^2 
		(\not{\hskip-5pt\Delta \hskip-2pt N})^{(p)ab}  
		=  -\sum_{k\in \hat{\mathbb{Z}}_{odd}} 2^{d/2-1}
		(N^{(k)}, \, \partial^2\Delta N ^{(k)})\,,    
		\ee
		where we used the expansion of the $N$ bispinor,  the last of~\refb{diffided},
		and~\refb{Cliffinte} noting that this trace is only over un-dotted indices, thus introducing
		a factor of one-half in this formula.   Working out other terms in the same manner, we arrive at the action in terms of the component fields
		\be
		\begin{split}
			2^{1-d/2} S_2\bigl|_{\RR}  =  & \ \  \frac{1}{32} \sum_{k\in \hat{\mathbb{Z}}_{odd}}  
			\Bigl[(\partial^2N^{(k)}, \triangle N^{(k)}) + 2(\partial^2N^{(k)}, \d P^{(k-1)}) - 2 (\partial^2N^{(k)},\d^\dagger P^{(k+1)})\\[-0.8ex]
			& \hskip50pt - 2(\partial^2N^{(k)},\d \Bar{P}^{(k-1)})- 2 (\partial^2N^{(k)}, \d^\dagger \Bar{P}^{(k+1)})\Bigr]\\[1.0ex]
			& + \frac{1}{32}\sum_{k\in \hat{\mathbb{Z}}_{even}}\Big[(P^{(k)}, \triangle P^{(k)})+(\Bar{P}^{(k)},\triangle \Bar{P}^{(k)})+2(\Bar{P}^{(k)}, \partial^2 P^{(k)})\Big]\\
			& +\frac{1}{4} \sum_{k\in \hat{\mathbb{Z}}_{odd}} \Bigl[(Q^{(k)}, - \partial^2N^{(k)}) - (Q^{(k)},  \d P^{(k-1)}) - (Q^{(k)},\d^\dagger P^{(k+1)}) \\[-0.8ex]
			& \hskip50pt + (Q^{(k)}, \d \Bar{P}^{(k-1)}) -  (Q^{(k)}, \d^\dagger\Bar{P}^{(k+1)})\Bigr]\,, 
		\end{split}
		\ee
		where $\triangle \equiv \d^\dagger\d-\d\d^\dagger$, is self adjoint in the inner product, just as $-\partial^2 = \d^\dagger\d+ \d\d^\dagger$ is.
		
		This action can be factorized in to the simpler form,
		\be
		\begin{split} 
			2^{1-d/2}   S_2\bigl|_{\RR}  = \sum_{k\in \hat{\mathbb{Z}}_{odd}}\Bigl[ & \   \tfrac{1}{32}
			\bigl(\, \d\, [\d^\dagger N^{(k)} - P^{(k-1)}+\bar P^{(k-1)}]\, , \d\, [\, \d^\dagger N^{(k)} - P^{(k-1)}+\bar P^{(k-1)}]\bigr)\\
			& \hskip-20pt - \tfrac{1}{32}\bigl(\, \d^\dagger[\d N^{(k)} -  P^{(k+1)}-\bar P^{(k+1)}]\, , \, \d^\dagger[\d N^{(k)} -  P^{(k+1)}-\bar P^{(k+1)}]\, \bigr) \\[1.0ex]
			&\hskip-20pt + \tfrac{1}{4}\, \bigl( \, Q^{(k)}\, , \d\, [\, \d^\dagger N^{(k)} - P^{(k-1)} + \bar P^{(k-1)}] + \d^\dagger[\d N^{(k)} -  P^{(k+1)}-\bar P^{(k+1)}]\bigr)\Bigr] \,. 
		\end{split}
		\ee 
		By making the field redefinitions,
		\be
		\label{defpmP} 
		\begin{split}
			& P_-^{(k-1)} = \tfrac{1}{4}\, \bigl( \,  P^{(k-1)}-\bar P^{(k-1)}  -\d^\dagger N^{(k)}  
			\bigr)\, , \\[0.5ex]
			& P_+ ^{(k+1)} = \tfrac{1}{4}\, \bigl(\,  P^{(k+1)} +\bar P^{(k+1)}-\d N^{(k)} \, \bigr)  \,, 
		\end{split}
		\ee
the action becomes
\be\label{finaction}
\begin{split} 
			\hskip-10pt 2^{-{d\over 2}}
			S_2\bigl|_{\RR}  = \hskip-5pt \sum_{k\in \hat{\mathbb{Z}}_{odd}} \hskip-3pt\Bigl[& 
			\tfrac{1}{4}(\d P_-^{(k-1)}, \d P_-^{(k-1)})  
			- \tfrac{1}{4}(\d^\dagger P_+^{(k+1)}, \d^\dagger P_+^{(k+1)}) 
			-\tfrac{1}{2}(Q^{(k)}, \d P_-^{(k-1)}+ \d^\dagger P_+ ^{(k+1)})\Bigr].
\end{split}
\ee
This is the `democratic' formulation of the kinetic terms, using all forms in the theory. 
The component fields here are not all independent; the duality relations between them is worked out in~\refb{dl} and take the form
		\be
		\label{dualitiesPQ} 
		* P_\pm^{(k)} = -(-1)^{k(k-1)/2}P_\mp^{(10-k)},  \hskip40pt  *Q^{(k)} = (-1)^{k(k-1)/2}Q^{(10-k)}\,. 
		\ee
		Explicitly, for the $Q$'s we have 
		\be
		* Q^{(1)}= Q^{(9)} \,,   \ \ * Q^{(3)}=- Q^{(7)}   \,, \ \ \
		* Q^{(5)}= Q^{(5)}   \,, \ \ \  * Q^{(7)}= - Q^{(3)}  \,, \ \ \  * Q^{(9)}= Q^{(1)} \,.
		\ee
		Note that $P_-^{(10)}$ and  $P_+^{(0)}$ (fields dual to one another), 
		do not appear in the action.  We can express the action in terms of independent fields: $Q^{(1)}, Q^{(3)}, Q^{(5)}$, as well as $P_-^{(0)}, P_-^{(2)}, P_-^{(4)}$, and
		$P_+^{(2)}, P_+^{(4)}$.  The action  is then
		\be
		\label{quadSRRIIB}
		\begin{split}
			2^{-d/2}  S_2\bigl|_{\RR}  
			&=\ \tfrac{1}{2}(\d P_-^{(0)}, \d  P_-^{(0)})
			-\tfrac{1}{2}(\d^\dagger  P_+^{(2)}, \d^\dagger P_+^{(2)})
			- (Q^{(1)},  \d P_-^{(0)} + \d^\dagger  P_+^{(2)})\\[2.5ex]
			& +\ \tfrac{1}{2}(\d  P_-^{(2)}, \d  P_-^{(2)})
			-\tfrac{1}{2}(\d^\dagger P_+^{(4)}, \d^\dagger P_+^{(4)})
			-(Q^{(3)},  \d P_-^{(2)} + \d^\dagger P_+^{(4)}) \\[2.5ex]
			&+\ \tfrac{1}{2}(\d P_-^{(4)}, \d P_-^{(4)})
			-(Q^{(5)},  \d P_-^{(4)}) \,.
		\end{split}
		\ee
		Note that the independent fields 
		and their duals give the same contribution to the action~\refb{finaction},  hence 
		in passing to the new expression we get a factor of two.  On each line above we include terms in which fields mix; there is no extra mixing across lines.
		The fields that couple on each line are
		\be
		(P_-^{(0)}  , P_+^{(2)},  Q^{(1)}  )\,,  \ \  (P_-^{(2)}  , P_+^{(4)}, Q^{(3)}   )\,,  \ \ 
		(P_-^{(4)} , Q^{(5)} )\,.
		\ee
		The pattern on the first two groups actually holds for the third, but here $P_+^{(6)}= - * P_-^{(4)}$, is not independent.                
		The part $S_2\bigl|_5$ of the quadratic action describing the self-dual five form is given by   
		\be
		2^{- d/2}\,  S_2|_5  =   \tfrac{1}{2}(\d P, \d P) - (\d P, Q)\,, 
		\ee
		where we wrote $P \equiv P_-^{(4)} = *P_+^{(6)}$, and $Q\equiv Q^{(5)}$.
		It agrees with the action anticipated in~\cite{Sen:2015nph}.
		
		\subsection{Physical states of the RR SFT} 
		In this section, we first examine the linearized gauge transformations in terms of the component fields. We will then analyze the linearized equation of motion, which, together with the linearized gauge transformation, forms a cohomology problem.  The cohomology classes describe the spectrum of the theory. 
		
		The BRST action on the ghost number one states in the gauge parameter $\wt\Lambda_{{}_\RR}$ in~\refb{gaugepara} is given in~\refb{brstforgt}. With this
		result one quickly finds that 
		\begin{equation}
			\begin{split}
				\delta \wt \Psi_{{}_\RR} = 	
				Q\wt \Lambda_{{}_\RR}  
				= \int\frac{d^Dp}{(2\pi)^D}\Big( &-\tfrac{1}{2}(n\slashed{p} + \slashed{p}^T\Bar{n})^{\Dot{a}\Dot{b}}\, c \Bar{c} \  
				e^{-3\phi/2}\Theta_{\dot a}  \, e^{-3\bar \phi/2}\Bar{\Theta}_{\dot b}\, e^{ip\cdot X}\\ 
				&+ \tfrac{1}{4}(p^2n+\slashed{p}^T l)^{\Dot{a}b}\, (\partial c + \bar \partial \bar c)\, c  \Bar{c}\, \bar \partial \bar\xi \, 
				e^{-3\phi/2}\Theta_{\dot a}  \,  e^{-5\bar \phi/2}\Bar{\Theta}_{ b}\, e^{ip\cdot X}\\
				& + \tfrac{1}{4}(p^2\Bar{n}-l\slashed{p})^{a\Dot{b}}\ (\partial c + \bar \partial \bar c)\, c  \Bar{c}\, \partial \xi\ e^{-5\phi/2}\Theta_{ a}  \, e^{-3\bar \phi/2}\Bar{\Theta}_{\dot b}\, e^{ip\cdot X}\Big)\,. 
			\end{split}
		\end{equation}
		We can now read that  the component fields of  
		$ \wt\Psi_{{}_\RR}$, 
		as given in~\refb{auxRRfield2}, transform as follows: 
		\be
		\begin{split}
			\delta N^{\Dot{a}\Dot{b}}(p) &= -\tfrac{1}{2} \, \bigl( \, n\slashed{p}+\slashed{p}^T\Bar{n} \bigr)^{\Dot{a}\Dot{b}}\,, \\[0.6ex]
			\delta P^{\Dot{a}b}(p) &= \ \tfrac{1}{2}\, \bigl(\, p^2n+\slashed{p}^Tl \bigr)^{\Dot{a}b} \, ,  \\[0.5ex]
			\delta\Bar{P}^{a\Dot{b}}(p) &= \ \tfrac{1}{2}\, \bigl(\, 
			p^2\Bar{n}-\, l\slashed{p}\bigr)^{a\Dot{b}}\,. 
		\end{split}
		\ee
		The gauge parameter bispinors can be expanded in terms of component 
		gauge parameters as follows
		\be
		\begin{split}\label{comopgne}
			& \bar n^{a\Dot{b}} = \sum_{k\in \hat{\mathbb{Z}}_{even}}\tfrac{1}{k!}\, 
			\bar n^{(k)}_{\mu_1\cdots\mu_k}(C\Gamma^{\mu_1\cdots\mu_k})^{a\Dot{b}}\,, \\
			& n^{\Dot{a}b} = \sum_{k\in \hat{\mathbb{Z}}_{even}}\tfrac{1}{k!}
			n^{(k)}_{\mu_1\cdots\mu_k}(C\Gamma^{\mu_1\cdots\mu_k})^{\Dot{a}b}
			\, , \\
			&  l^{ab} = \sum_{k\in \hat{\mathbb{Z}}_{odd}}\tfrac{1}{k!}il^{(k)}_{\mu_1\cdots\mu_k}(C\Gamma^{\mu_1\cdots\mu_k})^{ab}\, . 
		\end{split}
		\ee
		The gauge transformations for the component fields are then, 
		with $q \in \hat{\mathbb{Z}}_{odd}$  and  $k \in \hat{\mathbb{Z}}_{even}$,
		\be
		\begin{split}
			& \delta N^{(q)} = \tfrac{1}{2}\, \bigl( \d n^{(q-1)}+\d^\dagger n^{(q+1)} - \d\Bar{n}^{(q-1)} + \d^\dagger\Bar{n}^{(q+1)}\, \bigr)  \,, \\[0.5ex]
			& \delta P^{(k)} = \tfrac{1}{2}\, \bigl( \, -\partial^2n^{(k)} - \d l^{(k-1)} + \d^\dagger l^{(k+1)}\, \bigr) \,, \\[0.6ex]
			& \delta \Bar{P}^{(k)} = \tfrac{1}{2}\, \bigl( \, -\partial^2\Bar{n}^{(k)} + \d l^{(k-1)} + \d^\dagger l^{(k+1)}\, \bigr) \,.  
		\end{split}
		\ee
		Ultimately, the fields that appear in the simplified action are $P_\pm$, as
		defined in~\refb{defpmP}.  We find a much simpler result for their gauge
		transformations:   
		\begin{equation}
			\delta P_-^{(k-1)} = \d m_-^{(k-2)}\,, \ \ \ \ 
			\delta P_+^{(k+1)} = \d^\dagger m_+^{(k+2)}\,.   
		\end{equation}
		Here the new parameters $m_\pm$, composed by odd degree forms, 
		are given by 
		$$ m_-^{(k-2)} = \tfrac{1}{8}[\d^\dagger n^{(k-1)}
		- \d^\dagger\Bar{n}^{(k-1)} - 2 l^{(k-2)}]\, , \ \ \ 
		m_+^{(k+2)}= \tfrac{1}{8}[\, \d n^{(k+1)} + \d\Bar{n}^{(k+1)}  + 2 l^{(k+2)}]\,. $$ 
		
		The equations of motion following from the action~\refb{finaction} by variation with respect to 
		$P_-^{(k-1)},  P_+^{(k+1)}$, and $ Q^{(k)}$, respectively, are
		\be\label{eommm}
		\begin{split}
			\d^\dagger\d P_-^{(k-1)} - \d^\dagger Q^{(k)} = & \  0\,, \\
			\d\d^\dagger P_+^{(k+1)} +\d Q^{(k)} \ = & \  0\,, \ \quad \ k \in \hat{\mathbb{Z}}_{odd}\, .\\
			\d P_-^{(k-1)} +\d^\dagger P_+^{(k+1)}=& \ 0\,, 
		\end{split}
		\ee
		They are in fact equivalent to
		\be\label{eommm1}
		\begin{split}
			\d^\dagger Q^{(k)} = & \  0\,,\\
			\d \, Q^{(k)} =& \  0\,,\\
			\d P_-^{(k-1)} + \d^\dagger P_+^{(k+1)}= & \ 0\,.
		\end{split}
		\ee
		The first two follow by acting with $\d^\dagger$ and $\d$ on the third equation in~\refb{eommm},  showing that both in the first and second equations the first
		term vanishes.   The first two equations then show that $Q^{(k)}$ indeed
		propagates the degrees of freedom of a degree $(k-1)$ gauge potential. 
		It is actually possible to display the field strength that is associated with the extra degrees of freedom.  Following~\cite{Hull:2023dgp}, where this is done for the 
		extra five-form, we
		introduce a $G^{(k)}$  as follows
		\be
		G^{(k)} \equiv  \ \d P_-^{(k-1)}   - \d^\dagger P_+^{(k+1)} - Q^{(k)} \,. 
		\ee	
		The first and second equations in~\refb{eommm} imply, respectively, that:
		\be
		\d^\dagger G^{(k)}  = 0 \,, \ \ \ \   \d\, G^{(k)}  = 0\,,
		\ee
		demonstrating that $G^{(k)}$ is a field strength for a degree $(k-1)$ gauge potential. 
		Additionally, $G^{(k)}$ must satisfy the same duality condition as 
		$Q^{(k)}$, otherwise we have more field strengths than expected.  Indeed,
		using the duality property~\refb{dualitiesPQ}  of the $P_\pm$ gauge fields,  one finds that as expected
		\be
		\label{dualitiesPQX} 
		*G^{(k)} = (-1)^{k(k-1)/2}\, G^{(10-k)}\,. 
		\ee	
		
		\medskip	
		These extra degrees of freedom also correspond to BRST cohomology classes 
		at picture $(-{3\over 2} , -{3\over 2})$. The equations for states killed by the
		BRST operator arise from the variation of the $Q^{(k)}$'s in the quadratic action 
		and correspond to the last equation in~\refb{eommm1}:
		\be \label{eom4}
		\d P_-^{(k-1)} + \d^\dagger P_+^{(k+1)} = 0 \,,  \ \ k \in \hat {\mathbb{Z}}_{\rm odd}\,. 
		\ee
		Let us quickly sketch the argument that establishes the count of degrees of freedom
		from this equation.
		Applying $\d^\dagger$ to the equation of motion, we get
		\be
		\d^\dagger \d P_-^{(k-1)} = 0\,. 
		\ee
		We now use the gauge transformation $\delta P_-^{(k-1)} = \d m_-^{(k-2)}$
		to 
		set $\d^\dagger P_-^{(k-1)}  = 0$ (this is the analog of using the gauge invariance $\delta A_\mu = \partial_\mu \epsilon$ for an abelian gauge field to set
		$\partial \cdot A = 0$). 
		The equation of motion is then
		\be
		(\d^\dagger \d +  \d \d^\dagger )  P_-^{(k-1)} = -\partial^2  P_-^{(k-1)} = 0\,. 
		\ee
		We see that that $P_-^{(k-1)}$ fields describe massless $(k-1)$ form gauge fields. As for $P_+^{(k+1)}$, the part not killed by $\d^\dagger$ is completely determined by $P_-^{(k-1)}$ through the equation of motion~\refb{eom4}. The part that is killed by $\d^\dagger$ can be  gauged away using the transformation $\delta P_+^{(k+1)} = \d^\dagger m_+^{(k+2)} $.

		\medskip	
		As an illustration of the above argument, we do a light-cone analysis 
		of the $k=1$ equation in~\refb{eom4}: $\d^\dagger P_+^{(2)} + \d P_-^{(0)} = 0$. 
		Using $P\equiv   P_-^{(0)}$ and $\widehat P =  P_+^{(2)}$, we write this equation as
		\be
		\label{unklo}
		\d^\dagger \widehat P  =  - \d P \,,
		\ee
		which implies $\partial^2 P  = 0$, namely,  
		$P$  is a gauge invariant massless scalar. 
		In components the above equation gives 
		\be
		\label{subsc}
		\partial^\mu \widehat P_{\mu\nu}  =  \partial_\nu P\,,  
		\ee 
		with gauge symmetry $\delta 
		\widehat P_{\mu\nu}=
		-\partial^\alpha \ell_{\alpha\mu\nu}$. 
		In light-cone\footnote{We use the light-cone metric
			$a\cdot b= -a^+ b^-  - a^- b^+ + a^I b^I$, with $I$ for transverse components, and take $p^+ \not= 0$.   } this transformation gives  $\delta P^{\mu\nu} = p^+ \ell^{- \mu \nu} + \cdots$, 
		where the dots are terms without $p^+$. 
		Due to the antisymmetry of $\ell$, we can gauge away all components of 
		$\widehat P$ except those
		with a minus index. Therefore,  $\widehat P^{+-}$ and $\widehat P^{I-}$ cannot be gauged away.  
		But then, they are determined from $P$ by the equation of motion~\refb{subsc},
		which reads
		\be
		-p^+\widehat P^{-\nu} - p^- \widehat P^{+ \nu} + p^I \widehat P^{I\nu} = p^\nu P\,. 
		\ee 
		Recalling that the components
		of $\widehat P$ without a minus index have been gauged away, we find that
		the $\nu= +$ equation fixes $\widehat P^{+-} = P$ and 
		the $\nu= J$ equation  fixes $\widehat P^{J-} =  p^J P /p^+$.  
		The $\nu= -$ equation reproduces the mass-shell condition $p^2=0$ in light-cone 
		form: 
		$- 2p^+ p^- + p^\mu p^\mu = 0 $.
		
		\medskip
		The equation of motion~\refb{eom4} for $k =5$ only
		involves $P_-^{(4)}$, since $*P_+^{(6)}= P_-^{(4)}$.  It then reads 
		\be
		\d P_-^{(4)} -*\d P_-^{(4)} = 0\,, 
		\ee
		showing that $P_-^{(4)}$ is a gauge potential whose field strength
		$\d P_-^{(4)}$  is self-dual.  The equation of motion also implies
		that this field strength is killed by $\d^\dagger$.  This is just as for 
		the field strength $Q^{(5)}$ coming from the $-1/2$ picture string field equation.

		The zero momentum, space-time constant modes 
		of the fields   
		$P_\pm^{(k)}$ satisfy the equations of motion since all terms in
		these equations have derivatives. 
		As the gauge transformations also all involve derivatives, none of the zero modes can be gauged away. Hence, the cohomology is nontrivial at zero momentum. In particular, the zero momentum cohomology at picture $(-{3\over 2},-{3\over 2})$ is not isomorphic to the $(-{1\over 2},-{1\over 2} )$ picture cohomology.  Such a failure
		of the isomorphism was  noted by Berkovits and one of us in~\cite{Berkovits:1997mc}. 
		
		In order to show that the degrees of freedom described by the extra fields are
		of negative norm, it is best to use the action and couple the fields to sources.  One then fixes the gauge and eliminates the fields to leave the action in terms of the sources, from which one can read what degrees of freedom propagate and tell if they are of positive or negative norm.  
		We do this analysis in Appendix~\ref{sourcemethod}.

		\subsection{Kinetic term for NSNS fields}\label{kinnsnsn}
		
		Let us now construct the kinetic terms in the string field action for the NSNS sector of the theory.  These kinetic terms are valid both for type IIA and type IIB. 
		To begin we need the classical NSNS string field, which is a 
		Grassmann even vertex operator with picture $(-1, -1)$ and ghost number two.  It is given by
		\be
		\label{nsnsfield}
		\begin{split}
			\Psi_{{}_\NS}= & \int\frac{d^Dp}{(2\pi)^D}
			\biggl(\, 
			\tfrac{1}{2}\, 
			e_{\mu\nu}(p)\ c\Bar{c}\, \psi^\mu\Tilde{\psi}^\nu e^{-\phi}e^{-\bar \phi} - e(p)
			\, c\Bar{c}\, \eta\bar \partial\Bar{\xi}\, e^{-2\bar \phi}
			- \bar e(p)	\, c\Bar{c}\, \partial\xi\Bar{\eta} \, e^{-2\phi} 
			\\
			& -\tfrac{i}{\sqrt{8}}  \,  \Bigl(\, 
			f_\mu(p)(\partial c+\bar \partial \bar c)c\Bar{c}\, \psi^\mu\bar\partial\Bar{\xi}\, 
			e^{-\phi}e^{-2\bar \phi}
			+\Bar{f}_\mu(p)(\partial c+\bar \partial \bar c)c\Bar{c}\, 
			\partial\xi\Tilde{\psi}^\mu\,  e^{-2\phi}e^{-\bar \phi} \Bigr) \biggr)e^{ip\cdot X}\,. 
		\end{split} 
		\ee
		As we confirm below, $e_{\mu\nu}$ describes the linearized gravity field and the Kalb-Ramond field,  
		the dilaton is
		a linear combination of $e$ and $\bar e$, with the other combination being pure gauge,
		and $(f_\mu , \bar f_\mu)$ are auxiliary fields. 
		The gauge parameter string field in the NSNS sector takes the form
		\begin{equation}
			\label{vertexNSgp} 
			\begin{split} 
				\Lambda_{{}_\NS}  = \int{d^Dp\over (2\pi)^D}  & \ \Bigl(-\tfrac{i}{\sqrt{2}}\lambda_\mu(p) \ 
				c\Bar{c}\, \psi^\mu\, \bar \partial\bar \xi \ e^{-\phi}e^{-2\bar \phi} \  + \tfrac{i}{\sqrt{2}}\Bar{\lambda}_\mu(p)\, c\Bar{c}\, \partial\xi \,\tilde\psi^\mu \ e^{-2\phi}e^{-\bar \phi} \\
				& \hskip10pt 
				- \tfrac{1}{2} \mu(p) 
				(\partial c +\bar{\partial}\bar{c})\ c\Bar{c} \ \partial\xi\Bar{\partial}\xi\ e^{-2\phi}e^{-2\bar \phi}\ \Bigr)\ e^{ip\cdot X}\,. 
			\end{split} 
		\end{equation}
		The BRST action on the gauge parameter gives
		\begin{equation}
			\label{brstNSgp} 
			\begin{split} 
				Q \Lambda_{{}_\NS} = & \int\frac{d^Dp}{(2\pi)^D}
				\Bigl(\, 
				\tfrac{i}{2}(p_\nu\lambda_\mu+p_\mu\bar \lambda_\nu)\ c\Bar{c}\, \psi^\mu\Tilde{\psi}^\nu e^{-\phi}e^{-\bar \phi}  \\
				& \hskip40pt  + (\tfrac{i}{2}p\cdot \lambda -\mu)\, c\Bar{c}\, 
				\eta\bar \partial\Bar{\xi}\, e^{-2\bar \phi}-(\tfrac{i}{2}p\cdot \bar \lambda + \mu)\, c\Bar{c}\, \partial\xi\Bar{\eta} \, e^{-2\phi}
				\\[0.5ex]
				& \hskip40pt    -\tfrac{i}{\sqrt{8}} 
				(\tfrac{1}{2}p^2\lambda_\mu + ip_\mu \mu)(\partial c+\bar \partial \bar c)c\Bar{c}\, \psi^\mu\bar\partial\Bar{\xi}\, 
				e^{-\phi}e^{-2\bar \phi}\\
				&   \hskip40pt        -\tfrac{i}{\sqrt{8}} (-\tfrac{1}{2}p^2\bar \lambda_\mu + ip_\mu \mu)(\partial c+\bar \partial \bar c)c\Bar{c}\, 
				\partial\xi\Tilde{\psi}^\mu\,  e^{-2\phi}e^{-\bar \phi}  \Bigr)e^{ip\cdot X}\,. 
			\end{split} 
		\end{equation}
		We can now read off the gauge transformations of the component fields
		\be
		\begin{split}
			\delta e_{\mu\nu}&  = ip_\nu\lambda_\mu + ip_\mu\bar \lambda_\nu\,, \\
			\delta e &  =  -\tfrac{i}{2}p\cdot \lambda + \mu\,, \\
			\delta  \bar e & = \tfrac{i}{2}p\cdot \bar \lambda +\mu\,, \\
			\delta f_\mu & = \tfrac{1}{2}p^2\lambda_\mu + ip_\mu \mu\,, \\
			\delta \bar f_\mu & = -\tfrac{1}{2}p^2\bar \lambda_\mu + ip_\mu \mu\,.
		\end{split}
		\ee
		We have the BRST action on the NS string field
		\be
		\begin{split}
			Q \Psi_{{}_\NS}= & \int\frac{d^Dp}{(2\pi)^D}
			\biggl(\, 
			\tfrac{1}{4}(\tfrac{1}{2}\, p^2e_{\mu\nu}(p) - ip_\nu f_\mu +									ip_\mu\bar f_\nu)\ (\partial c +\bar \partial \bar c)\ c\Bar{c}\, \psi^\mu\Tilde{\psi}^\nu e^{-\phi}e^{-\bar \phi}\\
			&-\tfrac{1}{4}(p^2 e(p)+ip\cdot f)\ (\partial c +\bar \partial \bar c) \,c\Bar{c}\, \eta\bar \partial\Bar{\xi}\, e^{-2\bar \phi} -\tfrac{1}{4}(p^2 \bar e(p)+ip\cdot \bar f)\ (\partial c +\bar \partial \bar c) \, c\Bar{c}\, \partial\xi\Bar{\eta} \, e^{-2\phi} \\
			&+  \tfrac{1}{\sqrt{2}}  \,  \bigl(\, \tfrac{1}{2}p^\nu e_{\mu\nu} - p_\mu \bar e -i  f_\mu\big)
			c\Bar{c}\, \psi^\mu\bar\eta\, 
			e^{-\phi} -  \tfrac{1}{\sqrt{2}}  \,  \bigl(\,\tfrac{1}{2}p^\mu e_{\mu\nu} + p_\nu  e +i \bar f_\nu\big)
			c\Bar{c}\,\eta\, \tilde  \psi^\nu\,   e^{-\bar \phi}\biggr)e^{ip\cdot X}\,. 
		\end{split} 
		\ee
		The quadratic action for the massless fields in the NS sector is then
		\be
		\begin{split}
			&   \hskip-90pt S_2\bigl|_\NS = \tfrac{1}{2} \langle\Psi_{{}_\NS}, Q\Psi_{{}_\NS}\rangle \\
			= \tfrac{1}{2}  \int\frac{d^Dp}{(2\pi)^D}
			\biggl(\, & -\tfrac{1}{16} e^{\mu\nu}(-p)p^2e_{\mu\nu}(p) + \tfrac{i}{8}e^{\mu\nu}(-p)p_\nu f_\mu(p) - \tfrac{i}{8}e^{\mu\nu}(-p)p_\mu \bar f_\nu(p) \\
			&-\tfrac{1}{4}\big[ e(-p)p^2\bar e(p) +i\bar e(-p)p\cdot f (p) + \bar e(-p)p^2 e(p) + i e(-p)p\cdot \bar f(p) \big]\\
			&- \tfrac{i}{8} f^\mu(-p)p^\nu e_{\mu\nu}(p) + \tfrac{i}{4} f^\mu(-p)p_\mu \bar e(p)-\tfrac{1}{4} f^\mu(-p)f_\mu(p) \\
			&+ \tfrac{i}{8} \bar f^\nu(-p)p^\mu e_{\mu\nu}(p) + \tfrac{i}{4} \bar f^\mu(-p)p_\mu e(p)-\tfrac{1}{4} \bar f^\mu(-p)\bar f_\mu(p) \biggr)\\
			= \tfrac{1}{8}  \int d^Dx
			\biggl(\, & \tfrac{1}{4} e^{\mu\nu}\partial^2 e_{\mu\nu} +\tfrac{1}{2}e^{\mu\nu}\partial_\nu f_\mu -  \tfrac{1}{2}e^{\mu\nu}\partial_\mu \bar f_\nu+e\partial^2\bar e-\bar e\partial\cdot f  + \bar e\partial^2 e - e\partial\cdot \bar f \\
			&-\tfrac{1}{2} f^\mu\partial^\nu e_{\mu\nu} + f^\mu\partial_\mu \bar e - f^\mu f_\mu + \tfrac{1}{2} \bar f^\nu\partial^\mu e_{\mu\nu} + \bar f^\mu\partial_\mu e - \bar f^\mu\bar f_\mu\biggr)\,. 
		\end{split}
		\ee    
		Upon partial integration, the action simplifies to
		\be
		S_2\bigl|_\NS   = \tfrac{1}{8}  \int d^Dx\left( \tfrac{1}{4} e^{\mu\nu}\partial^2 e_{\mu\nu} + 2 e\partial^2\bar e - f^\mu(\partial^\nu e_{\mu\nu}  - 2                                                                        \partial_\mu \bar e) + \bar f^\nu(\partial^\mu e_{\mu\nu}   + 2\partial_\mu  e) - f^\mu f_\mu - \bar f^\mu\bar f_\mu\right)\,. 
		\ee    
		Integrating out the auxiliary fields $f_\mu, \bar f_\mu$, we get the action
		\be
		\begin{split}
			S_2\bigl|_\NS   = \tfrac{1}{8}  \int d^Dx
			\Bigl(  &\  \tfrac{1}{4} e^{\mu\nu}\partial^2 e_{\mu\nu}
			+\tfrac{1}{4}(\partial^\nu e_{\mu\nu})^2
			+\tfrac{1}{4}(\partial^\nu e_{\nu\mu})^2 \\
			& \  - e\partial^\mu\partial^\nu e_{\mu\nu} + \bar e\partial^\mu\partial^\nu e_{\mu\nu} + 2 e\partial^2\bar e - e\partial^2 e - \bar e\partial^2\bar e \Bigr)\,.
		\end{split} 
		\ee    
		Letting  $d = \tfrac{1}{2}(e-\bar e)$,  one finds
		\be
		S_2\bigl|_\NS   = \tfrac{1}{8}  \int d^Dx
		\left( \tfrac{1}{4} e^{\mu\nu}\partial^2 e_{\mu\nu}+\tfrac{1}{4}(\partial^\nu e_{\mu\nu})^2+ \tfrac{1}{4}(\partial^\nu e_{\nu\mu})^2 - 2 d\, \partial^\mu\partial^\nu e_{\mu\nu}  - 4 d\, \partial^2 d  \right)\,. 
		\ee
		The gauge transformations are
		\be
		\label{linnsgt}
		\delta e_{\mu\nu}  = \partial_\nu\lambda_\mu+\partial_\mu\bar \lambda_\nu\,, \ \ \ \ 
		\delta d  = -\tfrac{1}{4}(\partial\cdot \lambda +  \partial\cdot \bar \lambda)\,. 
		\ee
		With the graviton $h_{\mu\nu}$ and Kalb-Ramond field $b_{\mu\nu}$ appearing 
		as
		\be
		e_{\mu\nu}  = h_{\mu\nu}  + b_{\mu\nu}  , 
		\ee
		the action becomes
		\be
		\label{nsnsquadratic3476}
		S_2\bigl|_\NS  = \tfrac{1}{8}  \int d^Dx
		\left( \tfrac{1}{4} h^{\mu\nu}\partial^2 h_{\mu\nu}+\tfrac{1}{2}(\partial^\nu h_{\mu\nu})^2+\tfrac{1}{4} b^{\mu\nu}\partial^2 b_{\mu\nu}+\tfrac{1}{2}(\partial^\nu b_{\mu\nu})^2 - 2 d\, \partial^\mu\partial^\nu h_{\mu\nu}  
		- 4 d\, \partial^2 d  \right)\,.
		\ee
		This NSNS kinetic term is identical to the one arising  from the bosonic closed string field theory~\cite{Sen:2024nfd}.
	Indeed, the bosonic SFT massless sector uses exactly the same fields, and
		the gauge parameters are also the same.  
		In fact, we adjusted the above NSNS string field
		and the gauge parameter string field so that also the gauge transformations take exactly the same form as well. 
		
		We parameterize $\lambda$ and $\bar \lambda$ in terms 
		$X_\mu$ and $\e_\mu$ defined by  
		\be\label{xepsilon} 
		X_\mu =  \tfrac{1}{2} ( \lambda_\mu + \bar \lambda_\mu) \,, \ \ \ 
		\e_\mu = \tfrac{1}{2} ( \lambda_\mu - \bar \lambda_\mu)\,. 
		\ee
		The gauge transformations then appear as 
		\be
		\delta h_{\mu\nu}  = \partial_\mu X_\nu + \partial_\nu X_\mu \,, 
		\ \ \ \   \delta b_{\mu\nu}  =  - \partial_\mu \e_\nu + \partial_\nu e_\mu \,, 
		\ \ \ \delta d = - \tfrac{1}{2} \partial \cdot X \,. 
		\ee
		Note that for diffeomorphisms we have $\epsilon_\mu=0$ and for Kalb-Ramond gauge transformations we have $X^\mu = 0$. Therefore,
		\be\label{nsgaugee}
		\begin{split}
			\hbox{Diffeomorphisms:} &  \ \ \ X_\mu= \lambda_\mu = \bar \lambda_\mu   \, 
			\\
			\hbox{Kalb-Ramond:} &  \ \ \ \e_\mu= \lambda_\mu = -\bar \lambda_\mu \,.      
		\end{split}
		\ee
The antisymmetric tensor $b_{\mu\nu}$ is best thought of as a two form 
\be
\label{bformKR}
b \equiv \tfrac{1}{2} b_{\mu\nu} \, \d x^\mu \wedge \d x^\nu\,.
\ee  
Using form language to express the kinetic term of the $b$ field, the quadratic
action~\refb{nsnsquadratic3476} becomes 
\be
		\label{nsnsquadratic3466}
		S_2\bigl|_\NS  = \tfrac{1}{8}  \int d^Dx
		\left( \tfrac{1}{4} h^{\mu\nu}\partial^2 h_{\mu\nu}+\tfrac{1}{2}(\partial^\nu h_{\mu\nu})^2
		- 2 d\, \partial^\mu\partial^\nu h_{\mu\nu}  
		- 4 d\, \partial^2 d  \right)  \ - \, \tfrac{1}{16} ( \d b , \d b )  \,.
		\ee	 
		
		\subsection{NSNS-RR-RR couplings from SFT}
		
		The aim of this subsection is to compute from string field theory the three-point couplings
		that represent the interactions of the NSNS field with the RR fields.  Such a cubic term involves
		one NSNS field and two RR fields.  We will only focus on fields in the massless sector.
		
		The three-point vertex coupling two RR fields and one NSNS field is, from~\refb{S2+S3},
		\be
		\label{actioncubicnsrr}
		S|_{\rm cubic} \ = \  \tfrac{1}{2} \, \{\Psi_{{}_\NS}
		, \Psi_{{}_\RR}
		, \Psi_{{}_\RR}
		\} \,,
		\ee
		a multilinear function that requires no picture changing insertion.  The NSNS
		string field $\Psi_{{}_\NS}$ was given in~\refb{nsnsfield}.
		Only the state associated with $e_{\mu\nu}$ contributes;
		when inserted on the correlator the other states states give no contribution
		because they do not yield the total exponentials $e^{-2\phi} e^{-2\bar\phi}$ 
		that are required. 
		The multilinear function is thus
		\be
		\label{multnsrr}
		S|_{\rm cubic} \ 
		=  \  \frac{1}{4} \int\frac{d^Dp_1}{(2\pi)^D}\frac{d^Dp_2}{(2\pi)^D}\frac{d^Dp_3}{(2\pi)^D} \,  e_{\mu\nu}(p_1)Q^{ab}(p_2)Q^{cd}(p_3) \, Z_{abcd}^{\mu\nu}\,,  
\ee
where $Z$ represents the correlator  
\be
Z_{abcd}^{\mu\nu} = 
			\expval{c\bar{c}\, \psi^\mu\bar{\psi}^\nu e^{-\phi}e^{-\bar \phi} e^{ip_1\cdot X}(z_1)\  c\bar{c}\, e^{-\phi/2}\Theta_{a}\, e^{-\bar \phi/2} \bar{\Theta}_be^{ip_2\cdot X}(z_2)\,  c\bar{c}\, e^{-\phi/2}\Theta_{c}  e^{-\bar \phi/2} 
				\bar{\Theta}_de^{ip_3\cdot X}(z_3)}\,.
				\ee
It is useful to note that, strictly speaking,  this correlator uses the three-string vertex with its off-shell data;  the operators are to be inserted with specific local coordinates at the special points $z_1, z_2,$ and~$z_3$.  But except for corrections due to the momenta, the operators are dimension zero primaries, and thus, to leading
order of derivatives 
 (which here is no derivatives), the off-shell data is irrelevant. 
The correlator, in fact, does not depend on the $z_i$. 
  Rearranging the operators, we have
\be
\begin{split}
			Z_{abcd}^{\mu\nu}=  &
			\ \expval{c\bar{c}(z_1)c\bar{c}(z_2)c\bar{c}(z_3)}
			\expval{e^{ip_1\cdot X}(z_1)e^{ip_2\cdot X}(z_2)e^{ip_3\cdot X}(z_3)}\\[0.7ex]
			& \ \ \expval{e^{-\phi} \psi^\mu(z_1) \,      e^{-\phi/2} \Theta_a (z_2)\, 
				e^{-\phi/2} \Theta_c (z_3) }   
			\bigl\langle e^{-\bar\phi} \bar\psi^\nu (\bar z_1)  \, e^{-\bar\phi/2} \bar\Theta_b (\bar z_2)\, 
e^{-\bar\phi/2} \bar\Theta_d (\bar z_3) \bigr\rangle \,.
\end{split}
\ee
There is no sign rearranging the $c, \bar c$ ghosts, and there are two minus signs from
		rearranging the $e^{-\phi}, e^{-\bar \phi}, \psi, \bar \psi$ and the Grassmann odd
		$\phi$-dressed spin fields. 
		The correlator involving spin fields was given in~\refb{useful3pt}, with an
		analogous equation for the antiholomorphic sector.  It follows that $Z$, to leading order in momenta,  is now given by 
		\be
		\begin{split}
			Z_{abcd}^{\mu\nu} =  & \ 
			-  \abs{z_{12}z_{23}z_{31}}^2  (2\pi)^D\delta^D(\sum p) \cdot \tfrac{1}{2} 
			\abs{z_{12}z_{23}z_{31}}^{-2}  \,  (\Gamma^\mu C^{-1} )_{ac}  
			\,  (\Gamma^\nu C^{-1} )_{bd}  \\
			=  & \ 
			-  \tfrac{1}{2}  (2\pi)^D\delta^D(\sum p)  \,  (\Gamma^\mu C^{-1} )_{ac}  \,  (\Gamma^\nu C^{-1} )_{bd}\,.
\end{split}
\ee
Having determined $Z$, we can now return to the calculation of cubic coupling
		in~\refb{multnsrr}. To leading zeroth order in momenta:  
		\be
		\begin{split}
			S|_{\rm cubic} \  =\  & -2^{-3} \int\frac{d^Dp_1}{(2\pi)^D}\frac{d^Dp_2}{(2\pi)^D}
			\frac{d^Dp_3}{(2\pi)^D} \ e_{\mu\nu}(p_1)Q^{ab}(p_2)Q^{cd}(p_3)
			\\  & \ \ \cdot 
			(2\pi)^D\delta^D(\sum p) (\Gamma^\mu C^{-1})_{ac}(\Gamma^\nu C^{-1})_{bd}\,.
		\end{split}
		\ee
		In position space, this is
		\be
		\begin{split}
			S|_{\rm cubic} \  =  \ &  -2^{-3}  \int \d^Dx \  e_{\mu\nu}(x)Q^{ab}(x)Q^{cd}(x)(\Gamma^\mu C^{-1})_{ac}(\Gamma^\nu C^{-1})_{bd}\\
			= \ & -2^{-3} 
			\int \d^Dx \ e_\mu^{\ \, \nu}(x)  R^\mu_{\ \, \nu} (x) \,, \ \ \ \ 
			\hbox{with}  \ \ R^\mu_{\ \, \nu} \equiv  \Tr{\Gamma^\mu C^{-1} Q {C^{-1}}^T{\Gamma_\nu}^TQ^T}\,.
		\end{split}
		\ee
Manipulating the trace using $C\Gamma_\nu C^{-1} = - \Gamma_\nu^T$, we find
		\be
		R^\mu_{\ \, \nu}= - \Tr{\Gamma^\mu C^{-1} Q \Gamma_\nu{C^{-1}}^TQ^T} = - \Tr{\Gamma^\mu (C^{-1} Q) \Gamma_\nu(QC^{-1})^T}\,.
		\ee
Using the expansion of $Q^{ab}$ in~\refb{comopne} we find the partial results
		\be
		\begin{split}
			& (C^{-1} Q)_{\dot{a}}^{\hskip5pt b} = \sum_k \frac{1}{k!}\, i\, Q^{(k)}_{\mu_1\cdots\mu_k}(\Gamma^{\mu_1\cdots\mu_k})_{\dot{a}}^{\hskip5pt b}\,, \\
			& (QC^{-1})^a_{\hskip5pt \dot b}= \sum_{k}\frac{1}{k!}\, i\, Q^{(k)}_{\mu_1\cdots\mu_k}(C \Gamma^{\mu_1\cdots\mu_k} C^{-1})^a_{\hskip5pt \dot b} = \sum_{k}(-1)^{k(k+1)/2}\frac{1}{k!}\, i\, Q^{(k)}_{\mu_1\cdots\mu_k}(\Gamma^{\mu_1\cdots\mu_k T})^a_{\hskip5pt \dot b}\,, \\
			& ((QC^{-1})^T)_{\dot{a}}^{\hskip5pt b} = \sum_{k} (-1)^{k(k+1)/2} \frac{1}{k!}\, i\, Q^{(k)}_{\mu_1\cdots\mu_k}(\Gamma^{\mu_1\cdots\mu_k})_{\dot{a}}^{\hskip5pt b}\,. 
		\end{split}
		\ee
		From these it follows that  
		\be 
		R^\mu_{\ \, \nu}=  
		\hskip-5pt\sum_{k,p \, \in \, \hat{\mathbb{Z}}_{odd }} (-1)^{p(p+1)/2}\frac{1}{k!p!}Q^{(k)}_{\mu_1\cdots\mu_k}\, Q^{(p)\nu_1\cdots\nu_p} \cdot \tfrac{1}{2} \Tr{ \Gamma^\mu \Gamma^{\mu_1\cdots\mu_k} \Gamma_\nu \Gamma_{\nu_1\cdots\nu_p} }\,,
		\ee
		with the extra $\tfrac{1}{2}$ arising because the trace is only implemented on one type of
		index, due to the presence of the $Q$ bispinors.  The above trace is evaluated by
		using the second and the last equation in~\refb{property}.
		One finds four terms, two coupling $Q$ fields of the same rank, and two coupling $Q$
		fields with ranks differing by two. The later two terms are quickly seen to be the same.  
		The computation gives, for the action,
		\be
		\begin{split}\label{RRNS}
			S|_{\rm cubic}  
			& =  - 2^{{d\over 2}-4}\hskip-5pt \int \d^Dx
			\hskip-5pt \sum_{k\in \hat{\mathbb{Z}}_{odd}}\, {1\over k!} 
			\Bigl[
			(k+1) Q^{(k)}_{\mu_1\cdots\mu_k}e_{\mu}^{\hskip5pt [\mu}
			Q^{(k)\mu_1\cdots\mu_k]} - k\, Q^{(k)\mu}_{\ \ \ \  \mu_1\cdots\mu_{k-1}}e_{\mu \nu}
			Q^{(k)\nu\mu_1\cdots\mu_{k-1}} \\
			& \hskip120pt  \ +2 Q^{(k)}_{\mu_1\cdots\mu_k}e_{\mu\nu}Q^{(k+2)\mu\nu\mu_1\cdots\mu_k}  
			\Bigr]\,.  
		\end{split}
		\ee
		The first and second terms above actually give equal contributions.  
		This can be verified
		by using the following identity, derived using the duality relations for the $Q$'s: 
		\be \label{cubicdual}
		\frac{(k+1)}{k!} Q^{(k)}_{\mu_1\cdots\mu_k}e_{\mu}^{\hskip5pt [\mu}
		Q^{(k)\mu_1\cdots\mu_k]}  = -\frac{1}{(9-k)!} Q^{(10-k)\mu}_{ \hskip30pt  \mu_1\cdots\mu_{9-k}}e_{\mu \nu}
		Q^{(10-k)\nu\mu_1\cdots\mu_{9-k}}.
		\ee
		The action is then compactly written as
		\be
		\begin{split}\label{actionnnnnn}
			S|_{\rm cubic} & =   2^{{d\over 2}-3} \int \d^Dx \hskip-5pt\sum_{k\in \hat{\mathbb{Z}}_{odd}} 
			\biggl[ \frac{1}{(k-1)!}\, 
			Q^{(k)\mu}_{\ \ \ \  \mu_1\cdots\mu_{k-1}}h_{\mu \nu}\, 
			Q^{(k)\nu\mu_1\cdots\mu_{k-1}}  \, -\frac{1}{k!}
			Q^{(k)}_{\mu_1\cdots\mu_k}b_{\mu\nu}\, Q^{(k+2)\mu\nu\mu_1\cdots\mu_k}  
			\, \biggr]
		\end{split}
		\ee
		noting that with $e_{\mu\nu} = h_{\mu\nu} + b_{\mu\nu}$, only the gravity field
		contributes in the first term, and only the antisymmetric field contributes in the 
		second term.  We now write the above action in terms of the independent $Q$ fields
		$Q^{(1)}, Q^{(3)}, $ and $Q^{(5)}$.  One can see, using~\refb{cubicdual} again, that the contribution from $Q^{(9)}$, for example, gives the same coupling of $Q^{(1)}$ to
		$h_{\mu\nu}$ {\em plus} a coupling of $Q^{(1)}$ to $h = h_\mu^{\ \mu}$.  We then find
		\be
		\label{cubicnsnsrrrr}
		\begin{split}
			2^{-\frac{d}{2}}S|_{\rm cubic}  =   2^{-3} \int \d^Dx \,   
			\Bigl[ \, &  \,\, \,   2\,  Q^{(1)\mu}h_{\mu \nu}\, Q^{(1)\nu} - h \, Q^{(1)\mu}\, Q^{(1)}_\mu 
			\\[0.5ex] 
			&\hskip-6pt  + Q^{(3)\mu}_{\ \ \ \  \mu_1\mu_2}h_{\mu \nu}\, 
			Q^{(3)\nu\mu_1\mu_2}- \tfrac{1}{3!}  h\,  Q^{(3)}_{\mu_1\mu_2\mu_3}
			Q^{(3)\mu_1\mu_2\mu_2}\\[1.0ex]
			& \hskip-8pt   +  \tfrac{1}{4!} Q^{(5)\mu}_{\ \ \ \ \mu_1\cdots\mu_{4}} h_{\mu\nu} Q^{(5)\nu\mu_1\cdots\mu_{4}} \\[0.7ex]
			& \hskip-10pt  - 2\,  Q^{(1)}_{\mu_1} b_{\mu\nu}\, Q^{(3)\mu\nu\mu_1} -  \tfrac{2}{3!} Q^{(3)}_{\mu_1\mu_2\mu_3}b_{\mu\nu}\, Q^{(5)\mu\nu\mu_1\mu_2\mu_3} \, \Bigr]\,,\\[2.0ex]
			& \hskip-70pt =  \  \, 
			\tfrac{1}{2}\, (Q^{(1)}, h\,\hd\, Q^{(1)}) +  \tfrac{1}{2}\, (Q^{(3)}, h\,\hd\,Q^{(3)})  + \tfrac{1}{4}\,  (Q^{(5)}, h\,\hd\, Q^{(5)}) \\[1.4ex]
			& \hskip-60pt   - \tfrac{1}{2}\, (Q^{(1)}\wedge b\, , Q^{(3)}) - \tfrac{1}{2}   (Q^{(3)}\wedge b\, , Q^{(5)})\,. 	\\[2.0ex]
			& \hskip-70pt =  \  \,
			\sum_{k\in \hat{\mathbb{Z}}_{odd}} 
			\tfrac{1}{4}\, (Q^{(k)}, h\,\hd\, Q^{(k)})   - \tfrac{1}{4}\, (Q^{(k)}\wedge b\, , Q^{(k+2)}) \,. 	
		\end{split}
		\ee
In passing to the second form of the action we used the notation introduced in~\refb{haction}.  To get the last form of the action we used the duality relations
for $Q$ and the second relation  in~\refb{hactidentity}.	
		
\sectiono{Diffeomorphisms in the NSNS and RR sectors} \label{diffinthensnsrr} 
		In this section, we use the SFT to determine how diffeomorphisms act on the RR fields. Finding this action on the extra field at picture $ (-{3\over 2},-{3\over 2})$ requires evaluating the string field product of the NSNS gauge parameter and the {\em physical} RR string field. 
		The action of diffeomorphisms on 
		the physical field at picture	$(-{1\over 2},-{1\over 2})$ is then obtained by picture changing the transformations of the extra fields.

		\subsection{Diffeomorphisms in the RR sector from SFT}  \label{diffffff}

		The diffeomorphisms of the RR sector fields are the gauge transformations in~\refb{30erdlk}, where we only consider the NSNS gauge parameter
		\be\label{gauge}
		\begin{split}
			\delta \wt\Psi_{{}_\RR}  =\ & \  [ \Lambda_{{}_\NS}   , \Psi_{{}_\RR}]  + \cdots \, , \ \ \ \ \\
			\delta \Psi_{{}_\RR}  =\ & \  {\cal G}   [ \Lambda_{{}_\NS}  , \Psi_{{}_\RR}]   + \cdots = \, 
			{\cal X}_0 \bar{\cal X}_0 \, 
			\delta \wt\Psi_R \,.  
		\end{split}  
		\ee
		The dots represent terms quadratic and higher order in fields; our computations
		will only involve terms linear in the fields.  The 
		NSNS gauge parameter was
		given in~\refb{vertexNSgp} and includes parameters for diffeomorphisms as well as Kalb-Ramond transformations.  We will compute both, though our analysis will
		focus mostly on diffeomorphisms.     
		We will begin by computing the gauge
		transformation of the extra field $\wt\Psi_{{}_\RR}$.  With this field defined
		in~\refb{auxRRfield2} we  write the variations as follows:  
		\be
		\label{auxRRfield26}
		\begin{split}
			\delta \wt\Psi_{{}_\RR}  = \int\frac{d^Dp}{(2\pi)^D}\Big( &\ \ \ \  \delta 
			N^{\Dot{a}\Dot{b}}(p)\, c \Bar{c} \  
			e^{-3\phi/2}\Theta_{\dot a}  \, e^{-3\bar \phi/2}\Bar{\Theta}_{\dot b}\, e^{ip\cdot X}\\ 
			&+ \tfrac{1}{2}\, \delta P^{\Dot{a}b}(p)\,
			(\partial c + \bar \partial \bar c)c  \Bar{c}\, \bar \partial \bar\xi \, 
			e^{-3\phi/2}\Theta_{\dot a}  \,  e^{-5\bar \phi/2}\Bar{\Theta}_{ b}\, e^{ip\cdot X}\\
			& + \tfrac{1}{2}\, \delta\Bar{P}^{a\Dot{b}}(p)\, 
			(\partial c + \bar \partial \bar c)c  \Bar{c}\, \partial \xi\ 
			e^{-5\phi/2}\Theta_{ a}  \, e^{-3\bar \phi/2}\Bar{\Theta}_{\dot b}\, e^{ip\cdot X}\Big)\,. 
		\end{split}
		\ee
		We pick the variations by defining a set of dual states relative to the inner product:
		\be
		\begin{split}
			{\cal O}^{\dot a \dot b} (p)  = & \, -\, (\partial c+\bar \partial\bar c) \, 
			c\bar{c}\  e^{- \phi/2}\Theta^{\dot a}\, 
			e^{-\bar \phi/2}\Bar{\Theta}^{\dot b}\,  e^{ip\cdot X}\,, \\[0.5ex]
			{\cal O}^{\dot a b} (p)= & \   -   2\,  c\bar{c}\, \bar \eta\,  e^{-\phi/2}\Theta^{\dot a}\ 
			e^{\bar \phi/2}\Bar{\Theta}^{ b}\  e^{ip\cdot X} \, ,  \\[0.5ex]
			{\cal O}^{a \dot b} (p) =& \   - 
			2\, c\bar{c}\,  \eta\,  \ e^{\phi/2}\Theta^{a}\ 
			e^{-\bar \phi/2} \Bar{\Theta}^{\dot b}\  e^{ip\cdot X}\, \,,
		\end{split}
		\ee
		where the spin fields index is raised with the $C$ matrix:  $\Theta^\alpha = C^{\alpha\beta} \Theta_\beta$, in both holomorphic and anti-holomorphic sectors. 
		Calculating the overlap with the variation $\delta\wt\Psi{{}_\RR}$, we confirm that the above states
		select the desired variations 
		\be
		\label{30ridlk}
		\begin{split}
			\expval{   {\cal O}^{\dot a \dot b} (p) \, ,\, \delta \wt\Psi_{{}_\RR}} = & \ 
			\delta N^{\dot a \dot b} (-p) \,, \\
			\expval{  
				{\cal O}^{\dot a b}(p)\, ,\delta \wt \Psi_{{}_\RR}} =  & \ \delta P^{\dot a b} (-p) \,, \\
			\expval{
				{\cal O}^{a \dot b}(p) \, ,\delta \wt\Psi_{{}_\RR}} = &   \ \delta {\bar P}^{a\dot b}(-p)\,. 
		\end{split}
		\ee
		We can now find the gauge transformations in~\refb{gauge}.  The RR string field vertex operator   is given in \refb{auxRRfield2}, and using~\refb{30ridlk} we have 
		\be
		\delta N^{\dot a \dot b} (-p)  =   \{  {\cal O}^{\dot a \dot b}(p) , \Lambda_{{}_\NS}, 
		\Psi_{{}_\RR} \} = 0 \,, \ee
		due to $bc$-ghost number conservation: there are no $b$-antighosts,  the first operator
		contains $(\partial c+ \bar \partial c)c \bar c$,  the last contains $c \bar c$, and the middle one (see~\refb{vertexNSgp}), two or three $c$ factors.  Thus we have one or two factors of $c$ too many. 
		
		For the variation of $P$ we have 
		\be
		\begin{split}
			\delta P^{\dot a b} (-p)=  \ & \{
			{\cal O}^{\dot a b}(p) , \Lambda_{{}_\NS}, \Psi_{{}_\RR}\} =  \,  i \sqrt{2} 
			\int {d^Dp_2\over (2\pi)^D}  {d^Dp_3\over (2\pi)^D } 
			\, \lambda_\mu(p_2)   
			Q^{cd}(p_3)\\
			& \hskip-60pt  \expval{c\bar{c} 
				\, \bar\eta   
				e^{-\phi/2}\Theta^{\dot a} \, e^{\bar \phi/2} \Bar{\Theta}^{b} e^{ip\cdot X}(z_1) \ 
	c\Bar{c}\, 
	\psi^\mu    \, \bar\partial\bar\xi   
	\ e^{-\phi}e^{-2\bar \phi}e^{ip_2\cdot X} (z_2)  
	c\bar{c}\,  e^{-\phi/2} \Theta_c \, e^{-\bar \phi/2} \Bar{\Theta}_d 
				e^{ip_3\cdot X}  (z_3)}\,. 
		\end{split}
		\ee
Once again, the off-shell data of the three-string vertex does not enter
this computation to the leading order order 
in derivatives (zeroth order). 
		After doing the ghost correlator, the $\bar \eta$, $\bar \xi$ correlator, and moving
		some operators around we have
		\be
		\begin{split}
			\delta P^{\dot a b} (-p)=  \ &
			-\,  i \sqrt{2}\,  { |z_{12} z_{13} z_{23} |^2 \over
				\bar z_{12}^2}  
			\int {d^Dp_2\over (2\pi)^D}  {d^Dp_3\over (2\pi)^D } 
			\, \lambda_\mu(p_2)   
			Q^{cd}(p_3)  (2\pi)^D \delta^D(p+ p_2+p_3) \\
			&  
			\expval{
				e^{-\phi/2}\Theta^{\dot a}(z_1) \, \psi^\mu  e^{-\phi} (z_2)  e^{-\phi/2} \Theta_c (z_3)}
			\expval{\, e^{-2\bar \phi} (\bar z_2) \, 
				e^{\bar \phi/2} \Bar{\Theta}^{b} (\bar z_1)
				\, e^{-\bar \phi/2} \Bar{\Theta}_d  (z_3)}\,. 
		\end{split}
		\ee
		The first and second correlators are given by
		\be
		\begin{split}
			\expval{
				e^{-\phi/2}\Theta^{\dot a}(z_1) \, \psi^\mu  e^{-\phi} (z_2)  e^{-\phi/2} \Theta_c (z_3)}
			= & \  -{1\over \sqrt{2}} {1\over z_{12} z_{13} z_{23}} ( \Gamma^{\mu T} )^{\dot a}_{\ c} \,, \\[1.0ex] 
			\expval{
				\, e^{-2\bar \phi} (\bar z_2) \, 
				e^{\bar \phi/2} \Bar{\Theta}^{b} (\bar z_1)
				\, e^{-\bar \phi/2} \Bar{\Theta}_d  (z_3)}
			= & \  {\bar z_{12} \over  \bar z_{13}
				\, \bar z_{23}  }  \, \delta_d^b\,,  
		\end{split}
		\ee
		and therefore we get:
		\be
		\delta P^{\dot a b} (-p)=  \ 
		i  \int {d^Dp_2\over (2\pi)^D}  {d^Dp_3\over (2\pi)^D } 
		(2\pi)^D \delta^D(p+ p_2+p_3) 
		\, \lambda_\mu(p_2)   (\Gamma^{\mu T} Q(p_3))^{\dot a b}  \,. 
		\ee
		The  variation  $\delta\bar P$ is computed analogously
		\be
		\begin{split}
			\delta \bar P^{a \dot b} (-p) =  & \ \{   {\cal O}^{ a \dot b}(p)\, , 
			\Lambda_{{}_\NS}, \Psi\} 
			= - {i}{\sqrt{2}}\int {d^Dp_2\over (2\pi)^D}  {d^Dp_3\over (2\pi)^D } \, 
			\bar \lambda_\mu(p_2) Q^{cd}(p_3)\\
			&  \hskip-60pt 
			\expval{c\bar{c} \,\eta \, 
				e^{\phi/2}\Theta^a\, e^{-\bar \phi/2}\Bar{\Theta}^{\dot b}  e^{ip\cdot X} (z_1) \, c\Bar{c}\, \partial\xi \,\tilde\psi^\mu \ e^{-2\phi}e^{-\bar \phi}e^{ip_2\cdot X} (z_2)  \ 
				c\bar{c} \, e^{-\bar \phi/2}\Theta_c\, e^{-\phi/2}\Bar{\Theta}_d   
				e^{ip_3\cdot X} (z_3)}\,,
		\end{split}
		\ee
		and the result is 
		\be
		\delta \bar P^{ a \dot b} (-p)=  \  i 
		\int {d^Dp_2\over (2\pi)^D}  {d^Dp_3\over (2\pi)^D } 
		(2\pi)^D \delta^D(p+ p_2+p_3) 
		\, \bar\lambda_\mu(p_2)   (Q(p_3) \Gamma^\mu)^{a \dot b}  \,.
		\ee
		In position space, these are
		\be
		\begin{split}
			\delta P^{\dot a b}=  i  \, \lambda_\mu   (\Gamma^{\mu T} Q)^{\dot a b}\,,\hskip40pt
			\delta \bar P^{ a \dot b} = i \,  \bar\lambda_\mu   (Q \Gamma^\mu)^{a \dot b}  \,.
		\end{split}
		\ee
		Using the $\Gamma$ matrix identities \refb{diffcliffiden}, we get the transformation of the component fields 
		\be
		\begin{split}
			& \delta  P^{(k)} =  \lambda \wedge Q^{(k-1)} +  i_{ \lambda^\flat} Q^{(k+1)}\,,\\[0.5ex]
			& \delta \bar P^{(k)} = \bar \lambda \wedge Q^{(k-1)} - i_{\bar \lambda^\flat} Q^{(k+1)}\,. 
		\end{split}
		\ee
		In terms of the  fields $P^{(k)}_\mp$ defined in equation~\refb{defpmP} and recalling that $\delta N^{(k)} = 0$, we have
		\be
		\delta P^{(k)}_{\mp} =   \tfrac{1}{2} [\lambda_\mp \wedge Q^{(k-1)} + i_{ \lambda_\pm} Q^{(k+1)}]\,, \qquad \lambda_\pm \equiv \tfrac{1}{2}(\lambda\pm\bar\lambda)\,.
		\ee
		Recall the writing of $\lambda_\mu$ and  $\bar\lambda_\mu$ in terms
		of a vector $X^\mu$ and a one-form $\e_\mu$~(\refb{xepsilon}).  This
		gives, in the language of forms 
		\be   
		\begin{split}
			\lambda = & \ X^\sharp + \e\,, \ \ \ \ \  \lambda_+ =  X^\sharp\\
			\bar\lambda = & \  X^\sharp - \e\,, \ \ \ \ \  \lambda_- = \e \,. 
		\end{split} 
		\ee
		The diffeomorphism transformation is then written as
		\be\label{Pdiff}
		\delta_X P^{(k)}_- =   \tfrac{1}{2}i_X Q^{(k+1)}\, ,\hskip60pt 
		\delta_X P^{(k)}_+ =   \tfrac{1}{2} X^\sharp \wedge Q^{(k-1)}\, , 
		\ee
		while the Kalb-Ramond gauge transformation is 
		\be\label{PKR}
		\delta_\epsilon P^{(k)}_- =   \tfrac{1}{2} \epsilon \wedge Q^{(k-1)}  ,\hskip60pt 
		\delta_\e P^{(k)}_+ =  \tfrac{1}{2}i_{\e^\flat} Q^{(k+1)}\,. 
		\ee
		To find the way the standard RR fields $Q^{(k)}$ transform under diffeomorphisms we use~\refb{gauge}, which instructs us to act with picture changing operators on the variations of the extra fields.  The computation gives
		\be
		\delta\Psi_{{}_\RR}  = {\cal X}_0\Bar{{\cal X}}_0\delta \wt\Psi_{{}_\RR} 
		= \tfrac{1}{4}\left[\slashed{p}^T\delta N\slashed{p})(p) 
		+ \slashed{p}^T\delta P(p)
		+ \delta \Bar{P}\slashed{p}(p)\right]^{ab} c \Bar{c}\, e^{- \phi/2}\Theta_a\, e^{-\bar \phi/2} \Bar{\Theta}_b   e^{ip\cdot X}
		\ee
		from this and $\delta N =0$,  we get
		\be
		\delta Q^{ab}(p) = \tfrac{1}{4}\left[\slashed{p}^T\delta P(p)
		+ \delta \Bar{P}\slashed{p}(p)\right]^{ab}\,.
		\ee
		In position space this reads
		\be
		\delta Q = -\tfrac{i}{4}\left[(\slashed{\partial}^T\delta P)
		+ (\delta \Bar{P}\overleftarrow{\slashed{\partial}})\right]  = \tfrac{1}{4}\left[(\slashed{\partial}^T(\slashed{\lambda}^T Q)
		+ (Q\bar{ \slashed{\lambda}})\overleftarrow{\slashed{\partial}})\right]\,.
		\ee
		We then have from~\refb{diffcliffiden} and~\refb{derivvvaa }, that in components, the above transformations give  
		\be   
		\delta Q^{(k)} = \tfrac{1}{4}\, \d \, \Bigl(  (\lambda- \bar\lambda) \wedge Q^{(k-2)}  + i_{\lambda + \bar\lambda} Q^{(k)}  \Bigr) 
		- \tfrac{1}{4}\, \d^\dagger \, \Bigl(  (\lambda+ \bar\lambda) \wedge Q^{(k)}  + i_{\lambda - \bar\lambda} Q^{(k+2)}  \Bigr) \,,
		\ee
		which in terms of $X$ and $\e$ read
		\be\label{Qdiff}
		\delta Q^{(k)} = \tfrac{1}{2}\Bigl(
		\d (\e\wedge Q^{(k-2)}) + \d i_X Q^{(k)} - \d^\dagger( X^\sharp \wedge Q^{(k)} 
		)-\d^\dagger i_{\e^\flat} Q^{(k+2)} \Bigr) \,. 
		\ee
		For diffeomorphisms and Kalb-Ramond transformations, separately, we find 
		\be 
		\label{QKR}
		\begin{split}
			\delta_X Q^{(k)}  =  & \ \tfrac{1}{2}  \Bigl(  \ \d i_X Q^{(k)} - \ \d^\dagger( X^\sharp \wedge Q^{(k)} )  \
			\Bigr) \, = \tfrac{1}{2}  \Bigl(  \d i_X  Q^{(k)} - \d ^\dagger  i^\dagger_{X^\sharp} Q^{(k)}  \Bigr) \,, \\[0.4ex]
			\delta_\e Q^{(k)}  =  & \ \tfrac{1}{2}\Bigl(
			\d (\e\wedge Q^{(k-2)}) -\d^\dagger i_{\e^\flat} Q^{(k+2)}\Bigr) = \tfrac{1}{2}\Bigl(
			\d i^\dagger_\e Q^{(k-2)} -\d^\dagger i_{\e^\flat} Q^{(k+2)}\Bigr) \,. 
		\end{split}
		\ee
		The action of diffeomorphisms is unusual not only for the self-dual five form but
		for all other forms. The structure of the transformation preserves 
		duality relations,  including the self duality of the five form.
		%
		

		\sectiono{Collecting IIB SFT results and extension to IIA SFT }	
		
		For easy reference, we collect here the type IIB action and NSNS type gauge transformations of the fields, as derived in the previous sections.  
		We also give the corresponding results for the type IIA theory, skipping most of the computations, but giving the explicit string fields and noting how the computations differ from the previous ones in the IIB theory.

		\subsection{IIB SFT}

		The full cubic action (except for NSNS interactions)  is given by 
		combining the quadratic action in~\refb{quadSRRIIB} and the cubic action in~\refb{cubicnsnsrrrr}: 
		\be
		\label{IIBSFTaction}
		\begin{split}
			2^{-{d\over 2}} S_{{}_{\rm IIB}}
			&=\ \tfrac{1}{2}(\d P_-^{(0)}, \d  P_-^{(0)})
			-\tfrac{1}{2}(\d^\dagger  P_+^{(2)}, \d^\dagger P_+^{(2)})
			- (Q^{(1)},  \d P_-^{(0)} + \d^\dagger  P_+^{(2)})\\[2.0ex]
			& +\ \tfrac{1}{2}(\d  P_-^{(2)}, \d  P_-^{(2)})
			-\tfrac{1}{2}(\d^\dagger P_+^{(4)}, \d^\dagger P_+^{(4)})
			-(Q^{(3)},  \d P_-^{(2)} + \d^\dagger P_+^{(4)}) \\[2.0ex]
			&+\ \tfrac{1}{2}(\d P_-^{(4)}, \d P_-^{(4)})
			-(Q^{(5)},  \d P_-^{(4)}) \\[1.0ex]
			& + 2^{-{d\over 2} - 3}  \int d^Dx
			\left( \tfrac{1}{4} h^{\mu\nu}\partial^2 h_{\mu\nu}+\tfrac{1}{2}(\partial^\nu h_{\mu\nu})^2+\tfrac{1}{4} b^{\mu\nu}\partial^2 b_{\mu\nu}+\tfrac{1}{2}(\partial^\nu b_{\mu\nu})^2 - 2 d\, \partial^\mu\partial^\nu h_{\mu\nu}  - 4 d\, \partial^2 d  \right)\\[0.5ex]
			& +   
			\tfrac{1}{2}\, (Q^{(1)}, h\, \hd\,  Q^{(1)}) 
			+  \tfrac{1}{2}\, (Q^{(3}, h\, \hd\, Q^{(3)})  
+ \tfrac{1}{2}\,  (Q^{(5)}, h\, \hd\,  Q^{(5)}) \\[1.4ex]
			&   - \tfrac{1}{2}\, (Q^{(1)}\wedge b\, , Q^{(3)}) - \tfrac{1}{2}   (Q^{(3)}\wedge b\, , Q^{(5)})\,. 	
		\end{split}
		\ee
In democratic notation for the RR forms and their interactions and using the three-form $\d b$ to 
write the Kalb-Ramond kinetic term, we have		
\be\label{finXXXXaction}
\begin{split} 
 2^{-{d\over 2}} S_{{}_{\rm IIB}} \, =\,  
 &\    2^{-{d\over 2} - 3}  \int d^Dx
\left( \tfrac{1}{4} h^{\mu\nu}\partial^2 h_{\mu\nu}
+\tfrac{1}{2}(\partial^\nu h_{\mu\nu})^2 
- 2 d\, \partial^\mu\partial^\nu h_{\mu\nu}  - 4 d\, \partial^2 d  \right) 
\  -   2^{-{d\over 2} - 4}  ( \d b , \d b ) \\[1.0ex]
&+  \sum_{k\in \hat{\mathbb{Z}}_{odd}} \Bigl[\  \tfrac{1}{4}(\d P_-^{(k-1)}, \d P_-^{(k-1)}) 
 - \tfrac{1}{4}(\d^\dagger P_+^{(k+1)}, \d^\dagger P_+^{(k+1)})  
		- \tfrac{1}{2} (Q^{(k)}, \d P_-^{(k-1)}+ \d^\dagger P_+ ^{(k+1)})\\
		&  \hskip36pt +\tfrac{1}{4}\, (Q^{(k)}, h\,\hd\, Q^{(k)})   - \tfrac{1}{4}\, (Q^{(k)}\wedge b\, , Q^{(k+2)}) 	 \Bigr]. \\[-1.0ex]
\end{split}
\ee
Consider now the diffeomorphisms generated by the vector field $X = X^\mu\partial_\mu$. From ~\refb{nsgaugee},~\refb{Pdiff} and~\refb{Qdiff} 
		\be
		\begin{split}
			\delta_X P^{(k)}_-&  =   \tfrac{1}{2} i_X Q^{(k+1)}\, ,\\[0.5ex]
			\delta_X P^{(k)}_+ &  =   \tfrac{1}{2} X^\sharp \wedge Q^{(k-1)} 
			=   \tfrac{1}{2}\,  i_{ X^\sharp}^\dagger Q^{(k-1)}\, , \\[0.6ex]
			\delta_X Q^{(k)} &  = \tfrac{1}{2}\, \d i_X Q^{(k)} - \tfrac{1}{2}\, 
			\d^\dagger(X^\sharp \wedge Q^{(k)} ) 
			=  \tfrac{1}{2}\, \d i_X Q^{(k)} + \tfrac{1}{2}\, 
			* \d i_X *Q^{(k)} \\[0.5ex]
			& = \tfrac{1}{2}\, \d i_X Q^{(k)} - \tfrac{1}{2}\, 
			\d^\dagger i_{ X^\sharp}^\dagger Q^{(k)}\, , \\[0.6ex]
			\delta_X h_{\mu\nu}  & = \partial_\mu X_\nu + \partial_\mu X_\nu\, , \\[0.5ex]
			\delta_X d\  &  = -\tfrac{1}{2}\partial\cdot X\,. 
		\end{split}
		\ee
		In some of the equations above we have included equivalent forms of the
		transformations that make duality properties more manifest. 
		The Kalb Ramond gauge parameter is the one form $\e = \e_\mu\, \d x^\mu$, and from~\refb{nsgaugee},~\refb{PKR} and~\refb{QKR} we have 
		\be
		\begin{split}
			\delta_\e P^{(k)}_-&  =   \tfrac{1}{2}\, \e \wedge Q^{(k-1)}  
			= \, \tfrac{1}{2}\,  i_{\e }^\dagger \, Q^{(k-1)}\, , \\[0.5ex]
			\delta_\e P^{(k)}_+&  =   \tfrac{1}{2} \, i_{\e^\flat } \, Q^{(k+1)}\, , \\[0.5ex]
			\delta_\e Q^{(k)} &  = \tfrac{1}{2}\, 
			\d (\e \wedge Q^{(k-2)})
			-  \tfrac{1}{2} \d^\dagger i_{\e^\flat} Q^{(k+2)}
			=\tfrac{1}{2}\, 
			\d (\e \wedge Q^{(k-2)})-  \tfrac{1}{2} *\d (\e \wedge *Q^{(k+2)} )  \\[0.6ex]
			& = \tfrac{1}{2}\, 
			\d \, i_{\e}^\dagger  Q^{(k-2)}
			-  \tfrac{1}{2} \d^\dagger i_{\e^\flat} Q^{(k+2)}\, , \\[0.6ex]
			\delta_\e b \  & =   -\d \e \,. 
		\end{split}
		\ee
		The first two equations are read to mean that $\delta_\e P_-^{(0)} = \delta_\e P_+^{(10)} = 0$.
		As noted before, the field $P_+^{(0)}$ and its dual $P_-^{(10)}$ do not appear in the action. They are inert under $\delta_X$, but have nontrivial gauge $\delta_\e$ transformations. 
		
		The equations of motion for the $P$'s and $Q$'s following from the effective action~\refb{finXXXXaction} are
		\be
		\begin{split}
			0\ = & \ \d^\dagger \d P_-^{(k-1)} - \d^\dagger Q^{(k) }\, ,\\
			0\ = & \ \d \d^\dagger P_+^{(k+1)} + \d Q^{(k) }  \, , \\
			0\ = & \	\d P_-^{(k-1)} + \d^\dagger P_+^{(k+1)} - h\, \hd\,  Q^{(k)}+\tfrac{1}{2} b\wedge Q^{(k-2)} +\tfrac{1}{2} *(b\wedge *Q^{(k+2)}) \, .
		\end{split}
		\ee
		Eliminating the extra fields  $P$, these equations lead to equations involving only the physical $Q$ fields.  One finds, 
		\be \label{eomsftb}
		\begin{split}
			\d \big( Q^{(k)} + h\, \hd\,  Q^{(k)} - \tfrac{1}{2} b\wedge Q^{(k-2)} -\tfrac{1}{2} *(b\wedge *Q^{(k+2)}) \big) = 0\, , \\[0.5ex]
			\d^\dagger \big( Q^{(k)} - h\, \hd \,  Q^{(k)} + \tfrac{1}{2} b\wedge Q^{(k-2)} + \tfrac{1}{2} *(b\wedge *Q^{(k+2)}) \big) = 0 \, .
		\end{split}	
		\ee

		\subsection{IIA SFT} 
		
		We will now construct the effective action for the massless fields of the IIA RR sector.  The GSO values for the spin operators were  
		given in~\refb{gsoassign}, and  
		we have the RR string field 
		at picture $(-\tfrac{1}{2} ,-\tfrac{1}{2})$ and
		ghost number two: 
		\begin{equation}
			\Psi_{{}_\RR} \ = \ \int\frac{d^Dp}{(2\pi)^D}\, Q^{a\dot b}(p) \, 
			c\bar{c}\  e^{-\phi/2}\Theta_{a}  \, e^{-\bar \phi/2}\Bar{\Theta}_{\dot b}\, e^{ip\cdot X}\,. 
		\end{equation}
		As in IIB, the component fields are encoded in the (momentum space) bispinor $Q^{a\dot b}$.
		As required, the string field is GSO even operator in both holomorphic
		and antiholomorphic sectors.  There is no gauge parameter at picture $(-\tfrac{1}{2} ,-\tfrac{1}{2})$ as was the case in IIB. Therefore, $	\Lambda_{{}_\RR} = 0 \,$, implying that the $Q$ fields are gauge-invariant field strengths.

		The additional string field $\tilde \Psi$ 
		at picture $(-{3\over 2},-{3\over 2})$ and
		ghost number two is
		\be
		\label{auxRRfieldA}
		\begin{split}
			\tilde\Psi_{{}_\RR}  = \int\frac{d^Dp}{(2\pi)^D} \ \Big( &  \ \  N^{\Dot{a}b}(p)\, c \Bar{c} \  
			e^{-3\phi/2}\Theta_{\dot a}  \, e^{-3\bar \phi/2}\Bar{\Theta}_{ b}\, e^{ip\cdot X}\\ 
			&+ \tfrac{1}{2}P^{\Dot{a}\dot b}(p)\, (\partial c + \bar \partial \bar c)c  \Bar{c}\, \bar \partial \bar\xi \, 
			e^{-3\phi/2}\Theta_{\dot a}  \,  e^{-5\bar \phi/2}\Bar{\Theta}_{ \dot b}\, e^{ip\cdot X}\\
			& + \tfrac{1}{2}\Bar{P}^{ab}(p)\, (\partial c + \bar \partial \bar c)c  \Bar{c}\, \partial \xi\ 
			e^{-5\phi/2}\Theta_{ a}  \, e^{-3\bar \phi/2}\Bar{\Theta}_{b}\, e^{ip\cdot X}\Big)\,. 
		\end{split}
		\ee
		The index type of spinors is in such a way that we get  a GSO even operators
		both in the holomorphic and antiholomorphic sector. 
		
		The gauge parameter $\wt \Lambda_{{}_\RR}$ for the $\wt \Psi_{{}_\RR}$ string field is
		\be\label{gaugeparaA}
		\begin{split}
			\widetilde \Lambda_{{}_\RR} =  \int\frac{d^Dp}{(2\pi)^D}\Big(&\ n^{\Dot{a}\dot b}(p)c  \Bar{c}\, \bar \partial \bar\xi \, 
			e^{-3\phi/2}\Theta_{\dot a}  \,  e^{-5\bar \phi/2}\Bar{\Theta}_{ b}\, e^{ip\cdot X}  + \Bar{n}^{ab}(p)c  \Bar{c}\, \partial \xi\ 
			e^{-5\phi/2}\Theta_{ a}  \, e^{-3\bar \phi/2}\Bar{\Theta}_{ b}\, e^{ip\cdot X}\\
			& + \tfrac{1}{2}l^{a \dot b}(p)(\partial c + \bar \partial \bar c)c  \Bar{c}\, \partial \xi\ \bar \partial \bar\xi \, 
			e^{-5\phi/2}\Theta_{ a}  \,  e^{-5\bar \phi/2}\Bar{\Theta}_{ \dot b}\, e^{ip\cdot X} \Big).
		\end{split}
		\ee
		The decomposition of the bispinors in terms of the basis of the Clifford algebra are
		\be
		\label{comopneA}
		\begin{split}
			& Q^{a\dot b} = \sum_{k\in \hat{\mathbb{Z}}_{even}}\tfrac{1}{k!}Q^{(k)}_{\mu_1\cdots\mu_k}(C\Gamma^{\mu_1\cdots\mu_k})^{ab}\, ,\\
			& \bar P^{a\Dot{b}} = \sum_{k\in \hat{\mathbb{Z}}_{odd}}\tfrac{i}{k!}\, 
			\bar P^{(k)}_{\mu_1\cdots\mu_k}(C\Gamma^{\mu_1\cdots\mu_k})^{a\Dot{b}}\, ,\\
			& P^{\Dot{a}b} = \sum_{k\in \hat{\mathbb{Z}}_{odd}}\tfrac{i}{k!}
			P^{(k)}_{\mu_1\cdots\mu_k}(C\Gamma^{\mu_1\cdots\mu_k})^{\Dot{a}b}\, ,\\
			& N^{\Dot{a}\Dot{b}} = \sum_{k\in \hat{\mathbb{Z}}_{even}}\tfrac{1}{k!}N^{(k)}_{\mu_1\cdots\mu_k}(C\Gamma^{\mu_1\cdots\mu_k})^{\Dot{a}\Dot{b}}\,. 
		\end{split}
		\ee
		By making the field redefinitions (different from those in type IIB),
		\be
		\label{defpmPIIA}   
		\begin{split}
			& P_+^{(k-1)} = \tfrac{1}{4}\, \bigl( \,  P^{(k-1)}+\bar P^{(k-1)}  - \d^\dagger N^{(k)}  
			\bigr) \, , \\[0.5ex]
			& P_- ^{(k+1)} = \tfrac{1}{4}\, \bigl(\,  P^{(k+1)} - \bar P^{(k+1)} - \d N^{(k)} \, \bigr)  \, ,
		\end{split}
		\ee
		we write the cubic action, ignoring the cubic NSNS interactions,
		as
		\be\label{IIAaction}
		\begin{split}
			2^{-{d\over 2}} S_{{}_{\rm IIA}}  
			=\
			& -\tfrac{1}{2}(\d^\dagger  P_-^{(1)}, \d^\dagger P_-^{(1)})
			+ (Q^{(0)}, \d^\dagger  P_-^{(1)})\\[2.0ex]
			&  + \ \tfrac{1}{2}(\d P_+^{(1)}, \d  P_+^{(1)})
			-\tfrac{1}{2}(\d^\dagger P_-^{(3)}, \d^\dagger P_-^{(3)})
			+(Q^{(2)},  \d P_+^{(1)} + \d^\dagger P_-^{(3)}) \\[2.0ex]
			&+\ \tfrac{1}{2}(\d  P_+^{(3)}, \d  P_+^{(3)}) -\ \tfrac{1}{2}(\d^\dagger P_-^{(5)}, \d^\dagger P_-^{(5})
			+ (Q^{(4)},  \d P_+^{(3)}+ \d^\dagger P_-^{(5)} ) \\[1.0ex]
			& + 2^{-{d\over 2} - 3}  \int d^Dx
			\left( \tfrac{1}{4} h^{\mu\nu}\partial^2 h_{\mu\nu}+\tfrac{1}{2}(\partial^\nu h_{\mu\nu})^2+\tfrac{1}{4} b^{\mu\nu}\partial^2 b_{\mu\nu}+\tfrac{1}{2}(\partial^\nu b_{\mu\nu})^2 - 2 d\, \partial^\mu\partial^\nu h_{\mu\nu}  - 4 d\, \partial^2 d  \right)\\
			& -  \tfrac{1}{2} \int \d^Dx \,   
			\Bigl[ \,   \,\,   (Q^{(0)}, h\, \hd\,  Q^{(0)}) + (Q^{(2)}, h\, \hd\,  Q^{(2)}) +(Q^{(4)}, h\, \hd\,  Q^{(4)})\Bigr]\ \\[0.7ex]
			&   + \tfrac{1}{2}\, (Q^{(0)} \wedge b\, , Q^{(2)}) + \tfrac{1}{2} (Q^{(2)}\wedge b\, , Q^{(4)}) - \tfrac{1}{4} (Q^{(4)}\wedge b\, , *Q^{(4)}) \,. 	
		\end{split}
		\ee
		In the democratic notation for the RR forms and their interactions, we have 
		\be\label{finactionA}
		\begin{split} 
			2^{-{d\over 2}} S_{{}_{\rm IIA}} \, =\,  
			&\    2^{-{d\over 2} - 3}  \int d^Dx
			\left( \tfrac{1}{4} h^{\mu\nu}\partial^2 h_{\mu\nu}
			+\tfrac{1}{2}(\partial^\nu h_{\mu\nu})^2 
			- 2 d\, \partial^\mu\partial^\nu h_{\mu\nu}  - 4 d\, \partial^2 d  \right) 
			\  -   2^{-{d\over 2} - 4}  ( \d b , \d b ) \\[1.0ex]
			&+  \sum_{k\in \hat{\mathbb{Z}}_{even}} \Bigl[\  \tfrac{1}{4}(\d P_+^{(k-1)}, \d P_+^{(k-1)}) 
			- \tfrac{1}{4}(\d^\dagger P_-^{(k+1)}, \d^\dagger P_-^{(k+1)})  
			+ \tfrac{1}{2} (Q^{(k)}, \d P_+^{(k-1)}+ \d^\dagger P_- ^{(k+1)})\\
			&  \hskip36pt -\tfrac{1}{4}\, (Q^{(k)}, h\,\hd\, Q^{(k)})   + \tfrac{1}{4}\, (Q^{(k)}\wedge b\, , Q^{(k+2)}) 	 \Bigr]. \\[-1.0ex]
		\end{split}
		\ee
		For  the diffeomorphisms generated by the vector field $X = X^\mu\partial_\mu$, we have: 
		\be
		\begin{split}
			\delta_X P^{(k)}_- &  =   -\tfrac{1}{2} X^\sharp \wedge Q^{(k-1)} 
			=  - \tfrac{1}{2}\,  i_{X^\sharp}^\dagger Q^{(k-1)}\, ,\\[0.6ex]
			\delta_X P^{(k)}_+&  =   -\tfrac{1}{2} i_X Q^{(k+1)}\, ,\\[0.5ex]
			\delta_X Q^{(k)} &  =  \tfrac{1}{2}\, \d i_X Q^{(k)} + \tfrac{1}{2}\, 
			* \d i_X *Q^{(k)}  = \tfrac{1}{2}\, \d i_X Q^{(k)} - \tfrac{1}{2}\, 
			\d^\dagger  i_{X^\sharp}^\dagger Q^{(k)}\, ,  \\[0.6ex]
			\delta_X h_{\mu\nu}  & = \partial_\mu X_\nu + \partial_\mu X_\nu\, , \\[0.5ex]
			\delta_X d\  &  = -\tfrac{1}{2}\partial\cdot X\,. 
		\end{split}
		\ee
		And for the Kalb Ramond gauge parameter generated by the one form $\e = \e_\mu\, \d x^\mu$, we have:
		\be
		\begin{split}
			\delta_\e P^{(k)}_-&  =   -\tfrac{1}{2} \, i_{\e^\flat } \, Q^{(k+1)}\, ,\\[0.5ex]
			\delta_\e P^{(k)}_+&  =   -\tfrac{1}{2}\, \e \wedge Q^{(k-1)}  
			= \, -\tfrac{1}{2}\,  i_{\e }^\dagger \, Q^{(k-1)}\, ,\\[0.5ex]
			\delta_\e Q^{(k)} &  = \tfrac{1}{2}\, 
			\d (\e \wedge Q^{(k-2)})
			-  \tfrac{1}{2} \d^\dagger i_{\e^\flat} Q^{(k+2)}
			=\tfrac{1}{2}\, 
			\d (\e \wedge Q^{(k-2)})-  \tfrac{1}{2} *\d (\e \wedge *Q^{(k+2)} )\, , \\[0.6ex]
			\delta_\e b \  & =   -\d \e \,. 
		\end{split}
		\ee
		The equations of motion following from the action~\refb{IIAaction}, are
		\be
		\begin{split}
			0\ = & \ \d^\dagger \d P_+^{(k-1)} + \d^\dagger Q^{(k) }\, ,\\
			0\ = & \ \d \d^\dagger P_-^{(k+1)} - \d Q^{(k) }  \, , \\
			0\ = & \	\d P_+^{(k-1)} + \d^\dagger P_-^{(k+1)} - h\, \hd\,  Q^{(k)}+\tfrac{1}{2} b\wedge Q^{(k-2)} +\tfrac{1}{2} *(b\wedge *Q^{(k+2)}) \, .
		\end{split}
		\ee
		Eliminating the  extra fields $P$, these equations lead to equations involving only the physical $Q$ fields.  One finds, 
		\be \label{eomsftbIIa}   
		\begin{split}
			\d \big( Q^{(k)} - h\, \hd\,  Q^{(k)} + \tfrac{1}{2} b\wedge Q^{(k-2)} + \tfrac{1}{2} *(b\wedge *Q^{(k+2)}) \big) = 0\, , \\
			\d^\dagger \big( Q^{(k)} + h\, \hd \,  Q^{(k)} - \tfrac{1}{2} b\wedge Q^{(k-2)} - \tfrac{1}{2} *(b\wedge *Q^{(k+2)}) \big) = 0 \, .
		\end{split}	
		\ee
		
\sectiono{Effective theory and relating string and supergravity fields}  
		In this section we explore the connection between SFT fields and 
		supergravity fields. 
		 We begin by presenting the diffeormorphism sector  
		  of the type II theory gauge algebra  
		  computed in~\cite{RMBZ}. 
		 We then examine how the effective theory\cite{Sen:2015nph}
		 of $Q= Q^{(5)}$ and $P = P_-^{(4)}$ realizes 
		 diffeomorphisms.   The algebra of diffeomorphisms follows
		 the expected SFT pattern, with  field-dependent
		 structure constants and extra terms that vanish when using the
		 equations of motion.  
		 Surprisingly, however, the field independent part of the 
		 structure constants is the Lie bracket precisely, 
		 instead of the more complicated structure
		 of the SFT bracket~\cite{RMBZ}. 
		 We explain this difference by showing how a redefinition of the
		 diffeomorphism parameter brings full agreement.

		 We then look at the type IIB supergravity with all the form fields. 
		 SFT gives two equations for the $Q^{(k)}$ physical fields, one of which resembles a Bianchi identity and can be used to relate the $Q$'s to the supergravity
		 field strengths.   We then explicitly check that the other SFT  equation of motion for the $Q$'s, with this field redefinition is consistent with the supergravity equations of motion. We also compare how the fields vary under diffeomorphism and Kalb-Ramond transformations and show that, as expected, they agree on shell. 
		
		\subsection{Diffeomorphisms Parameters}
{\bf Diffeomorphisms in Type II Theory:} We now examine the algebra of gauge transformations in the NSNS sector. The commutator of two such gauge transformations yields another gauge transformation with a field-dependent parameter $\Lambda_{12}$, along with additional terms that vanish when the equations of motion ${\cal E} = 0$ are satisfied~\refb{primedproducts2049u2}. Acting on the extra string field, we have:
\be
[\delta_{\Lambda_2}, \delta_{\Lambda_1}] \, \widetilde{\Psi} = \delta_{\Lambda_{12}} \widetilde{\Psi} + [\Lambda_1, \Lambda_2, \mathcal{E}]'\,,
\ee
where the gauge parameters are from the NSNS sector. The parameter $\Lambda_{12}$ on the right-hand side is defined by~\refb{leadingalgeX} as
\be
\Lambda_{12} = [\Lambda_1, \Lambda_2]' = [\Lambda_1, \Lambda_2] + {\cal O}(\Psi)\,,
\ee
where $[\,\cdot\,,\,\cdot\,]$ denotes the string field theory bracket. In what follows, we will focus on the field-independent part $[\Lambda_1, \Lambda_2]$, which we will denote by $\Lambda_{12}$ for simplicity.

Consider gauge parameters of the form
\be
\Lambda_1 = (\lambda_1, \bar{\lambda}_1, \mu_1 = 0)\,, \quad \Lambda_2 = (\lambda_2, \bar{\lambda}_2, \mu_2 = 0)\,,
\ee
so that the resulting parameter is $\Lambda_{12} = (\lambda_{12}, \bar{\lambda}_{12}, \mu_{12} = 0)$. At leading order in derivatives and in the absence of field dependence, $\lambda_{12}$ was calculated in~\cite{RMBZ} and the result is
\be
\label{bosstringalgdiff}
\begin{split}
	\lambda_{12}^\mu = & \ \tfrac{1}{2} \left( \lambda_1 \cdot \partial \lambda_2^\mu - \lambda_2 \cdot \partial \lambda_1^\mu \right)
	- \tfrac{1}{4} \left( \lambda_1 \cdot \partial^\mu \lambda_2 - \lambda_2 \cdot \partial^\mu \lambda_1 \right)
	+ \tfrac{1}{4} \left( \bar{\lambda}_1 \cdot \partial \lambda_2^\mu - \bar{\lambda}_2 \cdot \partial \lambda_1^\mu \right) \\[1.5ex]
	& - \tfrac{1}{2} \left( \lambda_1^\mu \, \partial \cdot \bar{\lambda}_2 - \lambda_2^\mu \, \partial \cdot \bar{\lambda}_1 \right)
	- \tfrac{1}{4} \left( \lambda_1^\mu \, \partial \cdot \lambda_2 - \lambda_2^\mu \, \partial \cdot \lambda_1 \right)\,.
\end{split}
\ee
The expression for $\bar{\lambda}^\mu$ is obtained by exchanging $\lambda_i \leftrightarrow \bar{\lambda}_i$ and replacing $\mu_i \to -\mu_i$ in the formula above.

Now consider the special case where the gauge transformations are pure  diffeomorphisms, i.e., when
\[
X_1 \equiv \bar{\lambda}_1 = \lambda_1\,, \quad X_2 \equiv \bar{\lambda}_2 = \lambda_2\,, \quad \mu_i = 0\,.
\]
In this situation, the resulting parameter also satisfies $\lambda_{12} = \bar{\lambda}_{12} \equiv X_{12}$, and the commutator algebra then simplifies to:
\be
\label{supertringalgdiffXX}
X_{12}^\mu = \tfrac{3}{4} \left( X_1 \cdot \partial X_2^\mu - X_2 \cdot \partial X_1^\mu \right)
- \tfrac{1}{4} \left( X_1 \cdot \partial^\mu X_2 - X_2 \cdot \partial^\mu X_1 \right)
- \tfrac{3}{4} \left( X_1^\mu \, \partial \cdot X_2 - X_2^\mu \, \partial \cdot X_1 \right)\,.
\ee
{\bf Diffeomorphisms in the SFT-based effective action:}
		In  the string-inspired effective action of~\cite{Sen:2015nph},
		the field $P$ is a four form and $Q$ is the self-dual five form.  The quadratic 		action was reproduced before
		and now we supplement it with the cubic term coupling the five form to gravity.
		This term arises from a term $\tfrac{1}{16}  \int  Q^T {\cal M} Q $ in the action ((eqn.(4.47) of \cite{Sen:2015nph}).
		It takes a bit of effort to show that this gives, for the total action  
		\be
		S = \tfrac{1}{2}(\d P, \d P) - (\d P, Q) +   \tfrac{1}{8} \tfrac{1}{4!} 
		\int  Q^{\mu \nu_1 \cdots \mu_4} h_{\mu\nu} Q^{\nu}_{\ \nu_1  \cdots \nu_4}\,. 
		\ee
		This $S$ coincides exactly with the action~\refb{IIBSFTaction} determined from the string field theory (recall that $P= P_-^{(4)}$ and $Q= Q^{(5)}$. 
		
		The diffeomorphism  is also given by a more involved expression in  \cite{Sen:2015nph} (eqn.(5.13)). Keeping only up to the quadratic terms of $Q$ and $h$ in the gauge transformation, we get
		\be \label{effectivediffeo}
		\begin{split}
			\delta_X P = & \   \tfrac{1}{2}  i_X(Q +U)\,,  \ \ \   U = h\, \hd\,  Q\,,  \\[1.0ex]
			\delta_X Q = & \ \d \delta_X P  + *  \d \delta_X P   = \tfrac{1}{2} ( \d + * \d) i_X (Q+U)  \, . 
		\end{split}
		\ee
		Note that $U$ is quadratic in fields. Our SFT computation confirmed 
		the first term in $\delta_X P$, linear in fields.  The term involving $U$ would require a more complicated calculation.  
		The algebra of gauge transformations on $P$ is now calculated keeping only terms linear in fields.  We have
		\be
		\begin{split}
			4 \, [\delta_Y, \delta_X] P & =  
			\  4 \delta_Y \delta_X P - ( X \leftrightarrow Y) 
			=  2 \, i_X ( \delta_Y Q + \delta_Y U) -  ( X \leftrightarrow Y)  \\[1.0ex]
			& = \     \,  i_X(  \d   +  * \d )  i_Y Q 
			+  ( i_X \,  \d i_Y U + i_X * \d i_Y U ) 
			+ 2 \, i_X \delta_Y U- ( X \leftrightarrow Y)   \\[0.3ex]
			& = \     \,  i_X(  \d   +  * \d )  i_Y Q    
			+ 2 \, i_X \delta_Y U- ( X \leftrightarrow Y)  \,.
		\end{split}
		\ee
		In passing to the last line, we dropped the terms where $U$ is not acted by 
		a gauge transformation, since such terms contribute to quadratic order in fields. 
		We have already given $\delta_Y U = \delta_Y( h\, \hd\,  Q) $ in equation \refb{hactiondiffe}.  Using this result in the above equation we find some cancellations and get 
		\be
		\begin{split}
			4 \, [\delta_Y, \delta_X] P \,  & =    \,   i_X ( 2 \d i_Y Q  + i_Y \d Q -*i_Y\d Q ) - ( X \leftrightarrow Y)  \, ,\\
			& = \,   i_X ( 2 \mathcal{L}_Y Q  -i_Y \d Q -*i_Y\d Q ) - ( X \leftrightarrow Y)  \, ,\\   
			& =   \,  2 [i_X, \mathcal{L}_Y] Q  + 2  \mathcal{L}_Y i_X Q - 2 i_Y \mathcal{L}_X Q -  2 i_X i_Y  \d  Q    - i_X *i_Y \d Q +  i_Y * i_X \d Q \, , \\
			& =   \,  2 i_{[X,Y] }Q  + 2 \,\d\,  i_Y i_X Q - (i_X *i_Y- i_Y * i_X )\d Q \,. 
		\end{split}
		\ee
		We have therefore found that 
		\be
		[\delta_Y, \delta_X] P \  = \  \delta_{[X,Y]} P  +\ \tfrac{1}{2}\, \d \, i_{Y} i_{X} Q  \,  + \tfrac{1}{4} \bigl(  i_X ( Y^\sharp \wedge \d^\dagger Q)  
		- i_Y ( X^\sharp \wedge \d^\dagger Q) \bigr)  \, .
		\ee
		or in fact, using that $i_X$ is a graded derivation of the wedge product
		\be
		\label{galgeffft}
		[\delta_Y, \delta_X] P \  = \  \delta_{[X,Y]} P  +\ \tfrac{1}{2}\, \d \, i_{Y} i_{X} Q  \,  + \tfrac{1}{4} \bigl(   X^\sharp \wedge i_Y   
		-   Y^\sharp \wedge i_X \bigr) \d^\dagger Q \, .
		\ee
		The first term on the right-hand side is the diffeomorphism with the 
		Lie bracket of the parameters.  The second term is a gauge transformation in
		the RR sector of the theory. Perhaps surprisingly, it arises from the field-dependent part of the $\wt \Lambda_{12}$ parameter, as was demonstrated in~\refb{3oirc,w48i}.  The last term vanishes on-shell to the required order since $\d^\dagger Q =0$ is the linearized field equation.

\medskip
\noindent
{\bf Relating diffeomorphisms parameters:} 
The result in~\refb{galgeffft} shows the Lie bracket appearing in the commutator of
two diffeomorphisms.  On the other hand, the SFT predicts the bracket \refb{supertringalgdiffXX}.  Here we show that agreement is 
obtained by a redefinition of the gauge parameters. 

		Assume therefore that the gauge parameters of the effective theory
		are field dependent.  Using tildes to signify field 
		dependence, as in $\tilde X$ and $\tilde Y$, we have now, from~\refb{effectivediffeo}, 
		\be
		\begin{split}
			\delta_{\tilde X} P = & \   \tfrac{1}{2}  i_{\tilde X}(Q + U)\,, \  \\[1.0ex]
			\delta_{\tilde X}Q = &  \tfrac{1}{2} ( \d \, i_{\tilde X} Q  + * \d i_{\tilde X}Q) + \tfrac{1}{2}  ( \d \, i_{\tilde X} U  + * \d 
			i_{\tilde X} U)\,.
		\end{split}
		\ee
		The algebra of gauge transformations on $P$, evaluated above is now seen to give 
		\be\label{algebramod}
		\begin{split}
			\, [\delta_{\tilde Y}, \delta_{\tilde X}] P & =   \   \delta_{\tilde Y} \delta_{\tilde X}P - ( {\tilde X} \leftrightarrow {\tilde Y})  \\[1.0ex]
			& = \ \bigl(  \delta_{\delta_{\tilde Y} \tilde X}P - ( {\tilde X} \leftrightarrow {\tilde Y}) \bigr) 
			 + \delta_{[\tilde X,\tilde Y]} P  +\ \tfrac{1}{2}\, \d \, i_{\tilde Y} i_{\tilde X} Q  \\[1.0ex]
			&\quad + \tfrac{1}{4} \bigl(   {\tilde X}^\sharp \wedge i_{\tilde Y}   
		-   {\tilde Y}^\sharp \wedge i_{\tilde X} \bigr) \d^\dagger Q\,. 	 
		\end{split}
		\ee 
		Every term in the above algebra has been calculated, except for that 
		involving the gauge parameter~$\delta_{\tilde Y} \tilde X$.
		For a gauge parameter $\tilde X$ depending on the metric only to first order we write 
		\be \label{gaugeredef}
		\tilde X^\mu  = X^\mu  + e\,  h X^\mu  + f h^\mu_{\ \ \nu} X^\nu\, ,
		\ee
		where $e$ and $f$ are numerical constants. We then have
		\be
		\begin{split}
			\delta_{\tilde Y} \tilde X^\mu & =  2e\,  \partial\cdot\tilde Y X^\mu 
			+ f  X \cdot  \partial^\mu \tilde Y 
			+ f  X^\nu \partial_\nu \tilde Y^\mu \\
			& = 2e \partial\cdot Y X^\mu + f  X \cdot \partial^\mu  Y + f  X^\nu \partial_\nu Y^\mu + \order{h}\, ,
		\end{split}
		\ee
		so that 
		\be
		\begin{split}  
		\delta_{\tilde Y} \tilde X^\mu -  \delta_{\tilde X} \tilde Y^\mu = & \  f[X,Y]^\mu+f( X^\mu \partial\cdot Y - Y^\mu \partial \cdot X)  + 2 e(X\cdot\partial^\mu Y - Y \cdot\partial^\mu X) +\order{h}\, , \\ 
		= & \  f[\tilde X,\tilde Y]^\mu+f( \tilde X^\mu \partial\cdot \tilde Y - \tilde Y^\mu \partial \cdot \tilde X)  + 2 e\, (\, \tilde X\cdot\partial^\mu \tilde Y - \tilde Y \cdot\partial^\mu \tilde X) +\order{h}\, .
\end{split}
		\ee
		The algebra \refb{algebramod} becomes
		\be\label{algebradefined}
		\begin{split}   
			[\delta_{\tilde Y}, \delta_{\tilde X}] P 
			&=   \ \tfrac{1}{2}  (1+f)   i_{[\tilde X, \tilde Y] } Q \ \ 
			+ \tfrac{1}{2} f \  i_{ \tilde X\cdot \vec{\partial} \tilde Y 
			- Y \cdot \vec{\partial} X } Q \ \ 
			+ \  e   i_{\vec{\tilde X} \partial\cdot \tilde Y - \vec{\tilde Y} \partial \cdot \tilde X } Q
			\ + \tfrac{1}{2} \, \d \, i_{\tilde Y} i_{\tilde X} Q   \\[1.0ex]
			& \ \ + \tfrac{1}{4} \bigl(   {\tilde X}^\sharp \wedge i_{\tilde Y}   
		-   {\tilde Y}^\sharp \wedge i_{\tilde X} \bigr) \d^\dagger Q+  \order{hQ} \\[1.0ex]
			& = \delta_{[[\tilde X, \tilde Y]]} P 
			+ \tfrac{1}{2} \, \d \, i_{\tilde Y} i_{\tilde X} Q 
			+ + \tfrac{1}{4} \bigl(   {\tilde X}^\sharp \wedge i_{\tilde Y}   
		-   {\tilde Y}^\sharp \wedge i_{\tilde X} \bigr) \d^\dagger Q  +  \order{hQ} \, .
		\end{split}
		\ee
	The algebra is still in the standard form, but now with the bracket $[[\tilde X, \tilde Y]]$ given by
		\be
		[[\tilde X, \tilde Y]] = (1+f)[\tilde X, \tilde Y] + e (\vec{\tilde X} \partial\cdot \tilde Y - \vec{\tilde Y} \partial \cdot \tilde X )
		\ \  + f ( \tilde X\cdot \vec{\partial} \tilde Y - \tilde Y \cdot \vec{\partial} \tilde X )  +  \order{hQ} \, .
		\ee
Comparing with the SFT bracket in~\refb{supertringalgdiffXX}
		\be \label{SFTbracket}
		[X, Y]_{\rm sft}^\mu \ = \ \tfrac{3}{4} [X , Y]^\mu  
		-  \tfrac{3}{4}(X^\mu \partial\cdot Y    - 
		Y^\mu \partial \cdot X )  
		-  \tfrac{1}{4}  ( X \cdot \partial^\mu Y \, 
		- \ Y \cdot \,    \partial^\mu X ) \, ,
		\ee
we see that agreement happens with $ f= -1/4$ and $e = -3/4$.  This shows that the SFT-inspired effective field theory diffeomorphism algebra is the one predicted by the SFT.

		\subsection{Supergravity fields and string fields}

		In order to relate the type IIB string fields to the IIB supergravity fields, we first review the action and the definition of the fields in the supergravity.  From section 12.1 of \cite{Polchinski:1998rr} we read 
		\be
		S_{_{\RR}} + S_{_{\NS\RR\RR}} \propto  (F^{(1)} , F^{(1)} )_g + (\widehat F^{(3)} , \widehat F^{(3)} )_g + \tfrac{1}{2}(\widehat F^{(5)} , \widehat F^{(5)} )_g + (*_g F^{(5)} , b \wedge F^{(3)} )_g\, , 
		\ee
		where the inner products are written with respect to the metric 
		$g_{\mu\nu}$ 
		and the field strengths are given~by
		\be
		\begin{split}
			& \widehat F^{(3)}  = F^{(3)}  + b\wedge F^{(1)} \, ,  \\
			& \widehat F^{(5)} = F^{(5)}  + b \wedge F^{(3)} \, ,  \\
			& F^{(1)}  = \d C'^{(0)} , \hskip20pt 	F^{(3)} = \d C'^{(2)} , \hskip20pt	F^{(5)}  = \d C'^{(4)} \, .
		\end{split}
		\ee
		The field $\widehat F^{(5)}$ is required to be self-dual: 
		\be
		*\widehat F^{(5)} = \widehat F^{(5)} \,,
		\ee
		as an extra constraint
		{\em after} the equations of motion are derived. 
		The gauge potentials here are related to the gauge potentials $C^{(k)}$ of~\cite{Polchinski:1998rr} as~$C'^{(0)} = C^{(0)}$,  $C'^{(2)} = C^{(2)} - C^{(0)}\wedge b$, and $C'^{(4)} = C^{(4)} - {1\over 2} C^{(2)}\wedge b $. 
The equations of motion for the gauge potentials following from the action are
		\be\label{sugraeom}
		\begin{split}
		& \d^\dagger_g\Bigl( F^{(1)}+ *_g(b \wedge 
		*_g F^{(3)} )\Bigr) = 0\, ,\\ 
		& \d^\dagger_g\Bigl(F^{(3)}+ b\wedge F^{(1)}  
		+ \tfrac{1}{2} *_g\bigl (b \wedge \bigl[F^{(5)}+ *_g F^{(5)}\bigr] \bigr) \Bigr)= 0\, ,  \\
		&  \d^\dagger_g \Bigl( F^{(5)}   +  
		b \wedge F^{(3)}  - *_g(b \wedge F^{(3)} )\Bigr) =0\, .
		\end{split}
		\ee
		For $g_{\mu\nu} = \eta_{\mu\nu} + h_{\mu\nu}$, we have $*_g = *(1-2 h\, \hd) +\order{h^2}$, where $*$ here is with respect to the flat metric and $h\hd$ is the action on the differential form defined in~\refb{haction}. Keeping only 
		quadratic order 
		in the fields, the equation of motion~\refb{sugraeom} is then
		\be\label{sugraeom1}
		\begin{split}
			& \d^\dagger\Bigl( F^{(1)} - 2 h\, \hd\,  F^{(1)} + *(b \wedge *F^{(3)} )\Bigr) = 0\, ,\\
			& \d^\dagger\Bigl( F^{(3)} - 2 h\, \hd\,  F^{(3)}+ b\wedge F^{(1)} + \tfrac{1}{2} *\bigl( b \wedge \bigl[ F^{(5)} + * F^{(5)}\bigr]  
			\bigr )\Bigr)= 0\, ,  \\
			&  \d^\dagger \Bigl( F^{(5)} - 2 h\, \hd\,  F^{(5)} +  b \wedge F^{(3)} - *(b \wedge F^{(3)} )\Bigr) =0\, .
		\end{split}
		\ee
		The self duality of $\widehat F^{(5)}$,  implies $*_gF^{(5)} = F^{(5)} + b \wedge F^{(3)} - *_g (b \wedge F^{(3)})$. With this, the last two terms of the second equation above sum into a single term to the second order in fields.

		\medskip
		\noindent   
		{\bf Relating string fields to supegravity fields.}	
		As we have seen above, the field strengths 
		$F^{(k)}$ are defined as exterior derivative of the gauge fields, and therefore they obey  Bianchi identities. In addition, they obey equations of motion. On the SFT side, we have two equations of motion~\refb{eomsftb} for the $Q^{(k)}$'s. As a guide to the correct field relations, we want to interpret the second equation as an equation of motion and the first equation as a Bianchi identity.  Thus the first equation allows us to identify supergravity field strengths $F^{(k)}$ that are of the form $F^{(k)} = \d C'^{(k-1)}$:
		\be\label{redefs}   
		F^{(k)} = \tfrac{1}{2}\big(Q^{(k)} + h \,\hd\,  Q^{(k)} - \tfrac{1}{2} b\wedge Q^{(k-2)} -\tfrac{1}{2} *(b\wedge *Q^{(k+2)}) \big) + \cdots \, ,   
		\ee
where the dots represent terms cubic and higher order in the fields that are not included in the calculations.
		Using the various duality relations, we have, explicitly 
		\be
		\begin{split}
			F^{(1)} =\  & \tfrac{1}{2}\big(Q^{(1)} + h  \,\hd\,  Q^{(1)} - \tfrac{1}{2} *(b\wedge *Q^{(3))}) \big)+ \cdots \, ,   \\[0.5ex]
			F^{(3)} =\  & \tfrac{1}{2}\big(Q^{(3)} + h \,\hd\,  Q^{(3)} - \tfrac{1}{2} b\wedge Q^{(1)} -\tfrac{1}{2} *(b\wedge Q^{(5)}) \big)+ \cdots \, ,  \\[0.5ex]
			F^{(5)} =\  & \tfrac{1}{2}\big(Q^{(5)} +h \,\hd\,  Q^{(5)} - \tfrac{1}{2} b\wedge Q^{(3)} + \tfrac{1}{2} *(b\wedge Q^{(3)}) \big) + \cdots \,. 
		\end{split}
		\ee
		These equations can be inverted to find  
\be\label{redefsFQ}
		Q^{(k)} = 2\big(\, F^{(k)} - h \,\hd\,  F^{(k)} + \tfrac{1}{2} b\wedge F^{(k-2)}          	+\tfrac{1}{2} *(b\wedge *F^{(k+2)}) \big) + \cdots \, ,   
		\ee
		which yield, explicitly, 		
		\be
		\begin{split}
			Q^{(1)} = & \ 2F^{(1)} - 2 h  \,\hd\,  F^{(1)} + * ( b \wedge * F^{(3)}) + \cdots\, ,   \\[0.5ex]
			Q^{(3)} = & \ 2F^{(3)} -2 h  \,\hd\,  F^{(3)} +  b\wedge F^{(1)} + *(b\wedge F^{(5)}) + \cdots \, ,  \\[0.5ex]
			Q^{(5)} = & \ 2F^{(5)}-2 h  \,\hd\,  F^{(5)}  + b\wedge F^{(3)} -  *(b\wedge F^{(3)})  + \cdots \,. 
		\end{split}
		\ee 
The above are the desired relations between the SFT and the supergravity 
variables.

		With such relations, the second equation of \refb{eomsftb}, the SFT
		equation of motion,  becomes
		\be
		\begin{split}
			& \d^\dagger(F^{(1)} -2\, h  \,\hd\,  F^{(1)} + * ( b \wedge * F^{(3)}) + \cdots )  = 0\, ,  \\[0.5ex]
			& \d^\dagger(F^{(3)} -2\,  h  \,\hd\,  F^{(3)}  +  b\wedge F^{(1)} + *(b\wedge F^{(5)}) + \cdots ) = 0 \, , \\[0.5ex]
			& \d^\dagger(F^{(5)} -2\,  h  \,\hd\,  F^{(5)} + b\wedge F^{(3)} -  *(b\wedge F^{(3)})  + \cdots ) = 0  \,. 
		\end{split}
		\ee
		We then see that the equation of motion from SFT agrees with the equation of motion from supergravity \refb{sugraeom1} to the quadratic terms in fields proving that \refb{redefs} indeed relate the string fields to the supergravity fields. 
		In terms of the self-dual five form $\widehat F^{(5)} = F^{(5)}  + b\wedge F^{(3)} $ of supergravity, 
		\be
		\label{q5fordual} 
		Q^{(5)} =  \ 2\widehat F^{(5)} - 2\,  h  \,\hd\,  \widehat F^{(5)}   - b\wedge F^{(3)} -  *(b\wedge F^{(3)})  \,.
		\ee
This also matches the redefinition given in equations (3.16) and (4.32) of~\cite{Sen:2015nph} with no gravity, and to the first order in $h_{\mu\nu}$, respectively. The expression is consistent with the self-duality of $Q^{(5)}$ to second order in fields.  To see this first note that  using~\refb{8ginflat}, to leading
order in $h$, 
\be
*_g \widehat F^{(5)} = \widehat F^{(5)}   \quad \to \quad   * \widehat F^{(5)} = \widehat F^{(5)}  + 2 * h\, \hd \, \widehat F^{(5)} \, = \,  \widehat F^{(5)}  +  * \, h\, \hd \, \widehat F^{(5)} - \, h\, \hd \, \widehat F^{(5)}\,, 
\ee		
recalling the anticommutation of Hodge duality and the $h \, \hd$ action. 
This shows that 
\be
 * \bigl( \widehat F^{(5)}- h\, \hd \, \widehat F^{(5)} \bigr)   =  \widehat F^{(5)} - \, h\, \hd \, \widehat F^{(5)}\,, 
\ee		
thus making clear that $Q^{(5)}$ in~\refb{q5fordual} is duality invariant to the requisite order.  
 
		As a consistency check, we can also look at the gauge transformations. Keeping up to only second order terms in fields, the Kalb Ramond gauge transformation is
		\be
		\begin{split}
			\delta_\e F^{(k)} & =  \tfrac{1}{2}\big( \d \e \wedge Q^{(k-2)} - \tfrac{1}{2}\e\wedge \d Q^{(k-2)}-  \tfrac{1}{2} * (\e \wedge \d *Q^{(k+2)}) \big)\, ,  \\[1.0ex]
			& =   \d \e \wedge F^{(k-2)} - \tfrac{1}{2}\e\wedge \d F^{(k-2)}-  \tfrac{1}{2} * (\e \wedge \d *F^{(k+2)}) \, . 
		\end{split}
		\ee
		This is in agreement with the gauge transformations of the supergravity fields up to equation of motion (i.e. both equations from SFT). Ignoring terms proportional to the equation of motion and higher order terms in fields,  this gives us
		\be
		\begin{split}
			\delta_\e F^{(1)} = &\  0\, ,\\
			\delta_\e F^{(3)} = &\   \d \e \wedge F^{(1)}\, , \\
			\delta_\e F^{(5)} = &\   \d \e \wedge F^{(3)} \, .
		\end{split}
		\ee
		The diffeomorphism on $h\, \hd\,  Q^{(k)}$,  to the leading order in fields, is given in~\refb{hactiondiffe}. Gathering the diffeomorphisms 
		of other terms in~\refb{redefs} as well, we get
		\be
		\begin{split}
			\delta_X F^{(k)} & = \tfrac{1}{2}\Bigl(\d i_X Q^{(k)} +\tfrac{1}{2}\big(i_X^\dagger \d ^\dagger Q^{(k)}+ i_X \d Q^{(k)}\big)\Bigr)\, , \\
			& = \d i_X F^{(k)}  + \tfrac{1}{2} ( i_X^\dagger \d ^\dagger F^{(k)}+ i_X \d F^{(k)}) \, .
		\end{split}
		\ee
		The second and third terms in the last right-hand side vanish
		on shell to first order in fields.  
		Thus, on shell, 
		the $F^{(k)}$ transform under diffeomorphism in the standard way,  $\delta_X F^{(k)} = {\cal L}_X F^{(k)}$.
		
		
		\sectiono{Discussion}

		In this paper we have done an explicit analysis of the type II SFT
		computing the action, gauge transformations, and gauge algebra
		to  nontrivial order.  The results confirm the work of~\cite{Sen:2015nph}:
		 our results
		map exactly to his work on the self-dual five form.  The present work
		shows explicitly how all forms in the theory are described with a 
		similar mechanism.

		The questions that remain, range from straightforward calculations
		to more conceptual and challenging issues. We have:
		
		\begin{enumerate}
			
			\item  Computing the cubic NSNS interactions and finding the field
			redefinitions needed to simplify the action and gauge transformations,
			as was done for bosonic strings in~\cite{Hull:2009mi}.
			
			\item  Computing the action for the RNS and NSR sectors of the theory, the supersymmetry transformations and their gauge algebra.

			\item  Recent work on boundary terms for the kinetic term in string field theory~\cite{Firat:2024kxq, Stettinger:2024uus, Maccaferri:2025orz} can be extended to superstring theory. In this case, the non-cyclicity of the picture-changing operator (PCO), which arises from the non-cyclicity of the BRST operator, adds a level of complication. It is interesting to ask what happens to the extra degrees of freedom in spacetimes with boundaries. 
			\item  Clarifying how this theory manages to have an action that
			is diffeomorphism
			invariant, as expected due to the presence of a physical gravity field, but
			at the same time containing propagating degrees of freedom that 
			are {\em not} coupled to gravity.   This seems counter to the intuition that every
			propagating field gravitates.  It would be useful to learn how general
			is the mechanism at play in the SFT and in its low energy limit.

			\item  The type II SFT has been seen to implement background independence at the level of equations of motion in~\cite{Sen:2017szq}, but as this work makes clear, background independence does not work for the action.  
			Still, the SFT action after a string field shift that implements a background shift 
			should be consistent,  suggesting
			that a reformulation of the SFT  could exist implementing background independence
			at the level of the action. 
			
			\item  The background independence analysis discussed in the previous item was done
			for NSNS backgrounds.  In fact even the type II action is formulated with trivial RR backgrounds.  It would be desirable (although challenging) to discuss the background independence for changes of RR background.  This is critical, as
			it seems clear that at this point type II SFT provides (via its equations of motion) a concrete and explicit
			approach to deal with RR backgrounds. 
			
			\item  It appears as if the string field theory works with two metrics-- the starting point  for an extensive analysis by Hull at the level of field theory~\cite{Hull:2023dgp}.  The quadratic terms involving the extra fields are written with the background metric $\bar g$ that is implied by the superconformal field theory.  The interaction terms are also written using this background metric $\bar g$ but include
			the gravity fluctuation $h_{\mu\nu}$ which is expected to assemble into a new metric
			$g = \bar g + h$.  The metric $g$ would be the second metric.
			It is conceivable that such facts could be implemented more clearly as one
			attempts to improve the analysis of background independence. 
			
		\end{enumerate}

		\bigskip
		\noindent
		{\bf Acknowledgements}.  We thank Olaf Hohm for a useful
		conversation and acknowledge helpful discussions
		with Atakan Hilmi Firat and Ashoke Sen.

		\appendix

		\sectiono{Clifford algebra and spinors}
		$\Gamma$-matrices satisfy the anticommutation relations
		\begin{equation}
			\{\Gamma^\mu,\Gamma^\nu\} = 2 \eta^{\mu\nu}\,.
		\end{equation}
		The Clifford algebra is generated by products
		involving the identity $1$ and the  $\Gamma^\mu$.  Symmetric products can be reduced using the Clifford algebra relation. As a result, a basis for the Clifford algebra is given by the set of completely antisymmetric products $\Gamma^{\mu_1\cdots\mu_k}$, formed with weight
		one: 
		\begin{equation}
			\Gamma^{\mu_1\cdots\mu_k} = \Gamma^{[\mu_1}\cdots\Gamma^{\mu_k]}\,. 
		\end{equation}
		The spinor indices on the Gamma matrices are of the form $\Gamma_\alpha^{\hskip5pt \beta}$. The (unitary) charge conjugation matrices $C_{\alpha\beta}$ and $C^{\alpha\beta}$ raise and lower the spinor indices 
		\be
		C^{\alpha\beta} \eta_\beta = \eta^\alpha \,,  \ \ \ \hbox{and} \ \ \ 
		\eta_\alpha =  C_{\alpha\beta} \eta^\beta \,.
		\ee
		These require
		\be
		C^{\alpha\beta} C_{\beta\gamma} =  \delta^\alpha_{\ \gamma} \,, \ \ \ \ 
		C_{\alpha\beta} C^{\beta\gamma} =  \delta_\alpha^{\hskip5pt\gamma}\,. 
		\ee
		We think the matrices $C$ and $C^{-1}$ as ones with upper indices and lower
		indices, respectively,
		\be
		[C] =  C^{\alpha\beta} \,, \ \ \ [C^{-1}] =  C_{\alpha\beta}\,.
		\ee 
		Any bispinor $F^{\alpha\beta}$ can be expanded in the Clifford algebra basis
		\begin{equation}
			F^{\alpha\beta} = \sum_{k=0}^{d-1}\frac{1}{k!}F_{\mu_1\cdots\mu_k}(C\Gamma^{\mu_1\cdots\mu_k})^{\alpha\beta}\,, 
		\end{equation}
		where $C$ raises the first index on the $\Gamma$ matrices.  Moreover, the matrix $C\Gamma^{\mu_1\cdots\mu_k}$ is either symmetric
		or antisymetric, depending on the value of $k$. 
		Within the Clifford algebra,  the Lorentz algebra is generated by the matrices
		\begin{equation}
			\Sigma^{\mu\nu} = -\tfrac{i}{4}\, [\Gamma^\mu,\Gamma^\nu]\,.
		\end{equation}
		Spinors are representations of the Lorentz algebra. 
		One also defines the matrix $\Gamma$  by
		\begin{equation}
			\Gamma \equiv  (-i)^{d/2+1}\Gamma_0\cdots\Gamma_{d-1} \quad \to \quad \Gamma^2 = 1\,. 
		\end{equation}
		The matrix $\Gamma$ anticommutes with $\Gamma^{\mu_1\cdots\mu_k}$ if $k$ is odd
		and commutes with it if $k$ is even.
		In particular, it commutes with the Lorentz algebra generators. Since $\Gamma^2 = 1$, its eigenvalues are $\pm1$.  Spinors therefore decompose into $\Gamma$ eigenspaces. We use the Latin letters $a, b, \cdots$ for the $\Gamma = 1$ subspace and the dotted latin letters $\Dot{a}, \Dot{b}, \cdots $ for the $\Gamma = -1$ subspace.
		
		We collect here properties of $\Gamma$ matrices in even dimension spacetime:
		\be
		\label{property}
		\begin{split}
			& \Gamma\Gamma^{\mu_1\cdots\mu_k} = (-1)^{k}\Gamma^{\mu_1\cdots\mu_k}\Gamma = -(-1)^{k(k+1)/2}(-i)^{d/2-1}\frac{1}{(d-k)!}\epsilon^{\mu_1\cdots\mu_k}_{\hskip25pt\nu_1\cdots\nu_{d-k}}\Gamma^{\nu_1\cdots\nu_{d-k}}\,, \\
			& \Gamma^\nu\Gamma^{\mu_1\cdots\mu_k} = \Gamma^{\nu\mu_1\cdots\mu_k} + k\eta^{\nu[\mu_1}\Gamma^{\mu_2\cdots\mu_k]}
			\,, 		\\[0.7ex]
			&\Gamma^{\mu_1\cdots\mu_k}\Gamma^\nu = (-1)^k\Gamma^{\nu\mu_1\cdots\mu_k} + (-1)^{k+1}k\eta^{\nu[\mu_1}\Gamma^{\mu_2\cdots\mu_k]}\,, \\[0.8ex]
			&\hskip-3pt \Tr(\Gamma^{\mu_1\cdots\mu_k}\Gamma_{\nu_1\cdots\nu_p}) = (-1)^{k(k-1)/2}2^{d/2}\, k!\, \delta_{k,p}\  
			\delta^{[\mu_1}_{\nu_1}\cdots\delta^{\mu_k]}_{\nu_k}\,. 
		\end{split}
		\ee
		For the interplay of the $C$ matrix with the $\Gamma$-matrices we have
		\be
		\label{CGamm}
		C\Gamma^\mu C^{-1} = -(\Gamma^\mu)^T \,, 
		\quad C\Gamma^{\mu_1\cdots\mu_k}C^{-1} = (-1)^{k(k+1)/2}{\Gamma^{\mu_1\cdots\mu_k}}^T \,, 
		\quad C\Gamma C^{-1} = (-1)^{d(d+1)/2}\Gamma^T\,. 
		\ee
		With the identification of bispinors and differential forms given in~\refb{cea}, the properties of $\Gamma$ matrices from~\refb{property} above give
		\be
		\label{diffcliffiden}
		\begin{split}
			\slashed{A}^{(k)} \ \longleftrightarrow & \ {A}^{(k)}\,, \\
			\slashed{q}^T\slashed{A}^{(k)} \ \longleftrightarrow & \  -q\wedge{A}^{(k)} - i_{q^\flat}{A}^{(k)}\,, \\[0.5ex]
			\slashed{A}^{(k)} \slashed{q}  \ \longleftrightarrow & \ (-1)^k\, q\wedge{A}^{(k)} + (-1)^{k+1} i_{q^\flat}A^{(k)}\,, \\[0.5ex]
			\slashed{A}^{(k)}\Gamma \ \longleftrightarrow & \ -(-1)^{k(k-1)/2}(-i)^{d/2-1}(-1)^{k(d-k)}*A^{(k)}\, ,
		\end{split}
		\ee
		where $q = q_\mu\d x^\mu$ is one form, and $q^\flat$ is the associated vector. For derivatives acting on the bispinors, we have
		\be \label{derivvvaa }
		\begin{split}
			\slashed{\partial}^T\slashed{A}^{(k)} & \ \longleftrightarrow \  -\d A^{(k)} + \d^\dagger A^{(k)}\,, \\
			\slashed{A}^{(k)}  \overleftarrow{\slashed{\partial} }
			& \ \longleftrightarrow \
			(-1)^k  \d A^{(k)}+ (-1)^k \d^\dagger A^{(k)}\,, \\[0.5ex]
			\slashed{\partial}^T\slashed{A}^{(k)}\overleftarrow{\slashed{\partial}} 
			& \ \longleftrightarrow \ (-1)^k(-\d\d^\dagger + \d^\dagger\d)A^{(k)}\,. 
		\end{split}
		\ee
		In the computation of the closed string action, we have the objects of the form,
		\be
		\begin{split}
			\slashed{A}^{(k)}_{\alpha\beta}\, 
			\slashed{B}^{(q)\alpha\beta} \  =  & \	
			\slashed{A}^{(k)\gamma\lambda}
			\slashed{B}^{(q)\alpha\beta}
			C_{\gamma\alpha}C_{\lambda\beta} =  \Tr{{C^{-1}}^T{\slashed{A}^{(k)}}^T C^{-1}\slashed{B}^{(q)}}\\
			= & \  \frac{1}{k!}\frac{1}{q!}\, A^{(k)}_{\mu_1\cdots\mu_k}
			B^{(q)}_{\nu_1\cdots\nu_q}\Tr{{C^{-1}}^T(C\Gamma^{{\mu_1\cdots\mu_k}})^T C^{-1}(C\Gamma^{{\nu_1\cdots\nu_q}})}
			\\[0.5ex]
			= & \ \frac{1}{k!}\frac{1}{q!}\, A^{(k)}_{\mu_1\cdots\mu_k}
			B^{(q)}_{\nu_1\cdots\nu_q}\Tr{{C^{-1}}^T{\Gamma^{{\mu_1\cdots\mu_k}}}^T C^T \Gamma^{{\nu_1\cdots\nu_q}}}\,. 
		\end{split}
		\ee
		We use the second equation in~\refb{CGamm}, to write
		\be
		{C^{-1}}^T{\Gamma^{\mu_1\cdots\mu_k}}^T C^T = (C\Gamma^{\mu_1\cdots\mu_k}C^{-1})^T = (-1)^{k(k+1)/2}({\Gamma^{\mu_1\cdots\mu_k}}^T)^T 
		= (-1)^{k(k+1)/2}{\Gamma^{\mu_1\cdots\mu_k}}\,,
		\ee
		so that using the last equation of \refb{property} we find
		\be \label{productsp}
		\begin{split}
			\slashed{A}^{(k)}_{\alpha\beta}\, 
			\slashed{B}^{(q)\alpha\beta} \ = & \
			(-1)^{k(k+1)/2}\, \frac{1}{k!}\frac{1}{q!} \, A^{(k)}_{\mu_1\cdots\mu_k}  
			B^{(q)\nu_1\cdots\nu_q}\Tr{\Gamma^{\mu_1\cdots\mu_k}\Gamma_{\nu_1\cdots\nu_q}}\\
			= & \  (-1)^{k(k+1)/2}\, \frac{1}{k!}\frac{1}{q!} A^{(k)}_{\mu_1\cdots\mu_k}
			B^{(q)\nu_1\cdots\nu_q}
			\, (-1)^{k(k-1)/2}2^{d/2}k!\delta^{[\nu_1}_{\mu_1}\cdots\delta^{\nu_k]}_{\mu_k}\delta_{k,q} \\
			= & \  (-1)^k \, 2^{d/2} \frac{1}{k!}\, 
			A^{(k)}_{\mu_1\cdots\mu_k}
			B^{(k)\mu_1\cdots\mu_k}\delta_{k,q} \,. 
		\end{split}
		\ee
		Integrating against the volume form $\omega$, we find the useful identity 
		\be \label{innercliff}
		\int \slashed{A}^{(k)}_{\alpha\beta} \, \slashed{B}^{(q)\alpha\beta} \ \omega 
		= 2^{d/2}(-1)^k \, (A^{(k)},B^{(k)})\, \delta_{k,q} \,.
		\ee

		\sectiono{Bispinor decomposition for RR fields}\label{bispnrsdcmp}
		For the RR closed string fields we have the structure
		\begin{equation}
			\mathbf{F} = F^{\alpha\beta}\Theta_{\alpha}\Bar{\Theta}_{\beta}\,, 
		\end{equation}
		where $\Theta$ and $\Bar{\Theta}$ are spin fields. The spin fields 
		decompose into eigenspaces of $\Gamma$:  
		\be
		\Theta_\alpha = ( \Theta_a, \Theta_{\dot a})\ \ \ \hbox{and} \ \ 
		\tilde\Theta_\alpha = ( \tilde\Theta_a, \tilde\Theta_{\dot a})\,, 
		\ee  
		\begin{equation}\label{chiral}
			\begin{split}
				\Gamma_a^{\hskip5pt b}\Theta_b = \Theta_a\,, \ \ \  &
				\Gamma_{\Dot{a}}^{\hskip5pt \Dot{b}} \Theta_{\Dot{b}} = -\Theta_{\Dot{a}} \,, \\ 
				\Gamma_a^{\hskip5pt b}\tilde\Theta_b = \tilde\Theta_a\,, \ \ \  &
				\Gamma_{\Dot{a}}^{\hskip5pt \Dot{b}} \tilde\Theta_{\Dot{b}} = -\tilde\Theta_{\Dot{a}} \,. 
			\end{split}
		\end{equation}
		We are interested in the sectors  $(+,+), (+,-), (-,+)$ and $(-,-)$, where the signs indicate the $\Gamma$ eigenvalue. The structures we can have 
		are then of the form 
		\be
		\begin{split}
			{\bf A}  \ = \ & A^{ab}   \, 
			\Theta_a\Bar{\Theta}_b  \ \ \hbox{for} \ \   (+, +),  \\
			{\bf B} \ = \  &    
			B^{a\Dot{b}}\, \Theta_a\Bar{\Theta}_{\Dot{b}}\ \ \hbox{for} \ \   (+, -), \\
			{\bf C}  \ = \ &  C^{\Dot{a}b}  \, 
			\Theta_{\Dot{a}}\Bar{\Theta}_{b}  \ \ \hbox{for} \ \   (-, +),  \\
			{\bf D} \ = \ &    D^{\dot a\Dot{b}}\, \Theta_{\Dot{a}}\Bar{\Theta}_{\Dot{b}}\ \ \hbox{for} \ \  
			(-, -),. 
		\end{split}
		\ee
		Because the $\Theta$'s are constrained by the action of $\Gamma$ we have, for example
		\be
		A^{ab}  \Theta_a\Bar{\Theta}_b  = A^{ab} \Gamma_a^{\ c}  \Theta_c\Bar{\Theta}_b
		= (\Gamma^T)^c_{\ a} A^{ab} \Theta_c\Bar{\Theta}_b  = ( \Gamma^T  A)^{cb} \Theta_c\Bar{\Theta}_b = ( \Gamma^T  A)^{ab} \Theta_a\Bar{\Theta}_b \,, 
		\ee
		showing that $A = \Gamma^TA$.  Doing similarly on the `bar' 
		spinor we find $A = A \Gamma$.
		For the $B, C$ and $D$ bispinors we also find analogous identities.  All in all,
		\be
		A =   \Gamma^T A \ \ =  A \, \Gamma  \,, \ \ \ B =   \Gamma^T B  \ =  - B\,  \Gamma \,, \ \ \ 
		C =  - \Gamma^T C \, =    C \,  \Gamma \,,  \ \ \ 
		D  =    -\Gamma^T D =  - D \, \Gamma \,. 
		\ee
		For a bispinor $X$ we define a sign factor $\eta_X$ as
		\be
		\Gamma^T  X  =  \eta_X  \, X \,\Gamma\, .
		\ee
		Thus $\eta_A = \eta_D = 1$, and $\eta_B= \eta_{C} = -1 \,.$
		The bispinors are now expanded as follows:
		\begin{equation}
			X^{\alpha\beta} = \sum_{k=0}^d   
			\frac{1}{k!}X^{(k)}_{\mu_1\cdots\mu_k}(C\Gamma^{\mu_1\cdots\mu_k})^{\alpha\beta}\,. 
		\end{equation}
		The matrix $\Gamma$ either commutes or anticommutes with the basis elements $C\Gamma^{\mu_1\cdots\mu_k}$ for the expansion:
		\be
		\Gamma^T\bigl( C\Gamma^{\mu_1\cdots\mu_k}) =    (-1)^{k+1}  
		\bigl( C\Gamma^{\mu_1\cdots\mu_k}\bigr) \Gamma\, ,  \ \ \ \  d = 10\,, 
		\ee
		as can be verified using the last equation of~\refb{CGamm}, and  the first  in~\refb{property}. It follows that a bispinor $X$ with $\eta_X=1$ must have an expansion with odd $k$, and 
		a bispinor $X$ with $\eta_X = -1$ must have an expansion with even $k$.  Therefore
		\begin{subequations}\label{expansion}
			\begin{align}
				& A^{ab} = \sum_{k\in \hat{\mathbb{Z}}_{odd}}\frac{1}{k!}A^{(k)}_{\mu_1\cdots\mu_k}(C\Gamma^{\mu_1\cdots\mu_k})^{ab}\,,\\
				& B^{a\Dot{b}} = \sum_{k\in \hat{\mathbb{Z}}_{even}}\frac{1}{k!} 
				B^{(k)}_{\mu_1\cdots\mu_k}(C\Gamma^{\mu_1\cdots\mu_k})^{a\Dot{b}}
				\,,  \\
				& C^{\Dot{a}b} = \sum_{k\in \hat{\mathbb{Z}}_{even}} \frac{1}{k!}
				C^{(k)}_{\mu_1\cdots\mu_k}(C\Gamma^{\mu_1\cdots\mu_k})^{\Dot{a}b}\,, \\
				& D^{\Dot{a}\Dot{b}} = \sum_{k\in \hat{\mathbb{Z}}_{odd}}\frac{1}{k!}D^{(k)}_{\mu_1\cdots\mu_k}(C\Gamma^{\mu_1\cdots\mu_k})^{\Dot{a}\Dot{b}}\,. 
			\end{align}
		\end{subequations}
		To determine the duality conditions on the bispinor components consider
		the first equation in~\refb{property} with $d=10$ and $n\equiv d-k$: 
		\begin{equation}
			\Gamma^{\mu_1\cdots\mu_k}\Gamma = -(-1)^{k(k-1)/2}\frac{1}{n!}\, \epsilon^{\mu_1\cdots\mu_k}_{\hskip25pt\nu_1\cdots\nu_{n}}\Gamma^{\nu_1
				\cdots\nu_{n}}  \,, \ \   d = k+n = 10 \,. 
		\end{equation}
		We then have
		\be
		\begin{split}
			X \Gamma = & \  \sum_k -(-1)^{k(k-1)/2}
			\frac{1}{n!}\, \frac{1}{k!} \, \epsilon^{\mu_1\cdots\mu_k}_{\hskip25pt\nu_1\cdots\nu_{n}}
			X^{(k)}_{\mu_1\cdots\mu_k}\  C\Gamma^{\nu_1\cdots\nu_{n}}\\
			= & \  \sum_n  \frac{1}{n!}\,  \Bigl[\, (-1)^{n(n+1)/2}
			\frac{1}{(d-n)!} \, \epsilon^{\mu_1\cdots\mu_{d-n}}_{\hskip35pt\nu_1\cdots\nu_{n}}X^{(k)}_{\mu_1\cdots\mu_{d-n}} \Bigr] C\Gamma^{\nu_1\cdots\nu_{n}}\\
			= & \  \sum_n  \frac{1}{n!}\,  \Bigl[\,    (-1)^{n(n-1)/2}
			(* X^{(d-n)} )_{\nu_1 \cdots \nu_n} \Bigr] C\Gamma^{\nu_1\cdots\nu_{n}}\,,
		\end{split}
		\ee
		where we replaced the sum over $k$ by a sum over $n= d-k$ in the second equality. 
		Introducing the sign factor $\e_X$ as follows 
		\be
		X =  \e_X  \, X \, \Gamma \, ,
		\ee  
		we now see that
		\be
		X^{(n)}_{\nu_1 \cdots \nu_n}  =   (-1)^{n(n-1)/2} \epsilon_X
		(* X^{(d-n)} )_{\nu_1 \cdots \nu_n} \,.
		\ee
		Dropping the indices, we have the duality conditions
		\be
		\label{duality-forms}
		* X^{(n)}  =  (-1)^{n(n-1)/2} \epsilon_X \  X^{(d-n)}\,. 
		\ee
		These conditions are consistent with the application of duality for $d=10$.
		For our bispinors we have $\e_A = 1 \,,  \e_B = -1  \,,  \e_C = 1,\hbox{and} \ 
		\e_D = -1 $.
		We thus have
		\be\label{dl}
		\begin{split}
			* A^{(n)}  = \ & \ \ \ \   (-1)^{n(n-1)/2} \  A^{(d-n)}   \,, \hskip30pt    * B^{(m)}  = \   -  (-1)^{m(m-1)/2} \  B ^{(d-m)}\, ,\\[0.4ex]
			* C^{(m)}  = \  & \ \ \ \  (-1)^{m(m-1)/2}  \  C^{(d-m)} \,, \hskip30pt * D^{(n)}  = \   -    (-1)^{n(n-1)/2}  \  D^{(d-n)}.
		\end{split}
		\ee
		where $n\in \hat{\mathbb{Z}}_{odd}$ and $m\in \hat{\mathbb{Z}}_{even}$.
		In type IIB we had bispinors $Q, \bar P, P,$ and $N$ of type $A,B,C,$ and $D$, respectively. In type IIA, we instead have $Q, \bar P, P,$ and $N$ of type $B,A,D,$ and $C$ respectively.
		
		\sectiono{BRST and PCO action on states}
		In here we record BRST action on states, as required in the main text computations.
		\be
		\label{BRSTact01}
		\begin{split}
			Q(c e^{-3\phi/2}\Theta_\alpha  e^{ip\cdot X}) 	&=\tfrac{1}{4} p^2 \partial c\, c e^{-3\phi/2}\Theta_\alpha e^{ip\cdot X}\, ,\\
			Q(c e^{-\phi/2}\Theta_\alpha  e^{ip\cdot X}) 	&=\tfrac{1}{4} p^2\partial c\, c e^{-\phi/2}\Theta_\alpha e^{ip\cdot X} +\tfrac{1}{2}\slashed{p}_\alpha^{\hskip 5pt \beta} c\eta e^{\phi/2}\Theta_\beta  e^{ip\cdot X}\, ,\\
			Q(\partial c\, c e^{-3\phi/2}\Theta_\alpha e^{ip\cdot X}) &=0\, ,\\
			Q(c\partial \xi e^{-5\phi/2} \Theta_\alpha  e^{ip\cdot X}) &=\tfrac{1}{4} p^2\partial c\,c\partial \xi e^{-5\phi/2} \Theta_\alpha  e^{ip\cdot X} +\tfrac{1}{2}\slashed{p}_\alpha^{\hskip 5pt \beta} ce^{-3\phi/2}\Theta_\beta  e^{ip\cdot X}\, ,\\
			Q(\partial c\, c\partial \xi e^{-5\phi/2} \Theta_\alpha  e^{ip\cdot X}) &= -\tfrac{1}{2}\slashed{p}_\alpha^{\hskip 5pt \beta} \partial c\, ce^{-3\phi/2}\Theta_\beta  e^{ip\cdot X}\, .
		\end{split}
		\ee
		\be
		\label{3irfne}
		\begin{split}
			Q(c\Bar{c}e^{-\phi/2}\Theta_{a}  \, e^{-\bar \phi/2}\Bar{\Theta}_b\, e^{ip\cdot X}) 
			&= \tfrac{1}{4} p^2(\partial c+\bar\partial\bar c) c \bar c e^{-\phi/2}\Theta_{a}  
			\, e^{-\bar \phi/2}\Bar{\Theta}_b\, e^{ip\cdot X}\\
			&
			\hskip7pt 
			- \tfrac{1}{2}\slashed{p}_{a}^{\hskip 5pt \Dot{c}} c \bar c\,  \eta e^{\phi/2}\Theta_{\dot c}  \, e^{-\bar \phi/2}\Bar{\Theta}_b\, e^{ip\cdot X}\\
			&  \hskip7pt  - \tfrac{1}{2} \slashed{p}_{b}^{\hskip 5pt \Dot{d}}c\bar c \bar \eta e^{\phi/2}\Theta_{a}  \, e^{-\bar \phi/2}\Bar{\Theta}_{\dot d} \, e^{ip\cdot X}.
		\end{split}
		\ee
		
		\be
		\label{brstforgt} 
		\begin{split}
			Q(c  \Bar{c}\, \bar \partial \bar\xi \, 
			e^{-3\phi/2}\Theta_{\dot a}  \,  e^{-5\bar \phi/2}\Bar{\Theta}_{ b}\, e^{ip\cdot X} ) 
			& = \tfrac{1}{4}p^2(\partial c +\bar \partial \bar c)\, c  \Bar{c}\, \bar \partial \bar\xi \, 
			e^{-3\phi/2}\Theta_{\dot a}  \,  e^{-5\bar \phi/2}\Bar{\Theta}_{ b}\, e^{ip\cdot X} \\
			&  - \tfrac{1}{2}\slashed{p}_b^{\hskip 5pt \Dot{d}}c  \Bar{c}\,  
			e^{-3\phi/2}\Theta_{\dot a}  \,  e^{-3\bar \phi/2}\Bar{\Theta}_{ \dot d}\, e^{ip\cdot X} \, , \\[0.5ex]
			Q(c  \Bar{c}\,  \partial \xi \, 
			e^{-5\phi/2}\Theta_a  \,  e^{-5\bar \phi/2}\Bar{\Theta}_{ \dot b}\, e^{ip\cdot X} ) 
			& = \tfrac{1}{4}p^2(\partial c +\bar \partial \bar c)\, c  \Bar{c}\,  \partial \xi \, 
			e^{-5\phi/2}\Theta_a  \,  e^{-5\bar \phi/2}\Bar{\Theta}_{ \dot b}\, e^{ip\cdot X} \\
			&  - \tfrac{1}{2}\slashed{p}_a^{\hskip 5pt \Dot{c}}c  \Bar{c}\, 
			e^{-3\phi/2}\Theta_{\dot c}  \,  e^{-3\bar \phi/2}\Bar{\Theta}_{ \dot b}\, e^{ip\cdot X}\, , \\[0.5ex]
			Q(\tfrac{1}{2}(\partial c + \bar \partial \bar c)c  \Bar{c}\, \partial \xi\ \bar \partial \bar\xi \, 
			e^{-5\phi/2}\Theta_{ a}  \,  e^{-5\bar \phi/2}\Bar{\Theta}_{ b}\, e^{ip\cdot X} ) 
			&= \tfrac{1}{2}\slashed{p}_a^{\hskip 5pt \Dot{c}} (\partial c +\bar \partial \bar c)\, c  \Bar{c}\, \bar \partial \bar\xi \, 
			e^{-3\phi/2}\Theta_{\dot c}  \,  e^{-5\bar \phi/2}\Bar{\Theta}_{ b}\, e^{ip\cdot X} \\
			& - \tfrac{1}{2}\slashed{p}_b^{\hskip 5pt \Dot{d}}(\partial c +\bar \partial \bar c)\, c  \Bar{c}\,  \partial \xi \, 
			e^{-5\phi/2}\Theta_a  \,  e^{-5\bar \phi/2}\Bar{\Theta}_{ \dot d}\, e^{ip\cdot X}\, .
		\end{split}
		\ee
		A couple of picture changing actions on states:
		\be
		\label{dkjfiue=}
		\begin{split}
			{\cal X}_0(c e^{-3\phi/2}\Theta_\alpha  e^{ip\cdot X}) = &\ 
			\tfrac{1}{2}\, \slashed{p}_{\alpha}^{\hskip 5pt \beta}c \, e^{-\phi/2}
			\Theta_{\beta } e^{ip\cdot X}\, ,\\[0.8ex]
			{\cal X}_0(\partial c \, c e^{-3\phi/2}\Theta_\alpha  e^{ip\cdot X}) =& \ 
			\tfrac{1}{2}\, \slashed{p}_{\alpha}^{\hskip 5pt \beta} \partial c\, c e^{-\phi/2}
			\Theta_{\beta } e^{ip\cdot X}+ c \eta e^{\phi/2}\Theta_\alpha  e^{ip\cdot X} \,.
		\end{split}
		\ee
		
		\noindent
		We now give a set of two-point functions used in computing the free RR action. 
		For all of them one uses the normalization from~\refb{normalization}.
		\be\label{overlap}
		\begin{split}
			&\langle I \circ   c\Bar{c}
			e^{-3\phi/2}\Theta_{\dot a}  \, e^{-3\bar \phi/2}\Bar{\Theta}_{\dot b}
			\, e^{ip'\cdot X}(0) \,
			\partial c\bar\partial\bar c\,c\Bar{c}
			\,e^{-\phi/2}\Theta_c  \, 
			e^{-\bar \phi/2}\Bar{\Theta}_d\,e^{ip\cdot X}(0)\rangle 
			= C_{\Dot{a}c}C_{ \Dot{b} d}(2\pi)^D\delta^{(D)}(p+p')\, ,\\[1.0ex]
			&\langle I \circ \, \tfrac{1}{2} 
			(\partial c+\bar \partial\bar c) c\Bar{c}\,\bar \partial  \bar\xi\, 
			e^{-3\phi/2}\Theta_{\dot a}  
			\,  e^{-5\bar  \phi/2}\Bar{\Theta}_{ b}e^{ip'\cdot X})(0)
			(\partial c -\bar \partial \bar c) c\Bar{c}\bar \eta \,e^{-\phi/2}
			\Theta_a  \,  e^{\bar \phi/2}\Bar{\Theta}_{ \dot b}\, e^{ip\cdot X}(0)\rangle\\
			&\hskip320pt=  C_{\Dot{a}c}C_{b\Dot{d}}(2\pi)^D\delta^{(D)}(p+p')\, ,\\[1.0ex]
			& \langle I \circ \tfrac{1}{2} (\partial c+\bar \partial\bar c) c\Bar{c}\, \partial    \xi\, 
			e^{-5\phi/2}\Theta_a  \,  
			e^{-3\bar \phi/2}\Bar{\Theta}_{ \dot b} e^{ip'\cdot X}(0)
			(\partial c -\bar \partial \bar c) c\Bar{c}\eta \,e^{\phi/2}\Theta_{\dot a}  \,  e^{-\bar \phi/2}\Bar{\Theta}_{ b}\,  e^{ip\cdot X}(0)\rangle\\
			&\hskip320pt = C_{a \Dot{c}}C_{\Dot{b}d}(2\pi)^D\delta^{(D)}(p+p')\, ,\\[1.0ex]
			&\langle I \circ
			c\Bar{c} \  e^{-\phi/2}\Theta_a  \, e^{-\bar \phi/2}\Bar{\Theta}_b\,e^{ip'\cdot X}(0)\,  
			\partial c\bar\partial\bar c\,c\Bar{c}\,e^{-3\phi/2}\Theta_{\dot c}  \, e^{-3\bar \phi/2}\Bar{\Theta}_{\dot d}\, e^{ip\cdot X}(0)\rangle = C_{a \Dot{c}}C_{b \Dot{d}}(2\pi)^D\delta^{(D)}(p+p')\, .
		\end{split}
		\ee
		Consider the manipulations associated with the simplification  
		of~\refb{firstversionofaction}: 
		\be
		\label{cross-terms} 
		\begin{split}
			p^2 P_{a \dot b }(-p)(\slashed{p}^TN(p))^{a\dot b} & = p^2 P_{\dot c d}(-p)(\slashed{p}^TN(p))^{a\dot b}C_{\Dot{c}a}C_{d\Dot{b}}=  p^2 \trace\left[{P^T(-p)C^{-1}(\slashed{p}^TN(p)) C^{-1}}\right]\\
			& = -p^2 \trace\left[{P^T(-p)\slashed{p} C^{-1}N(p) (C^{-1})^T}\right] \\
			& = - p^2 \trace\left[{(\slashed{p}^T P(-p))^TC^{-1}N(p) (C^{-1})^T}\right]\\
			& =  - p^2 \trace\left[{(\slashed{p}^T P(-p))^T(C^{-1})^T N(p) C^{-1}}\right]\, .
		\end{split}
		\ee
		The first equality follows by commuting the charge conjugation across $\slashed{p}$ as in equation~\refb{CGamm}. The second equality is rewriting the transpose and the last inequality follows from the antisymmetry of the charge conjugation matrix. The last line is recognized to give the same contribution as the other $PN$ term in~\refb{firstversionofaction} upon changing $p\rightarrow -p$ in the integration variable. 
		\sectiono{Counting degrees of freedom}\label{sourcemethod}
		
		We will confirm now the nature of the spectrum of the theory, and the 
		statement that the doubled spectrum contains positive norm states that
		experience the interactions, and negative norm states.  
		
		As a warmup consider the case of a $C^{(k)}$ form with the familiar, positive
		norm action 
		\be
		S =  -\tfrac{1}{2}  ( \d  C^{(k)},\d C^{(k)})  +  (C^{(k)} , J^{(k)})	\,, 
		\ee
		where $J^{(k)}$ is the associated current.   Variying $C^{(k)}$ we get the equation
		of motion 
		\be
		\label{30rudlk} 
		\d^\dagger \d  C^{(k)} =  J^{(k)}\quad \to \quad    \d^\dagger J^{(k)} = 0\,.
		\ee
		The gauge invariance $\delta C^{(k)} = \d \Lambda^{(k-1)}$  allows
		us to impose the gauge fixing condition $\d ^\dagger C^{(k)} = 0$.  This implies
		that $\d^\dagger \d C^{(k)} = -\partial^2C^{(k)}$ and then
		the equation of motion is solved by 
		\be
		C^{(k)} =  - {1\over \partial^2} J^{(k)}\,.
		\ee	
		The action becomes
		\be
		\label{actionbecoes}
		S = -\tfrac{1}{2} \, \bigl(  J^{(k)},  \tfrac{1}{\partial^2}   J^{(k)}  \bigr) \,, \ \  \hbox{with} \ \  \ 
		\d^\dagger J^{(k)} = 0 \,.
		\ee
		When the current satisfies the subsidiary condition 
		$\d^\dagger J^{(k)} =0$,   
		this propagates the correct degrees of freedom.  Indeed, in
		momentum space, and with the explicit expression for the inner product, 
		\be
		\label{kformsac}
		S = \tfrac{1}{2}  \tfrac{1}{k!} \int \d^d p \ 
		J^{(k)\mu_1 \cdots \mu_k} (-p)\,     J^{(k)}_{\ \mu_1 \cdots \mu_k} (p)\,  \frac{1}{p^2}  \, \,  , 
		\ee
		the condition 
		$\d^\dagger J^{(k)} =0$     
		implies that near the pole $p^2=0$ and
		taking $p^\mu = (p^0, 0 , \cdots , p^0),$  we have 
		$J_{0, \mu_2 \cdots \mu_k} = - J_{d , \mu_2 \cdots \mu_k}$.  Given that indices $0$ and $d$ can only appear once in the currents, this shows that such contributions vanish
		in the contraction~\refb{kformsac}. As a result, only the transverse components of the current survive, and these describe the propagation mediated by the degrees of freedom
		of the massless $k$-form gauge field. 
		
		Now we can focus on the case of IIB SFT.   
		Coupling the fields $Q^{(k)}, P_+^{(k+1)}$ and $P_-^{(k-1)}$ to the current sources $J^{(k)}, J_+^{(k+1)}$ and $J_-^{(k-1)}$ respectively, we can identify the propagating degrees of freedom and see which ones have negative norm.  The action is  
		\be
		\begin{split} 
			\hskip-10pt 2^{1-d/2} S_2\bigl|_{\RR}&  = \hskip-2pt \sum_{k\in \hat{\mathbb{Z}}_{odd}} \hskip-3pt\Bigl[ \, \tfrac{1}{2}(\d P_-^{(k-1)}, \d P_-^{(k-1)})  - \tfrac{1}{2}(\d^\dagger P_+^{(k+1)}, \d^\dagger P_+^{(k+1)}) 
			-(Q^{(k)}, \d P_-^{(k-1)}+ \d^\dagger P_+ ^{(k+1)})\\
			& \hskip40pt  +  (Q^{(k)},J^{(k)} )+(P_+^{(k+1)}, J_+^{(k+1)})+ (P_-^{(k-1)},J_-^{(k-1)} )\Bigr].
		\end{split}
		\ee
		For consistency, the duality constraints on the $Q$'s and $P_\pm$'s require
		the sources to satisfy associated duality conditions.  
		We want to solve the equations of motion and write the action in terms of the currents.
		The equations of motion with the current source are
		\be
		\begin{split}
			\d^\dagger \d P_-^{(k-1)} - \d^\dagger Q^{(k)} + J_-^{(k-1)} & \, = 0\,,\\
			-  \d \d^\dagger P_+ ^{(k+1)} - \d Q^{(k)} + J_+^{(k+1)} & \, = 0\,, \\
			- \d P_-^{(k-1)}- \d^\dagger P_+ ^{(k+1)}\  + J^{(k)} \ & \, = 0\,. 
		\end{split}
		\ee
		From the first and second equations follows that $\d J_+^{(k+1)} = 0$ and $\d^\dagger J_-^{(k-1)} = 0$, confirming that the gauge fields $P_\pm$ couple to conserved currents. Applying $\d$ and $\d^\dagger$ on the third equation, we get, respectively,
		\be
		\begin{split}
			-\d^\dagger \d P_-^{(k-1)} + \d^\dagger J^{(k)} = & \ 0\,, \\
			-\d \d^\dagger P_+ ^{(k+1)} + \d J^{(k)}\,  = & \ 0\,. 
		\end{split}
		\ee
		Given the gauge symmetries $\delta P_-^{(k-1)} = \d \Lambda_-^{(k-2)}$
		and $\delta P_+^{(k+1)} = \d^\dagger \Lambda_+^{(k+2)}$, we
		can choose the gauges in which $\d^\dagger P_-^{(k-1)} =0$ and 
		$\d P_+ ^{(k+1)}=0$,  
		and solve for the gauge fields in terms of currents:
		\be
		\begin{split}
			\partial^2 P_-^{(k-1)} + \d^\dagger J^{(k)} = 0 \quad \to \quad &\   P_-^{(k-1)} = -\frac{1}{\partial^2}\d^\dagger J^{(k)} \,, \\
			\partial^2 P_+ ^{(k+1)} + \d J^{(k)} = 0\quad \to \quad& \   P_+^{(k+1)} = -\frac{1}{\partial^2}\d J^{(k)} \,. 
		\end{split}
		\ee
		Since the action is linear in $Q^{(k)}$, its equation of motion sets the terms 
		in the action involving $Q^{(k)}$ equal to zero. The action in terms of sources becomes
		\be
		\begin{split} 
			\hskip-10pt 2^{1-d/2} S_2\bigl|_{\RR}  = \hskip-7pt \sum_{k\in \hat{\mathbb{Z}}_{odd}} \hskip-3pt\Bigl[ &\tfrac{1}{2}\Bigl(\frac{1}{\partial^2}\d \d^\dagger J^{(k)},\frac{1}{ \partial^2}\d \d^\dagger J^{(k)}\Bigr)  - \tfrac{1}{2}\Bigl(\frac{1}{\partial^2}\d^\dagger \d J^{(k)} , \frac{1}{\partial^2}\d^\dagger \d J^{(k)} \Bigr)\\
			&  -\Bigl(\frac{1}{\partial^2}\d J^{(k)} , J_+^{(k+1)}\Bigl)- \Bigl(\frac{1}{\partial^2}\d^\dagger J^{(k)},J_-^{(k-1)} \Bigr)\Bigr]\,,   
		\end{split}
		\ee
		where we have used that $\partial^2$ commutes with both $\d$ and $\d^\dagger$.  We can further simplify the action. For example, the first term is simplified as follows
		\be
		\begin{split}
			\Bigl(\frac{1}{\partial^2}\d \d^\dagger J^{(k)},\frac{1}{ \partial^2}\d \d^\dagger J^{(k)})
			\Bigr)& \, = \Bigl(\frac{1}{\partial^2}\d^\dagger\d \d^\dagger J^{(k)},\frac{1}{ \partial^2} \d^\dagger J^{(k)}\Bigr)  = -(\frac{1}{\partial^2}\partial^2 \d^\dagger J^{(k)},\frac{1}{ \partial^2}\d \d^\dagger J^{(k)}) \\
			& \, = -  \Bigl( \d^\dagger J^{(k)},\frac{1}{ \partial^2} \d^\dagger J^{(k)}\Bigr) \,. 
		\end{split}
		\ee
		Doing likewise with the second term, the action becomes
		\be
		\begin{split} 
			\hskip-10pt 2^{1-d/2} S_2\bigl|_{\RR}  = \hskip-7pt \sum_{k\in \hat{\mathbb{Z}}_{odd}} \hskip-3pt\Bigl[& -\tfrac{1}{2} \Bigl( \d^\dagger J^{(k)},\frac{1}{ \partial^2} \d^\dagger J^{(k)}\Bigr) +  \tfrac{1}{2}\Bigl( \d J^{(k)} , \frac{1}{\partial^2} \d J^{(k)} 
			\Bigr)\\
			&  -\Bigl(\frac{1}{\partial^2}\d J^{(k)} , J_+^{(k+1)}\Bigr)- \Bigl(\frac{1}{\partial^2}\d^\dagger J^{(k)},J_-^{(k-1)} \Bigr)\Bigr]\, .
		\end{split}
		\ee
		We can further modify the action to a convenient form by using the identity $\d^\dagger\d = -\partial^2 - \d \d^\dagger$ for the second term on the first line above:
		\be
		\Bigl( \d J^{(k)} , \frac{1}{\partial^2} \d J^{(k)} \Bigr) = \Bigl( J^{(k)} , \frac{1}{\partial^2} \d^\dagger \d J^{(k)} \Bigr) ) = -\Bigl(J^{(k)} , \frac{1}{\partial^2} \partial^2J^{(k)}\Bigr) - \Bigl( J^{(k)} ,\frac{1}{\partial^2} \d \d^\dagger J^{(k)}\Bigr)\,.
		\ee
		The action becomes  
		\be
		\begin{split} 
			\hskip-10pt 2^{1-d/2} S_2\bigl|_{\RR}  = 
			\hskip-7pt \sum_{k\in \hat{\mathbb{Z}}_{odd}} \hskip-3pt
			\Bigl[& -\Bigl( \d^\dagger J^{(k)},\frac{1}{ \partial^2} \d^\dagger J^{(k)}\Bigr) -  \tfrac{1}{2}(J^{(k)} ,J^{(k)} )
			\\& 
			-\Bigl(\frac{1}{\partial^2}\d J^{(k)} , J_+^{(k+1)}\Bigr)- \Bigl(\frac{1}{\partial^2}\d^\dagger J^{(k)},J_-^{(k-1)} \Bigr)\Bigr].
		\end{split}
		\ee
		We had $\d J_+^{(k+1)} = 0$ and $\d^\dagger J_-^{(k-1)} = 0$ which implies that we can write locally $ J_+^{(k+1)} = \d \tilde J_+^{(k)}$ and 
		$J_-^{(k-1)} = \d^\dagger \tilde J_-^{(k)}$   
		for some $\tilde J_+^{(k)}$ and $\tilde J_-^{(k)}$ both $k$-forms. In terms of these new forms the action is
		\be
		\begin{split} 
			\hskip-10pt 2^{1-d/2} S_2\bigl|_{\RR}  = \hskip-7pt \sum_{k\in \hat{\mathbb{Z}}_{odd}} \hskip-3pt\Bigl[& -\Bigl( \d^\dagger J^{(k)},\frac{1}{ \partial^2} \d^\dagger J^{(k)}\Bigr) -  \tfrac{1}{2}(J^{(k)} ,J^{(k)} )\\
			&  -\Bigl(\frac{1}{\partial^2}\d J^{(k)} , \d \tilde J_+^{(k)}\Bigr)- 
			\Bigl(\frac{1}{\partial^2}\d^\dagger J^{(k)},\d ^\dagger \tilde J_-^{(k)} 
			\Bigr)\Bigr]\,. 
		\end{split}
		\ee
		The first term on the second line can be simplified
		\be
		\begin{split}
			-\Bigl(\frac{1}{\partial^2}\d J^{(k)} , \d \tilde J_+^{(k)}\Bigr)
			=   \Bigl(\frac{1}{\partial^2}\d^\dagger J^{(k)} , \d^\dagger \tilde J_+^{(k)}\Bigr)
			+ (J^{(k)} ,\tilde J_+^{(k)} ) \,.
		\end{split}
		\ee
		This results in 
		\be
		\begin{split} 
			\hskip-10pt 2^{1-d/2} S_2\bigl|_{\RR}  
			=  \hskip-7pt \sum_{k\in \hat{\mathbb{Z}}_{odd}} \hskip-3pt\Bigl[& -\Bigl( \d^\dagger J^{(k)},\frac{1}{ \partial^2} \d^\dagger J^{(k)}\Bigr) -  \tfrac{1}{2}(J^{(k)} ,J^{(k)} )\\& 
			+ \Bigl(\frac{1}{\partial^2}\d^\dagger J^{(k)} , \d^\dagger[\tilde J_+^{(k)}- \tilde J_-^{(k)} ]\Bigr) + (J^{(k)} ,\tilde J_+^{(k)} ) \Bigr]\,. 
		\end{split}
		\ee
		We can diagonalize the sources to find
		\be
		\label{woijweoirh}
		\begin{split} 
			\hskip-10pt 2^{1-d/2} S_2\bigl|_{\RR}  
			=   \hskip-7pt \sum_{k\in \hat{\mathbb{Z}}_{odd}} \hskip-3pt\Bigl[& -\Bigl(\,  \d^\dagger \bigl[J^{(k)} -  \tfrac{1}{2}(  \tilde J_+^{(k)}- \tilde J_-^{(k)})\bigr], \frac{1}{ \partial^2} \d^\dagger \bigl[J^{(k)} -  \tfrac{1}{2}(  \tilde J_+^{(k)}- \tilde J_-^{(k)})
			\bigr] \Bigr) + \tfrac{1}{2}(J^{(k)} ,J^{(k)} )\\
			&  + \tfrac{1}{4}\Bigl( \d^\dagger \bigl[  \tilde J_+^{(k)}- \tilde J_-^{(k)})\bigr]\, , \frac{1}{ \partial^2} \d^\dagger \bigl[\tilde J_+^{(k)}- \tilde J_-^{(k)}\bigr] \Bigr) + (J^{(k)} ,\tilde J_+^{(k)} ) \Bigr]\,. 
		\end{split}
		\ee
		Note that, consistent with the example worked out at the start of this section and
		equation~\refb{actionbecoes},
		the currents involved as residues at $\partial^2 = 0$ are of rank $k-1$ and 
		are manifestly
		killed by $\d^\dagger$.  Therefore, 
		apart from the regular terms that have no effect, the first line shows the positive
		norm states of a degree  $k-1$ form gauge field, and the second line shows 
		the negative norm states of a degree  $k-1$ form gauge field.  The duality constraints
		on the currents imply that the independent ones arise from 
		$k = 1,3,$ and $5$.  This is what was
		expected. 
		
		Coupling the IIA fields to sources and doing a similar computation, we get the same pole structure as in~\refb{woijweoirh} up to some sign factors but now with 
		$k \in \hat {\mathbb{Z}}_{even}$. For $k=0$, the terms involving $\d^\dagger$ are identically zero, and there is no propagation, implying that in type IIA neither 
		$Q^{(0)}$ nor $P_-^{(1)}$ are propagating.  For $k=2$ and $k=4$, we see that the propagating degrees of freedom are two one-form gauge fields and two three-form gauge fields, respectively.

	\end{document}